\documentclass[11pt,a4paper]{article}
\pdfoutput=1 

\usepackage{jheppub} 

\usepackage[T1]{fontenc} 
\usepackage{amsmath,mathrsfs,xspace,mathtools,bm}
\usepackage[dvipsnames]{xcolor}
\usepackage{comment,booktabs}
\usepackage{blindtext,multirow}
\usepackage[normalem]{ulem}

\usepackage{array}
\newcolumntype{L}{>{$}l<{$}}
\newcolumntype{C}{>{$}c<{$}}
\newcolumntype{R}{>{$}r<{$}}

\newcommand{\lagr}{\mathscr{L}}
\newcommand{\like}{\mathcal{L}}
\newcommand{\calM}{\mathcal{M}}
\newcommand{\ovr}{\overline}

\newcommand{\fermi}{\emph{Fermi}-LAT\xspace}
\newcommand{\planck}{\emph{Planck}\xspace}
\newcommand{\micro}{\textsf{micrOMEGAs v5.0.8}\xspace}

\newcommand{\multinestver}{\textsf{MultiNest v3.10.0}\xspace}
\newcommand{\multinest}{\textsf{MultiNest}\xspace}
\newcommand{\feynrules}{\textsf{FeynRules}\xspace}
\newcommand{\pippi}{\textsf{pippi v2.0}\xspace}
\newcommand{\HB}{\textsf{HiggsBounds}\xspace}
\newcommand{\HBver}{\textsf{HiggsBounds v5.3.2beta}\xspace}
\newcommand{\HS}{\textsf{HiggsSignals}\xspace}
\newcommand{\HSver}{\textsf{HiggsSignals v2.2.3beta}\xspace}

\newcommand{\tu}[1]{\begin{color}{Cyan} To be updated! \end{color}} 

\makeatletter
\newcommand\footnoteref[1]{\protected@xdef\@thefnmark{\ref{#1}}\@footnotemark}
\makeatother

\title{Global fit of pseudo-Nambu-Goldstone Dark Matter}

\author[a]{Chiara Arina,}
\author[a,1]{Ankit Beniwal,\note{ORCID ID: \href{https://orcid.org/0000-0003-4849-0611}{0000-0003-4849-0611}}}
\author[a,2]{C\'{e}line Degrande,\note{ORCID ID: \href{https://orcid.org/0000-0002-1605-4586}{0000-0002-1605-4586}}}
\author[a]{Jan Heisig}
\author[b,3]{and Andre Scaffidi\note{ORCID ID: \href{https://orcid.org/0000-0002-1203-6452}{0000-0002-1203-6452}}}

\affiliation[a]{Centre for Cosmology, Particle Physics and Phenomenology (CP3), \\
Universit\'{e} catholique de Louvain, B-1348 Louvain-la-Neuve, Belgium}
\affiliation[b]{ARC Centre of Excellence for Particle Physics at the Terascale (CoEPP) and CSSM, \\
Department of Physics, University of Adelaide, South Australia 5005, Adelaide, Australia}

\emailAdd{chiara.arina@uclouvain.be}
\emailAdd{ankit.beniwal@uclouvain.be}
\emailAdd{celine.degrande@uclouvain.be}
\emailAdd{jan.heisig@uclouvain.be}
\emailAdd{andre.scaffidi@adelaide.edu.au}

\abstract{We perform a global fit within the pseudo-Nambu-Goldstone dark matter (DM) model emerging from an additional complex scalar singlet with a softly broken global $U(1)$ symmetry.~Leading to a momentum-suppressed DM-nucleon cross section at tree level, the model provides a natural explanation for the null results from direct detection experiments. Our global fit combines constraints from perturbative unitarity, DM relic abundance, Higgs invisible decay, electroweak precision observables and latest Higgs searches at colliders. The results are presented in both frequentist and Bayesian statistical frameworks. Furthermore, post-processing our samples, we include the likelihood from gamma-ray observations of \emph{Fermi}-LAT dwarf spheroidal galaxies and compute the one-loop DM-nucleon cross section. We find two favoured regions characterised by their dominant annihilation channel: the Higgs funnel and annihilation into Higgs pairs. Both are compatible with current \emph{Fermi}-LAT observations, and furthermore, can fit the slight excess observed in four dwarfs in a mass range between about 30--300\,GeV. While the former region is hard to probe experimentally, the latter can partly be tested by current observations of cosmic-ray antiprotons as well as future gamma-ray observations.}


\arxivnumber{1912.04008}

\preprint{ADP-19-30/T1110, CP3-19-56}

\begin{document}

\maketitle

\flushbottom

\section{Introduction}
The true particle nature of dark matter (DM) continues to remain a mystery despite a plethora of astrophysical/cosmological evidence to support its existence \cite{Bertone:2010zza}.~Although the well-known Weakly Interacting Massive Particle (WIMP) offers a viable solution, most common models are strongly constrained by direct detection experiments \cite{Akerib:2016vxi,Cui:2017nnn,Aprile:2018dbl}.~This has forced us to either seek alternate particle DM candidates (e.g., axions \cite{Peccei:1977hh,Peccei:1977ur,Weinberg:1977ma,Wilczek:1977pj}, sterile neutrinos \cite{Dodelson:1993je,Shi:1998km}) or explore new ways of saving the canonical `WIMP paradigm'. 

A natural way of achieving the latter is to suppress the DM-nucleon interaction at tree level. For instance, in certain particle DM models, some parameter combinations can lead to blind spots in direct detection experiments or even a suppression of the DM-nucleon coupling \cite{Han:2016qtc,Choudhury:2017lxb,Han:2018gej,Altmannshofer:2019wjb}.~Alternatively, the DM-nucleon couplings could vanish due to symmetries \cite{Xiang:2017yfs,Wang:2017sxx}.~More commonly, however, particle DM models with a pseudoscalar mediator \cite{Fan:2010gt,Fitzpatrick:2012ix,DelNobile:2013sia,Boehm:2014hva,Bauer:2017ota,Tunney:2017yfp} leads to a momentum-suppressed DM-nucleon cross section.~Thus, this class of models can naturally evade the strong limits from direct detection experiments.

A popular example in this regard is the pseudo-Nambu-Goldstone (pNG) DM~\cite{Barger:2008jx,Barger:2010yn,Barducci:2016fue,Gross:2017dan,Balkin:2018tma,Huitu:2018gbc}.~It can be realised by adding a complex scalar singlet with a softly broken global $U(1)$ symmetry to the Standard Model (SM) particle content \cite{Barger:2008jx,Barger:2010yn,Barducci:2016fue,Balkin:2018tma}.~Due to the soft symmetry breaking, the resulting Goldstone becomes massive, i.e., a pNG boson.~An additional $CP$ symmetry ensures the stability of the pNG boson, which serves as a viable DM candidate.~The Goldstone nature of the DM particle implies that the pNG DM-nucleon cross section is momentum-suppressed at tree-level~\cite{Barger:2010yn}.\footnote{In fact, this suppression persists in the general case of $N$ scalars which are symmetric under a global $U(1) \otimes S_N$ symmetry \cite{Karamitros:2019ewv}.}~Thus, a pNG DM model offers a natural way of evading the strong direct detection limits \cite{Barger:2008jx,Barger:2010yn,Gross:2017dan}. A leading-order contribution to the pNG DM-nucleon cross section in the zero-momentum limit appears at the one-loop level \cite{Azevedo:2018exj,Ishiwata:2018sdi}. For typical DM velocities in our galaxy, $v_\chi \sim 10^{-3}$, it can easily dominate over the tree-level contribution.~The one-loop cross section can vary by several orders of magnitude in the allowed model parameter space. It has been shown that for parameter points which satisfy the relic density constraint, the one-loop cross section is typically below $\sim 10^{-50}$~cm$^2$ \cite{Azevedo:2018exj} and thus beyond the expected reach of future direct detection experiments, e.g., LUX-ZEPLIN (LZ) \cite{Akerib:2018lyp} and DARWIN \cite{Aalbers:2016jon}.

More recently, the pNG DM model was confronted against the constraints from perturbative unitarity, DM relic density, Higgs invisible decay, XENON1T, \fermi dwarf spheroidal (dSph) galaxies~\cite{Fermi-LAT:2016uux} and LHC searches at $\sqrt{s}$ = 13 TeV \cite{Azevedo:2018oxv,Alanne:2018zjm,Huitu:2018gbc,Jiang:2019soj,Ruhdorfer:2019utl}.~Projected limits from DARWIN were also imposed in ref.~\cite{Ishiwata:2018sdi}, although they were only slightly stronger than the perturbative unitarity constraint. The \fermi limits in ref.~\cite{Alanne:2018zjm} were computed in an approximate way by considering annihilation into $b\bar b$, (on-shell) $W^+ W^-$, $ZZ$ and $hh$, where the latter three channels were included by applying a re-scaling factor on the $b\bar b$ limit from refs.~\cite{Clark:2017fum,Boddy:2018qur}. In addition, the model has also been used as a testbed for fitting the galactic centre gamma-ray and cosmic-ray antiproton excess \cite{Cline:2019okt}.

The pNG DM model has also been studied in light of electroweak baryogenesis. In ref.~\cite{Kannike:2019wsn}, the authors found that the phase transition in this model is of second-order, and thus a sizable gravitational wave signal is not possible.~However, the situation definitely improves if the model is extended to possess a $\mathbb{Z}_3$ symmetry. In this case, both a strong first-order phase transition and a sizable gravitational wave signal is possible \cite{Kannike:2019mzk}.

In this paper, we perform a global fit of the pNG DM model.~Our likelihood include constraints from perturbative unitarity, DM relic abundance, Higgs invisible decay width, electroweak precision observables, and latest Higgs searches at colliders.~Our results are presented in both frequentist and Bayesian statistical frameworks.~We also post-process our samples by computing the gamma-ray flux and the resulting likelihood from \fermi observations of dwarf Spheroidal galaxies (dSphs). We consider a set of 41 and 45 dSphs, excluding and including, respectively, those that show slight excesses compatible with DM annihilation. We take into account all relevant annihilation channels including annihilation into $HH$ and $hH$ (where $h$ and $H$ are the 125\,GeV Higgs and new scalar, respectively) as well as those proceeding via off-shell vector bosons. In addition, we compute the one-loop pNG DM-nucleon cross section for our samples and compare the resulting values against  the current limits from XENON1T (2018), and projected future limits from LZ and DARWIN.~Our \feynrules \cite{Alloul:2013bka}, \textsf{UFO} \cite{Degrande:2011ua}, \textsf{CalcHEP} \cite{Belyaev:2012qa} and \textsf{FeynArts} \cite{Hahn:2000kx} model files are publicly available at the \feynrules database.\footnote{\url{https://feynrules.irmp.ucl.ac.be/wiki/pNG}}

The remainder of the paper is organised as follows.~In section~\ref{sec:model}, we introduce the pNG DM model. In section~\ref{sec:like_constraints}, we describe the various observables and likelihoods used in our global fit and in the post-processing.~Our numerical scan details, global fit results, including \fermi and direct detection constraints, are presented in section~\ref{sec:num_scans_results}.~We conclude in section~\ref{sec:conclusions}.~Appendices~\ref{app:dm-nucleon} and~\ref{app:oblique_pars} summarise analytic expressions used in this paper.

\section{Pseudo-Nambu-Goldstone Dark Matter}\label{sec:model}
We extend the SM Lagrangian by adding a new complex scalar field $S$ that couples to the SM particles via a \emph{Higgs portal} term, $\Phi^\dagger \Phi$ ($\Phi$ is the SM Higgs doublet).~The model Lagrangian is given by \cite{Gross:2017dan}
\begin{equation}
	\lagr = \lagr_{\textnormal{SM}} + \lagr_S + \lagr_{\textnormal{soft}}, \label{eqn:mdl_lagr} 
\end{equation}
where $\lagr_{\textnormal{SM}}$ is the SM Lagrangian, 
\begin{align}	
	\lagr_S &= (\partial_\mu S)^* (\partial^\mu S) + \frac{\mu_S^2}{2} |S|^2 - \lambda_{\Phi S}\,\Phi^\dagger \Phi |S|^2 - \frac{\lambda_S}{2} |S|^4, \label{eqn:S_part} \\
	\lagr_{\textnormal{soft}} &= \frac{\mu_S'^{2}}{4} (S^2 + S^{*2}). \label{eqn:soft_part}
\end{align}
Notice that eq.~\eqref{eqn:S_part} is invariant under a dark $U(1)$ global symmetry:
\begin{equation}
    S \rightarrow e^{i\alpha} S,    
\end{equation}
where $\alpha$ is a real, space-time independent parameter.~However, the $\mu_S'^2$ term in eq.~\eqref{eqn:soft_part} softly breaks this symmetry.~Thus, the model contains a \emph{massive} Goldstone boson, i.e., a pNG boson.~After this symmetry breaking, we are left with a residual $\mathbb{Z}_2$ symmetry, $S \rightarrow -S$, of the dark $U(1)$ group, which forbids a linear term in $S$ in the above Lagrangians. 

The parameter $\mu_S'^2$ can be made real and positive by the phase redefinition of $S$.~Thus, eq.~\eqref{eqn:mdl_lagr} is invariant under a dark $CP$ symmetry:
\begin{equation}\label{eqn:CP_sym}
    S \rightarrow S^*.
\end{equation}
This symmetry is unbroken by the $S$ vacuum expectation value (VEV) as for positive $\mu_S'^2$, the VEV is real.~Thus, the total symmetry of the model Lagrangian is $\mathbb{Z}_2 \otimes CP$.

With an extra scalar, the scalar potential becomes
\begin{equation}\label{eqn:pot}
	V = V_{\textnormal{SM}} + V_S + V_{\textnormal{soft}},
\end{equation}
where $V_S$ and $V_{\textnormal{soft}}$ can be read directly from eqs.~\eqref{eqn:S_part} and \eqref{eqn:soft_part} respectively. Meanwhile, the SM part of the potential reads
\begin{equation}
	V_{\textnormal{SM}} = -\frac{\mu_\Phi^2}{2} \Phi^\dagger \Phi + \frac{\lambda_\Phi}{2} (\Phi^\dagger \Phi)^2.
\end{equation}

After electroweak symmetry breaking, the model spectrum can be analysed by decomposing $\Phi$ and $S$ in the unitary gauge as
\begin{equation}
	\Phi = \frac{1}{\sqrt{2}}
	\begin{pmatrix}
		0 \\
		v_h + \phi
	\end{pmatrix},
	\quad S = \frac{v_s + s + i\chi}{\sqrt{2}},
\end{equation}
where $v_h$ is the SM Higgs VEV.~Under the dark $CP$ symmetry in eq.~\eqref{eqn:CP_sym}, $\chi \rightarrow -\chi$. This guarantees the stability of $\chi$ and makes it a viable DM candidate; the physical $\chi$ mass is $m_\chi^2 = \mu_S'^2$.

After imposing the stationary point conditions at $(\phi,\,s) = (0,\,0)$, we get
\begin{align}
	\mu_\Phi^2 &= \lambda_\Phi v_h^2 + \lambda_{\Phi S} v_s^2, \\[1mm]
	\mu_S^2 &= \lambda_S v_s^2 + \lambda_{\Phi S} v_h^2 - \mu_S'^2.
\end{align}

Given that the $S$ VEV is non-zero in general, the $\lambda_{\Phi S}$ term in eq.~\eqref{eqn:S_part} leads to a mixing between the $CP$-even interaction eigenstates $(\phi,\,s)$.~Thus, the squared mass matrix $\calM^2$ is non-diagonal, namely
\begin{equation}
	\calM^2 = 
	\begin{pmatrix}
		\lambda_\Phi v_h^2 & \lambda_{\Phi S} v_h v_s \\[1.5mm]
		\lambda_{\Phi S} v_h v_s & \lambda_S v_s^2
	\end{pmatrix}.
\end{equation}
As $\calM^2$ is real and symmetric, it can be diagonalised by a unitary transformation:
\begin{equation}
	 \mathcal{O}^T \calM^2 O = 
	 \begin{pmatrix}
	 	m_h^2 & 0 \\
	 	0 & m_H^2
	 \end{pmatrix},
\end{equation}
where 
\begin{equation}
	\mathcal{O} = 
	\begin{pmatrix}
		 \cos\theta & \sin\theta \\
		-\sin\theta & \cos\theta
	\end{pmatrix}.
\end{equation}
Here $\theta$ is a mixing angle that satisfies the following relation:
\begin{equation}
	\tan 2\theta = \frac{2 \lambda_{\Phi S} v_h v_s}{\lambda_S v_s^2 - \lambda_\Phi v_h^2}.
\end{equation}
The eigenvalues of $\calM^2$ correspond to the masses of the $CP$-even mass eigenstates $(h,\,H)$,
\begin{equation}\label{eqn:m1_m2}
	m_{h,\,H}^{2} = \frac{1}{2} \left[\lambda_\Phi v_h^{2} + \lambda_{S} v_{s}^{2} \mp \left(\frac{\lambda_{S} v_{s}^{2} - \lambda_\Phi v_h^{2}}{\cos 2\theta} \right)\right]\;.
\end{equation}
Given the discovery of a SM-like Higgs at the LHC \cite{Aad:2012tfa,Chatrchyan:2012xdj}, we identify $h$ as a SM-like Higgs boson with
\begin{equation}
	 m_h = 125\;\textnormal{GeV}, \quad v_h = 246
	 \;\textnormal{GeV}.	
\end{equation}
Thus, the pNG DM model contains 4 free parameters:
\begin{equation}\label{eqn:free_pars}
	\left\{ m_\chi, \, v_s, \, \theta, \, m_H \right\}.
\end{equation}
The remaining parameters in eqs.~\eqref{eqn:S_part} and \eqref{eqn:soft_part} can be expressed as
\begin{align}
	\lambda_\Phi &= \frac{1}{v_h^2} \left(m_h^2 \cos^2 \theta + m_H^2 \sin^2 \theta \right), & \mu_S'^2 &= m_\chi^2, \label{eqn:dep-Pars-start} \\[1.2mm]
	\lambda_S &= \frac{1}{v_s^2} \left(m_h^2 \sin^2 \theta + m_H^2 \cos^2 \theta \right), & \mu_\Phi^2 &= \lambda_\Phi v_h^2 + \lambda_{\Phi S} v_s^2, \label{eqn:lamS} \\[1.2mm]
	\lambda_{\Phi S} &= \frac{1}{v_h v_s} \left(m_H^2 - m_h^2 \right) \sin \theta \cos \theta, & \mu_S^2 &= \lambda_{S} v_s^2 + \lambda_{\Phi S} v_h^2 - \mu_S'^2. \label{eqn:lamPS}
\end{align}

\section{Observables and constraints}\label{sec:like_constraints}
In this section, we describe the set of constraints included in our global fit and post-processing of final samples.

\subsection{Theoretical bounds}
We require the model parameters to satisfy the following two theoretical bounds.
\begin{enumerate}
    \item \emph{Bounded tree-level potential}: The tree-level potential in eq.~\eqref{eqn:pot} must be bounded from below. This translates into the following lower bounds:
    \begin{equation}
    	\lambda_\Phi > 0, \quad \lambda_S > 0, \quad \lambda_{\Phi S} > - \sqrt{\lambda_\Phi \lambda_S}.
    \end{equation}

    \item \emph{Perturbative unitarity}: We require the perturbative unitarity of scattering amplitudes \cite{Lee:1977lhj}.~Using the $HH \rightarrow HH$ scattering process, we impose the following upper bound on the $S$ quartic coupling \cite{Chen:2014ask}:
    \begin{equation}\label{eqn:pert_unit}
        \lambda_S < 8\pi/3.
    \end{equation}
    Although this bound can vary with the exact scattering process of interest, we choose this form to maintain comparability with previous studies in literature~\cite{Gross:2017dan,Ishiwata:2018sdi,Huitu:2018gbc,Cline:2019okt}.
\end{enumerate}
Parameter points that do not fulfill these requirements are discarded from our scan. This is formally achieved by assigning a very small likelihood to such points.

\subsection{Thermal relic abundance}
The pNG boson $\chi$ is the DM candidate. Similar to the extended scalar singlet model \cite{Beniwal:2018hyi}, $\chi$ can annihilate into $f\ovr{f}$ (where $f$ = quarks/leptons), $W^+ W^-$, $Z Z$, $hh$, $h H$ and $HH$ final states via an $s$-channel $h/H$ exchange. In addition, $\chi$ annihilation into $h h$, $hH$ and $HH$ final states is also possible via $t$- and $u$-channels via $\chi$ exchange.

In our numerical scans, we require $\chi$ to make up \emph{all} of the observed DM relic abundance.\footnote{In general, $\chi$ can account for a subdominant component of the observed DM relic abundance, i.e., $f_{\textnormal{rel}} \equiv \Omega_\chi/\Omega_{\textnormal{DM}} < 1$. However, this choice (generally) leads to a larger allowed parameter space than the $f_{\textnormal{rel}} = 1$ case \cite{Beniwal:2018hyi}. We adopt the latter choice in our study.}~This is achieved using a Gaussian likelihood function for the DM relic density that is centered at the \planck (2018) measured value \cite{Aghanim:2018eyx}:
\begin{equation}
    \Omega_{\textnormal{DM}} h^2 = 0.120 \pm 0.001.    
\end{equation}
We also include a 5\% theoretical uncertainty and combine it in quadrature with the \planck measured uncertainty. This is done to reflect any uncertainties arising from the relic density calculation in \micro \cite{Belanger:2018mqt}.

With two neutral scalar mediators, the $\chi$ annihilation cross section is resonantly enhanced when $m_\chi \sim m_{h,\,H}/2$. To obtain the correct DM abundance, the $\chi$ annihilation cross section must be sufficiently suppressed. This is achieved for small values of $v_h/v_s$ (or large $v_s$); an expression for the DM-scalar coupling can be found in appendix~\ref{app:dm-nucleon}. Away from these resonances, large values of $v_h/v_s$ (or small $v_s$) generally saturates the $\chi$ relic density to the observed value.

\subsection{Higgs invisible decay width}
When $m_\chi \lesssim m_{h,\,H}/2$, the two scalars $\{h,\,H\}$ are kinematically allowed to decay into a pair of DM particles, i.e., $h,\,H \rightarrow \chi\chi$. This contributes to the following invisible decay widths \cite{Huitu:2018gbc}: 
\begin{align}
    \Gamma_{\textnormal{inv}} (h \rightarrow \chi \chi) &= \frac{1}{32\pi} \frac{m_h^3 \sin^2\theta}{v_s^2} \sqrt{1 - \frac{4 m_\chi^2}{m_h^2}}, \\
    \Gamma_{\textnormal{inv}} (H \rightarrow \chi \chi) &= \frac{1}{32\pi} \frac{m_H^3 \cos^2\theta}{v_s^2} \sqrt{1 - \frac{4 m_\chi^2}{m_H^2}}.
\end{align}

Recently, both the ATLAS \cite{Aaboud:2019rtt} and CMS \cite{Sirunyan:2018owy} experiments released new upper limits on the Higgs invisible branching ratio $\mathcal{BR}(h \rightarrow \chi \chi)$ for a SM-like Higgs from a combination of Run 1 and 2 analyses.~Here we adopt the conservative upper limit from the ATLAS experiment \cite{Aaboud:2019rtt}, namely
\begin{equation}\label{eqn:Gamma_up_limit}
    \mathcal{BR}(h \rightarrow \chi \chi) \equiv \frac{\Gamma_{\textnormal{inv}} (h \rightarrow \chi \chi)}{\Gamma_h^{\textnormal{tot}} (m_h)} \leq 0.26,
\end{equation}
where $\Gamma_h^{\textnormal{tot}} (m_h)$ is the total decay width of $h$ into SM and non-SM final states.

In the following, we apply the limit in eq.~\eqref{eqn:Gamma_up_limit} only on the scalar $h$ whose mass is fixed at 125 GeV. This experimental limit is derived from Higgs production in association with a weak gauge boson or through vector boson fusion (VBF), and assuming a SM-like Higgs except for the fact that it can decay to a pair of invisible particles, e.g., DM. The experimental signatures are large missing energy with either a weak boson or a pair of jets. Thus, the invariant mass of the invisible particles is not measured, and the second scalar can contribute as well to these processes.~However, this contribution is small when the mixing angle is small as the production of the second (first) scalar is suppressed by a factor $\sin ^2 \theta$ ($\cos^2 \theta$) compared to the SM production rate. The production rate also varies with the mass of the scalar but this effect is small for a light scalar ($m_H\sim 100$\,GeV) in the most constraining channel, i.e.,~VBF, as the cuts on the invariant mass of the two jets ($m_{jj}>1$\,TeV) are already requiring a very large partonic center-of-mass energy.~As the second scalar mass increases, its production rate is further reduced.~To sum up, our approximation is only valid for small mixing angles and for $m_H\neq m_h$.~However, as we show in the results section, the mixing angle is allowed to be large and even maximal when the two scalars are degenerate, i.e., when $m_H\sim m_h = 125$\,GeV. In this case, both scalars contribute to the process and interfere quite strongly. The amplitude can be written as
\begin{equation}
    \mathcal{A} \propto \frac{\cos\theta\sin\theta}{v_s}\left(\frac{m_h^2}{p^2-m_h^2+im_h\Gamma_h^\text{tot} }-\frac{m_H^2}{p^2-m_H^2+i m_H \Gamma_H^\text{tot}}\right),
    \label{eq:amp}
\end{equation}
where $\Gamma_h^\text{tot}\,(\Gamma_H^\text{tot})$ are the total decay width of scalars $h\,(H)$. The amplitude for the production of the scalar is multiplied by a factor $\cos\theta$ ($\sin\theta$) compared to the SM due to the modification of the couplings between the gauge bosons and $h$ ($H$). The remaining factors, besides the propagators, are due to the couplings with the DM particle $\chi$ (see appendix~\ref{app:dm-nucleon}). The missing pre-factor in eq.~\eqref{eq:amp} depends only on pure SM couplings and its exact expression depends on the process considered.~The two terms in the brackets can cancel exactly for $\theta=\pi/4$ if the two masses are identical, as in that case, the two widths are also identical. Thus, these points are unconstrained experimentally but many of them would be excluded by applying blindly the constraints on the invisible decay width, which is the dominant channel for low values of $m_\chi$ and $v_s$. However, the correct re-interpretation of the invisible width constraint goes beyond the scope of this paper and thus is left as a future work.

The upper limit in eq.~\eqref{eqn:Gamma_up_limit} constrains the $m_\chi \lesssim m_h/2$ region where $\Gamma_{\textnormal{inv}} (h \rightarrow \chi\chi)$ can be sizeable. In our numerical scans, we use a one-sided Gaussian likelihood function that is centered at the above measured value for $\mathcal{BR}(h \rightarrow \chi \chi)$.~Similar to the relic density likelihood, we add a 5\% theoretical uncertainty from our calculation of $\mathcal{BR}(h \rightarrow \chi \chi)$ and combine it in quadrature with the (expected) branching ratio uncertainty of 0.07 \cite{Aaboud:2019rtt}.

\subsection{Electroweak precision observables}
The extra scalar $S$ contributes to the gauge boson self-energy diagrams. Its effect on the electroweak precision observables (EWPO) can be parametrised by the oblique parameters $S$, $T$ and $U$ \cite{Peskin:1990zt,Peskin:1991sw}. As $S$ is electrically neutral, only the $W$ and $Z$ boson self-energies are modified.

Given that only the real component of $S$ acquires a non-zero VEV, the pNG DM $\chi$ (imaginary component of $S$) does not contribute to the $W/Z$ self-energies.~Thus, the oblique parameters in our model have the same functional dependency as in the extended scalar singlet model \cite{Beniwal:2018hyi}, namely, 
\begin{equation}\label{eqn:mod_oblique}
    \Delta \mathcal{O} \equiv \mathcal{O} - \mathcal{O}_{\textnormal{SM}} = (1 - \cos^2 \theta) \Big[\mathcal{O}_{\textnormal{SM}}(m_H) - \mathcal{O}_{\textnormal{SM}}(m_h)\Big],
\end{equation}
where $\mathcal{O} \in (S,\,T,\,U)$; for the analytical expressions, see appendix~\ref{app:oblique_pars}. From eq.~\eqref{eqn:mod_oblique}, it is clear that for large $m_H$, $\theta \sim 0$ is required, whereas large mixing angles $\theta$ are allowed for $m_H \simeq m_h$ (see e.g., ref.~\cite{Robens:2015gla}).

Using the SM reference as $m_{h}^{\textnormal{ref}} = 125$\,GeV and $m_{t}^{\textnormal{ref}} = 172.5$\,GeV, a recent global electroweak fit obtains \cite{Haller:2018nnx}
\begin{equation}\label{eqn:STU_pars}
    \Delta S = 0.04 \pm 0.11, \quad \Delta T = 0.09 \pm 0.14, \quad \Delta U = -0.02 \pm 0.11,
\end{equation}
and the following correlation matrix:
\begin{equation}\label{eqn:corr_mat}
    \rho_{ij} = 
    \begin{pmatrix}
        1 & 0.92 & -0.68 \\
          0.92 & 1 & -0.87 \\
          -0.68 & -0.87 & 1\\        
    \end{pmatrix}.
\end{equation}
In our numerical scans, we use the following EWPO likelihood function \cite{Profumo:2014opa}:
\begin{equation}
    \ln \like_{\textnormal{EWPO}} = -\frac{1}{2} \sum_{i,\,j} (\Delta \mathcal{O}_i - \ovr{\Delta \mathcal{O}}_i) \left(\Sigma^2 \right)_{ij}^{-1} (\Delta \mathcal{O}_j - \ovr{\Delta \mathcal{O}}_j),
\end{equation}
where $\ovr{\Delta \mathcal{O}}_i$ are the central values for the shifts in eq.~\eqref{eqn:STU_pars}, $\Sigma_{ij}^2 \equiv \sigma_i \rho_{ij} \sigma_j$ is the covariance matrix, $\rho_{ij}$ is the correlation matrix in eq.~\eqref{eqn:corr_mat} and $\sigma_i$ are the associated errors in eq.~\eqref{eqn:STU_pars}.

\subsection{Higgs searches at colliders}
In the narrow-width approximation, the signal strength $\mu_h$ for a SM-like Higgs $h$ \cite{Huitu:2018gbc} is
\begin{equation}\label{eqn:mu_H}
    \mu_h \equiv \sigma(p p \rightarrow h) \cdot \mathcal{BR} (h \rightarrow \textnormal{SM}).
\end{equation}
The inclusion of $H$ in our model leads to a universal suppression of couplings between $h$ and SM particles. Thus, the $h$ production cross section is
\begin{equation}
    \sigma (p p \rightarrow h) = \cos^2 \theta~ \sigma_{p p \rightarrow h}^{\textnormal{SM}} (m_h),
\end{equation}
where $\sigma_{p p \rightarrow h}^{\textnormal{SM}} (m_h)$ is the $h$ production cross section in the SM.~Similarly, the branching ratio of $h$ into SM particles is
\begin{equation}
    \mathcal{BR}(h \rightarrow \textnormal{SM}) \equiv \frac{\Gamma (h \rightarrow \textnormal{SM})}{\Gamma_h^{\textnormal{tot}}} = \frac{\cos^2 \theta ~ \Gamma_h^{\textnormal{SM}}(m_h)}{\cos^2 \theta ~ \Gamma_h^{\textnormal{SM}} (m_h) + \Gamma (h \rightarrow \chi \chi) + \Gamma (h \rightarrow HH) },
\end{equation}
where $\Gamma_h^{\textnormal{SM}} (m_h)$ is the total decay width of a SM-like Higgs $h$ with mass $m_h$ into SM final states.~The last two terms in the denominator correspond to the new decay modes of $h$, namely $h \rightarrow \chi \chi$ and $h \rightarrow HH$. 

Thus, the signal strength $\mu_h$ in eq.~\eqref{eqn:mu_H} becomes
\begin{equation}\label{eqn:mu_h}
    \mu_h = \frac{\cos^4 \theta ~ \mu_h^{\textnormal{SM}} }{\cos^2 \theta ~ \Gamma_h^{\textnormal{SM}}(m_h) + \Gamma (h \rightarrow \chi \chi) + \Gamma (h \rightarrow HH)},
\end{equation}
where $\mu_h^{\textnormal{SM}} \equiv \sigma_{p p \rightarrow h}^{\textnormal{SM}} (m_h) \cdot \Gamma_h^{\textnormal{SM}}(m_h)$ is the $h$ signal strength in the SM. From the above expression, it is clear that $\mu_h \neq \mu_h^{\textnormal{SM}}$ when the mixing angle $\theta \neq 0$, or when $h$ decay into non-SM final states  is kinematically allowed. 

To constrain the parameters of the pNG DM model in our scan, we take into account the following two contributions to the likelihood:
\begin{itemize}
    \item We consider Higgs searches performed at the LEP experiment by utilising the \HBver \cite{Bechtle:2019kbl,Bechtle:2013wla} package. It allows us to compute a $\chi^2_{\textnormal{LEP}}$ for the most-sensitive LEP analysis on the basis of the scalar masses, effective Higgs-SM couplings, total decay widths and branching ratios. The corresponding likelihood function reads:
    \begin{equation}
        \ln \like_{\textnormal{LEP}} = - \frac{1}{2} \chi^2_{\textnormal{LEP}}.
    \end{equation}
    \item We take into account constraints from the observed Higgs signal strengths and mass measurements performed for a SM-like Higgs at the LHC. This is achieved using the \HSver \cite{Bechtle:2013xfa} package. In practice, we compute three contributions to the $\chi^2$ that are based on \emph{i}) combined run 1 results $(\chi^2_{\textnormal{R1}})$; \emph{ii}) results from 13 TeV LHC analyses $(\chi^2_{\textnormal{13\,TeV}})$; and \emph{iii}) results in the form of Simplified Template Cross Sections (STXS) $(\chi^2_{\textnormal{STXS}})$. The final \HS likelihood that we use in our scans is
    \begin{equation}
        \ln \like_{\textnormal{HS}} = -\frac{1}{2} \left(\chi^2_{\textnormal{R1}} + \chi^2_{\textnormal{13\,TeV}} + \chi^2_{\textnormal{STXS}} \right).
    \end{equation}
    Each of the individual chi-squares are computed using the \emph{peak-centered} method with Gaussian probability density function and zero theoretical mass uncertainty for the two scalar masses. For more details, see ref.~\cite{Bechtle:2013xfa}.
\end{itemize}
The LEP analyses that we use are sensitive to parameter points with $m_H \lesssim 120$\,GeV only, whereas the observed Higgs signal strengths potentially constrain points with $m_H$ in the whole considered mass range.\footnote{Note that we do not consider constraints from direct searches for a second (heavy) Higgs performed at the LHC (also implemented in \HBver). Besides the technical limitation that \HB does not provide a likelihood for these searches, we found that the respective 95\% C.L. limits are weaker than those of other constraints applied in our analysis. In particular, the EWPO constraint excludes sizeable values of $\theta$ in the region $m_H\gtrsim 130\,$GeV.}

\subsection{\emph{Fermi}-LAT gamma-ray observations}\label{sec:FermiCon}
Gamma-ray observations of dSphs by \fermi provide robust limits on the DM annihilation flux.~To constrain our model, we use the publicly available energy-binned likelihood profiles\footnote{\url{https://www-glast.stanford.edu/pub_data/1203/}} from \fermi \cite{Fermi-LAT:2016uux}, as implemented in \textsf{MadDM v3.0}~\cite{Ambrogi:2018jqj}.~We consider the set of all 45 observed dSphs and use the measured $J$-factors based on spectroscopic observations~\cite{Fermi-LAT:2016uux} as adopted from ref.~\cite{Geringer-Sameth:2014yza}.~When measurements are not available, we use the values predicted from the distance scaling relationship with a nominal uncertainty of 0.6\,dex~\cite{Fermi-LAT:2016uux}. We profile over the $J$-factor of each dwarf galaxy according to its uncertainty and obtain a total likelihood function as described in ref.~\cite{Ackermann:2015zua}. 

The corresponding gamma-ray energy spectra are computed using \textsf{MadDM}, while showering and hadronisation is achieved using \textsf{Pythia v8.0}~\cite{Sjostrand:2014zea}.~We include all annihilation channels that contribute at least 1\% to the total DM annihilation rate today. In particular, besides the usual $2\to2$ processes in \textsf{MadDM}, we also include $2\to3$ annihilation processes, $\chi \chi \to V V^*$, where $V^{*}$ is an off-shell weak gauge boson. In addition, we include $2\to3$, $2\to4$, $2\to5$ and $2\to6$ processes such as $\chi \chi \to H h$, $H H$ where $H$ decays further into pairs of SM particles, including $V V^*$. The decay of all on-shell SM particles is performed within \textsf{Pythia}.

In four (Reticulum II, Tucana III, Tucana IV and Indus II) of the 45 dSphs, slight excesses have been found with a local significance of roughly $2\sigma$ each~\cite{Geringer-Sameth:2015lua,Li:2015kag,Fermi-LAT:2016uux}. Consequently, a combination of the likelihoods from all dwarfs favors a certain range of DM masses and annihilation cross sections that provide a flux compatible with the excess.~However, as the DM origin of these excesses is not yet established, we show our results in section~\ref{sec:num_scans_results} with and without including the respective four dwarfs.~The latter choice imposes an upper bound on the DM annihilation cross section only.

\subsection{Direct detection at one-loop level}\label{sec:direct}
The elastic spin-independent (SI) pNG DM-nucleon cross section is momentum-suppressed at tree-level (see appendix~\ref{app:dm-nucleon}). It is given by \cite{Azevedo:2018exj}
\begin{equation}\label{eqn:tree-xsection}
    \sigma_{\chi N}^{\textnormal{tree-level}} \approx \frac{4 \sin^2\theta \cos^2\theta \, f_N^2}{3\pi} \frac{m_N^2 \mu_{\chi N}^6}{m_\chi^2 v_h^2 v_s^2} \frac{(m_h^2 - m_H^2)^2}{m_h^4 m_H^4} v_\chi^4,
\end{equation}
where $\mu_{\chi N} \equiv m_\chi m_N/(m_\chi + m_N)$ is the DM-nucleon reduced mass, $f_N = 0.3$ is the effective Higgs-nucleon coupling \cite{Alarcon:2011zs,Cline:2013gha,Ling:2017jyz}, $m_N = 939$\,MeV is the averaged nucleon mass and $v_\chi$ is the DM velocity in the laboratory frame.~In the vicinity of the Earth, $v_\chi \sim 10^{-3}$.~Thus, the nuclear recoil rate is suppressed by a factor of $v_\chi^4 \sim 10^{-13}$.~For a typical choice of model parameters, the tree-level cross section in eq.~\eqref{eqn:tree-xsection} is too small to be experimentally observed at current or future planned experiments \cite{Azevedo:2018exj}.

The first non-vanishing contribution to the DM-nucleon cross section at zero velocity appears at one-loop level. It can be approximated by \cite{Gross:2017dan,Azevedo:2018exj}
\begin{equation}\label{eqn:one-loop-approx}
    \left.\sigma_{\chi N}^{\textnormal{1-loop}}\right|_{\textnormal{Approx}} \approx 
    \begin{cases}
        \dfrac{\sin^2\theta}{64 \pi^5} \dfrac{m_N^4 f_N^2}{m_h^4 v_h^2} \dfrac{m_H^4 m_\chi^2}{v_s^6}, & m_\chi \leq m_H, \\[4mm]
        \dfrac{\sin^2\theta}{64 \pi^5} \dfrac{m_N^4 f_N^2}{m_h^4 v_h^2} \dfrac{m_H^8}{m_\chi^2 v_s^6}, & m_\chi > m_H.
    \end{cases}
\end{equation}
In the limit of $m_\chi^2 \equiv \mu_S'^2 \rightarrow 0$, the DM particle $\chi$ becomes a true Goldstone boson of the spontaneously broken global $U(1)$ symmetry. Thus, the direct detection amplitude should vanish. This behavior is observed in eq.~\eqref{eqn:one-loop-approx}.

In ref.~\cite{Azevedo:2018exj}, the authors point out that the approximate one-loop cross section in eq.~\eqref{eqn:one-loop-approx} can under/overestimate the actual cross section by several orders of magnitude depending on the model parameters.~Thus, we use the full one-loop cross section from ref.~\cite{Azevedo:2018exj} in our study:
\begin{equation}\label{eqn:one-loop-exact}
    \sigma_{\chi N}^{\textnormal{1-loop}} = \frac{\mu_{\chi N}^2}{\pi} \frac{f_N^2 m_N^2}{v_h^2 m_\chi^2} \mathcal{F}^2,
\end{equation}
where the one-loop function $\mathcal{F}$ is
\begin{align}\label{eqn:one-loop-F}
    \mathcal{F} &= - \frac{\sin 2\theta \, (m_h^2 - m_H^2) m_\chi^2}{128 \pi^2 \, v_h v_s^3 \, m_h^2 m_H^2} \left[ \mathcal{A}_1 C_2 (0, m_\chi^2, m_\chi^2, m_h^2, m_H^2, m_\chi^2) \right. \nonumber \\
    &\hspace{4.5cm} \left. + \, \mathcal{A}_2 D_3(0, 0, m_\chi^2, m_\chi^2, 0, m_\chi^2, m_h^2, m_h^2, m_H^2, m_\chi^2) \right. \nonumber \\[2mm]
    &\hspace{4.5cm} \left. + \, \mathcal{A}_3 D_3(0, 0, m_\chi^2, m_\chi^2, 0, m_\chi^2, m_h^2, m_H^2, m_H^2, m_\chi^2) \right].
\end{align}
Here the $C$ and $D$ terms are Passarino-Veltman functions \cite{Passarino:1978jh,Denner:1991kt,Hahn:1998yk} which we compute using \textsf{LoopTools v2.14} \cite{vanOldenborgh:1989wn,Hahn:1998yk}.\footnote{\url{http://www.feynarts.de/looptools/}}~The coefficients $\mathcal{A}_i$ are defined as
\begin{align}
    \mathcal{A}_1 &\equiv 4 (m_h^2 \sin^2\theta + m_H^2 \cos^2\theta) (2 m_h^2 v_h \sin^2\theta + 2 m_H^2 v_h \cos^2\theta - m_h^2 v_s \sin 2\theta + m_H^2 v_s \sin 2\theta), \nonumber \\
    \mathcal{A}_2 &\equiv -2 m_h^4 \sin \theta \left[(m_h^2 + 5 m_H^2) v_s \cos\theta - (m_h^2 - m_H^2) (v_s \cos 3\theta + 4 v_h \sin^3 \theta) \right], \nonumber \\
    \mathcal{A}_3 &\equiv 2 m_H^4 \cos \theta \left[(5 m_h^2 + m_H^2) v_s \sin \theta - (m_h^2 - m_H^2) (v_s \sin 3\theta + 4 v_h \cos^3 \theta) \right].
\end{align}
Note that $\mathcal{F}$ is proportional to $m_\chi^2$, and the fact that both the $C_2$ and $D_3$ functions behave as constants in the limit of $m_\chi \rightarrow 0$ \cite{Gross:2017dan,Azevedo:2018exj}, the direct detection amplitude indeed vanishes in this limit.

By means of our global fit, we are able to check the conclusions of ref.~\cite{Azevedo:2018exj} more generally.~We post-process our final samples, compute the approximate and actual one-loop cross sections using eqs.~\eqref{eqn:one-loop-approx} and \eqref{eqn:one-loop-exact} respectively, and present results in the $\{m_\chi, \, \sigma_{\chi N}^{\textnormal{1-loop}}\}$-plane. This allows us to confront the allowed parameter space of the model against the current limits from XENON1T \cite{Aprile:2018dbl}, and projected sensitivities from LZ~\cite{Akerib:2018lyp} and DARWIN \cite{Aalbers:2016jon}.~These results are presented in section~\ref{sec:num_scans_results}.

The authors of ref.~\cite{Ishiwata:2018sdi} have also computed the full one-loop DM-nucleon scattering cross section.~Their results are consistent with ref.~\cite{Azevedo:2018exj} in most parts of the parameter space.~However, large deviations appear at small DM masses.~We have verified that for the parameter space that we consider, these deviations do not impact our conclusions. In regions where these differences can be sizeable, the overall cross section lies well below the experimental limits that we show in section~\ref{sec:num_scans_results}.

\section{Results}\label{sec:num_scans_results}
In this section, we discuss the statistical treatment of the various constraints included in our global fit and show the corresponding results.

\subsection{Statistical analysis}
To find the allowed regions in the model parameter space, we use \multinestver \cite{Feroz:2008xx} with $50,000$ live points\footnote{For our frequentist analysis, we combine results from multiple \multinest scans with 10k, 12k, 25k and 50k live points, see footnote \ref{fn:sets2} for more details.~For our Bayesian analysis, we instead rely on a single \multinest scan with 50k live points.\label{fn:sets1}} and a stopping tolerance of $0.01$. \multinest is based on an implementation of the Importance Nested Sampling algorithm. It is primarily a Bayesian inference tool designed to compute the Bayesian evidence $\mathcal{Z}$ (defined below).~As a by-product, it draws posterior samples from a distribution that may contain a high multiplicity of nodes and/or degeneracies.~Furthermore, \multinest is capable of sampling the profile likelihood ratio (defined below) for the purpose of frequentist analysis.

\begin{table}[t]
    \centering
    
    \begin{tabular}{lcc}
        \toprule
        \textbf{Parameters} & \textbf{Ranges} & \textbf{Priors} \\ \midrule
        $m_\chi$\,(GeV) & $[10,\, 10^3]$ & log \\[2mm]
        $v_s$\,(GeV) & $[10,\,10^6]$ & log \\[2mm]
        $\theta$\,(rad) & $[0, \,\pi/2]$ & flat \\[2mm]
        $m_H$\,(GeV) & $[10, \,10^3]$ & log \\
        \bottomrule
    \end{tabular}
    
    \caption{Ranges and priors for the free model parameters.}
    \label{tab:range_priors}
\end{table}

A key ingredient for both a frequentist and Bayesian analysis is the likelihood (or log-likelihood) function.~In our numerical scans, the total log-likelihood function is
\begin{equation}\label{eqn:totlike}
    \ln \like_{\textnormal{total}} (\bm{\theta}) = \ln \like_{\Omega_\chi h^2} (\bm{\theta}) + \ln \like_{\Gamma_{h \rightarrow \chi \chi}} (\bm{\theta}) + \ln \like_{\textnormal{EWPO}} (\bm{\theta}) + \ln \like_{\textnormal{LEP}} (\bm{\theta}) + \ln \like_{\textnormal{HS}} (\bm{\theta}),
\end{equation}
where $\bm{\theta} \equiv (m_\chi,\,v_s,\,\theta,\,m_H)$ are the free model parameters.~Each of the individual likelihood functions are described in section~\ref{sec:like_constraints}.

The range and prior types for our free model parameters are summarised in table~\ref{tab:range_priors}.~For the mixing angle $\theta$, we find that our results are symmetric under $\theta \rightarrow - \theta$. In addition, the case $\theta = \pi$ is analogous to $\theta = 0$, thus we only restrict $\theta \in [0, \pi/2]$. To cover the region close to the two resonances, $m_\chi \simeq m_{h,H}/2$, where the annihilation cross section is enhanced, we scan up to very large values for the second scalar VEV, i.e.,~$v_s \in [10,\,10^6]$\,GeV. The upper boundary corresponds to very small couplings $\lambda_S$ and $\lambda_{\Phi S}$, see eqs.~\eqref{eqn:lamS} and \eqref{eqn:lamPS} respectively.

\subsubsection{Profile likelihoods}
In a frequentist analysis, the statistical precision of a parameter estimate is represented by a confidence interval that encapsulates the frequentist `coverage probability'. Such an interval is dependent on the data $\bm{x}$, and thus changes upon each re-iteration of the experiment.~As proper frequentist coverage is usually not possible for complicated likelihoods and parameter spaces, approximate methods are often used \cite{Akrami:2010cz}.~One such method is the well-known \textit{profile construction} \cite{2006sppp.conf..112C}, which depends on the profile likelihood ratio (PLR):
\begin{equation}
    \Lambda (\theta_i,\,\theta_j) \equiv \frac{\like_{\textrm{total}} (\theta_i,\, \theta_j,\,\hat{\hat{\boldsymbol{\nu}}}(\theta_i,\,\theta_j))}{\like_{\textnormal{total}} (\hat{\bm{\theta}})}.
\end{equation}
Here $\hat{\hat{\boldsymbol{\nu}}}(\theta_i,\,\theta_j)$ are the parameter values $\{\theta_{k} | \, k \neq i, \, j\}$ that maximise $\like_{\rm{total}} (\bm{\theta})$ for a fixed $(\theta_i,\,\theta_j)$, whereas $\boldsymbol{\hat{\theta}}$ is the maximum likelihood estimate for $\bm{\theta}$, i.e., a `best-fit' point that maximises $\like_{\rm{total}} (\bm{\theta})$ \cite{Cowan:2010js,Workgroup:2017htr}.~To construct confidence intervals, we maximise $\Lambda$ in the relevant parameter planes of interest while profiling over the other parameters and construct iso-likelihood contours at fixed confidence level (CL), e.g., $68.3\%$ for $1\sigma$ and $95.4\%$ for $2\sigma$ CL.

\begin{table}[t]
    \centering

    \begin{tabular}{lccc}
        \toprule
        
        \multicolumn{1}{c}{} & \multirow{2}{*}{\textbf{Global fit}} & \multicolumn{2}{c}{\textbf{Post-processing with \fermi}} \\ \cmidrule(lr){3-4}
        
        & & \multicolumn{1}{c}{\textbf{41 dSphs}} & \multicolumn{1}{c}{\textbf{45 dSphs}} \\ \midrule
        
        \textbf{Parameters} & & & \\[2mm]
        
        $m_\chi$\,(GeV) & $62.573$ & $121.632$ & $62.598$ \\[2mm]
        
        $v_h/v_s$    & $0.0187$ & $3.46$ & $8.93 \times 10^{-3}$ \\[2mm]
        
        $\theta$\,(rad) & $1.53$ & $1.54$ & $1.49$ \\[2mm]
        
        $m_H$\,(GeV)    & $125.30$ & $125.30$ & $125.30$ \\[2mm] 
        
        \textbf{Observables} & & &\\[2mm]
        
        $\Omega_\chi h^2$ & $0.119$ & $0.119$ & $0.120$ \\[2mm]
        
        Dominant channel (FO) & $\chi \chi \rightarrow b\ovr{b}~(76\%)$ & $\chi \chi \rightarrow hh~(100\%)$ & $\chi \chi \rightarrow b\ovr{b}~(76\%)$ \\[2mm]
          
        Dominant channel (today) & $\chi \chi \rightarrow b\ovr{b}~(77\%)$ & $\chi \chi \rightarrow W W~(70\%)$ & $\chi \chi \rightarrow b\ovr{b}~(77\%)$ \\[2mm]
        
        $\langle \sigma v \rangle_0$\,(cm$^3$\,s$^{-1}$) & $8.6 \times 10^{-27}$ & $4.4 \times 10^{-31}$ & $1.1 \times 10^{-26}$ \\[2mm] 

        $\sigma_{\chi N}^{\textnormal{1-loop}}$\,(cm$^2$) & $6.3 \times 10^{-68}$ & $3.1 \times 10^{-54}$ & $3.3 \times 10^{-69}$ \\[2mm] \midrule 

        $\ln \like_{\textnormal{total}}^{\textnormal{BF}} (\bm{\theta})$ & $-91.568$ & $-91.569$ & $-87.620$ \\ 
        \bottomrule
    \end{tabular}
    
    \caption{A summary of the best-fit (BF) points, key DM observables (the DM relic abundance, the dominant annihilation channel during freeze-out (FO) and today, the DM annihilation cross section today and the one-loop DM-nucleon cross section) and total log-likelihood $\ln \like_{\textnormal{total}}^{\textnormal{BF}} (\bm{\theta)}$ from our global fit (\emph{column 1}), and after post-processing our samples with \fermi likelihood with 41 (\emph{column 2}) and 45 (\emph{column 3}) dwarf spheroidal (dSph) galaxies.}
    \label{tab:bestfit}
\end{table}

\begin{figure}[t]
    \raggedright
	
	\hspace{-0.66mm}
	\includegraphics[width=0.35814\textwidth,trim={0 1.8cm 0 0},clip]{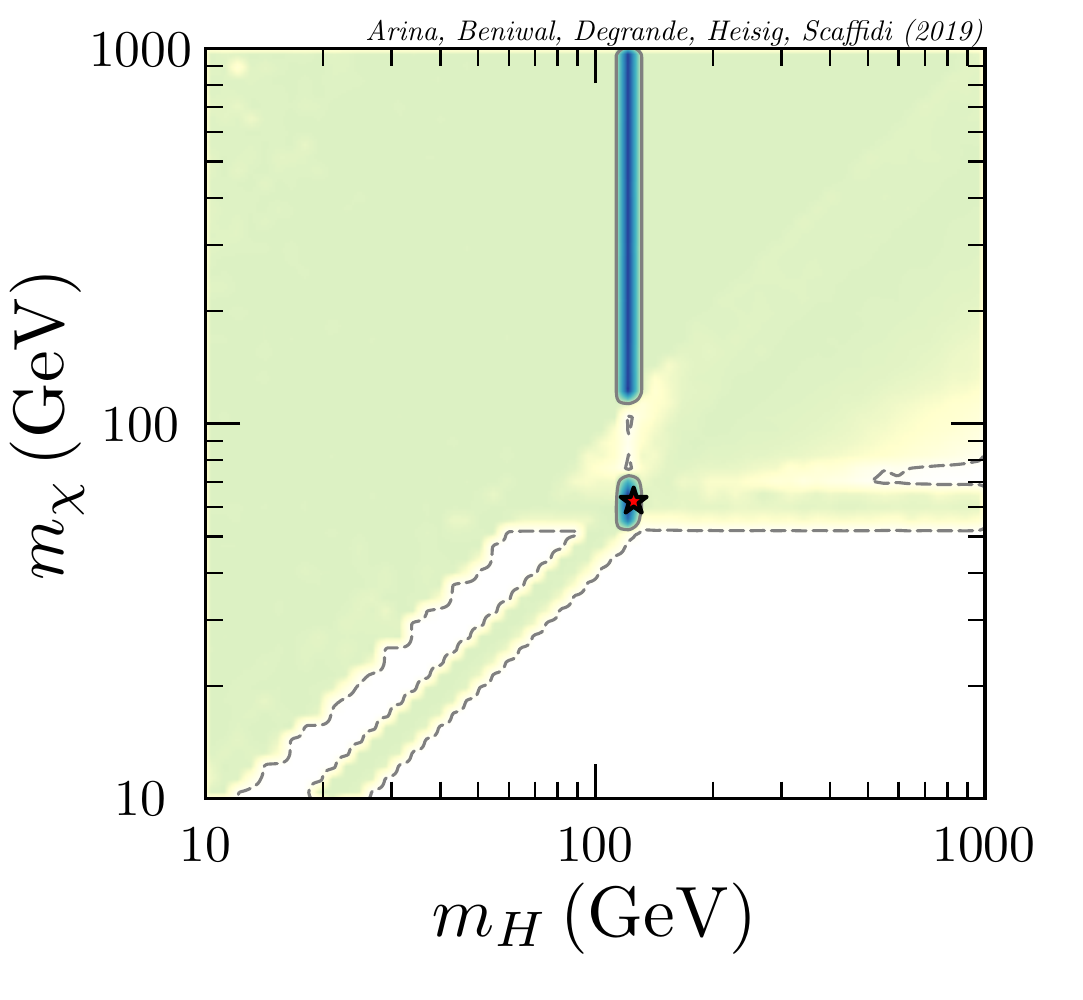} 
	
	\hspace{-0.66mm} 
	\includegraphics[width=0.35814\textwidth,trim={0 1.8cm 0 0.48cm},clip]{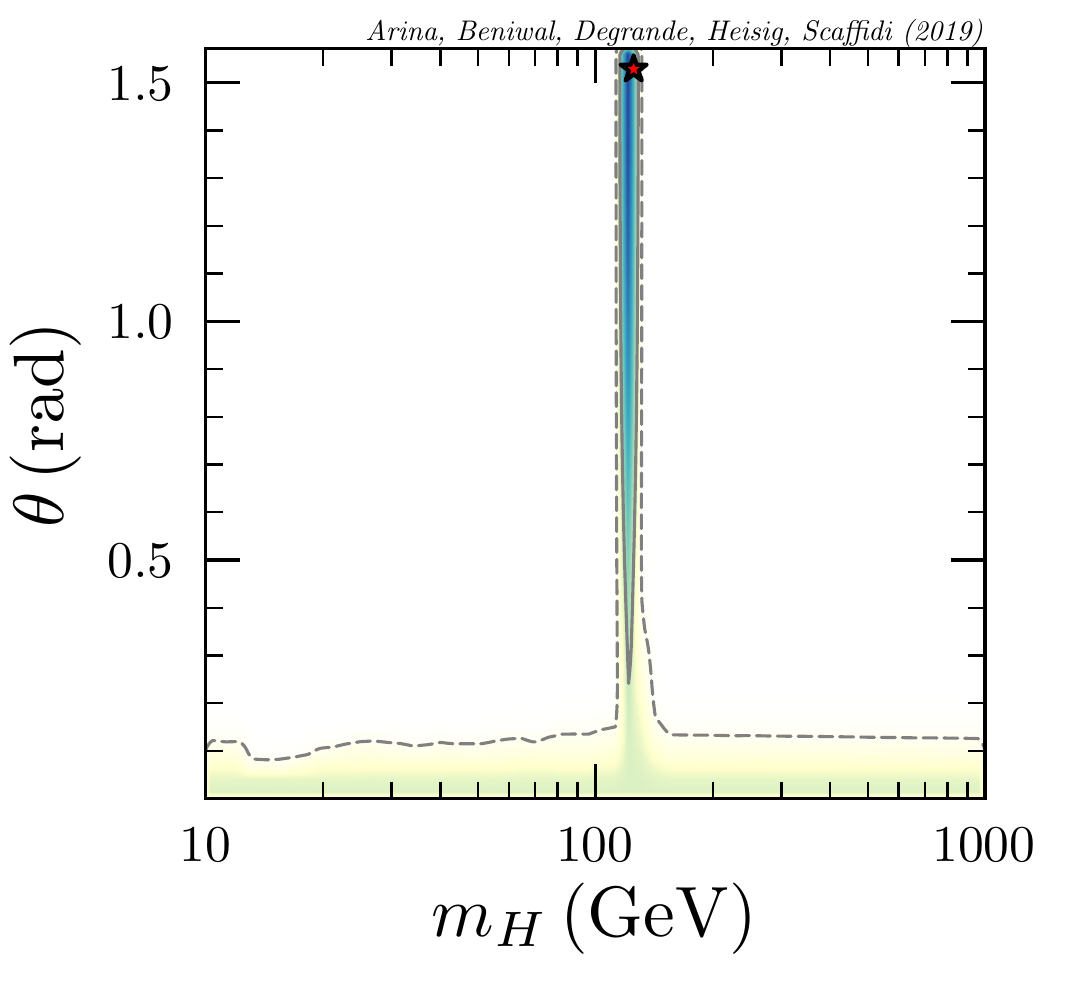} 
	\hspace{-4.0mm}
	\includegraphics[width=0.301176\textwidth,trim={1.8cm 1.8cm 0 0},clip]{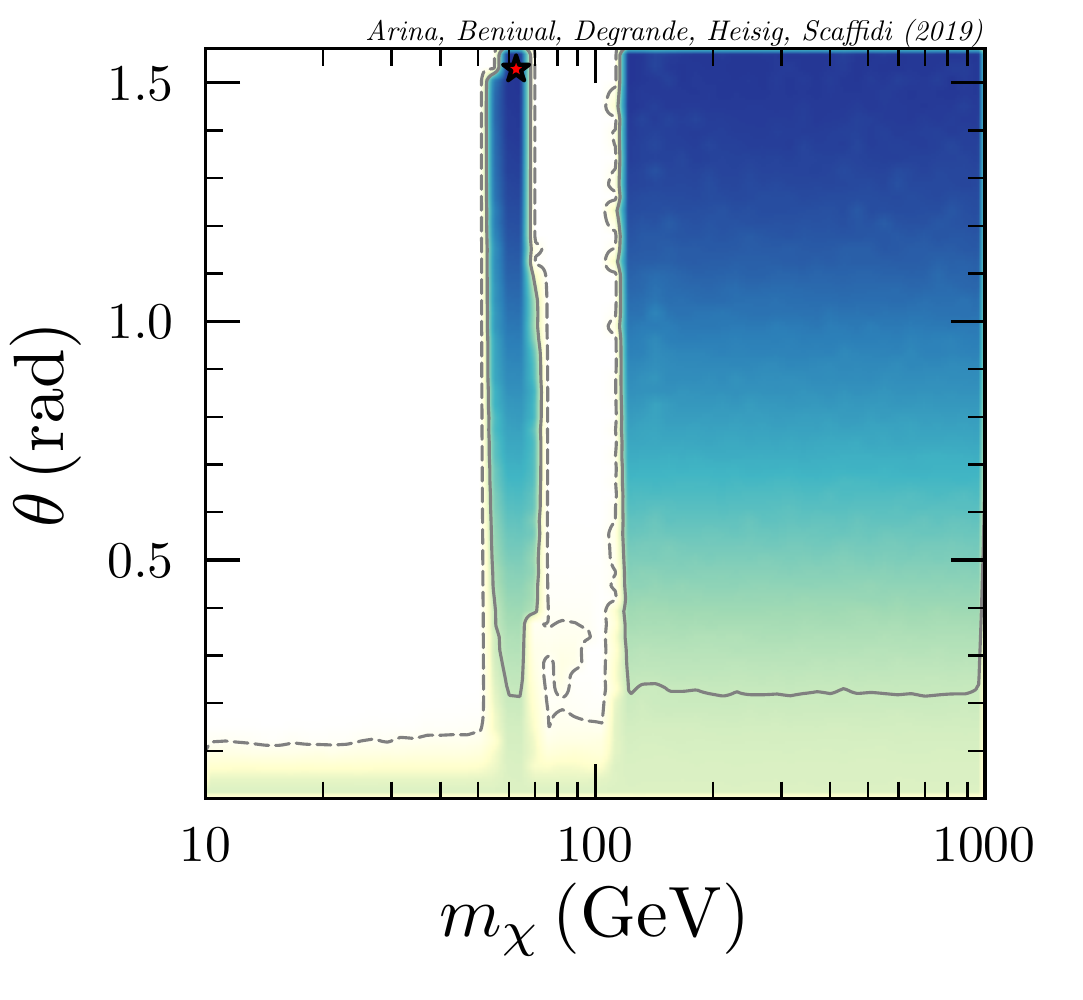}
	
	\hspace{-0.66mm} 
	\includegraphics[width=0.35814\textwidth,trim={0 0 0 0.48cm},clip]{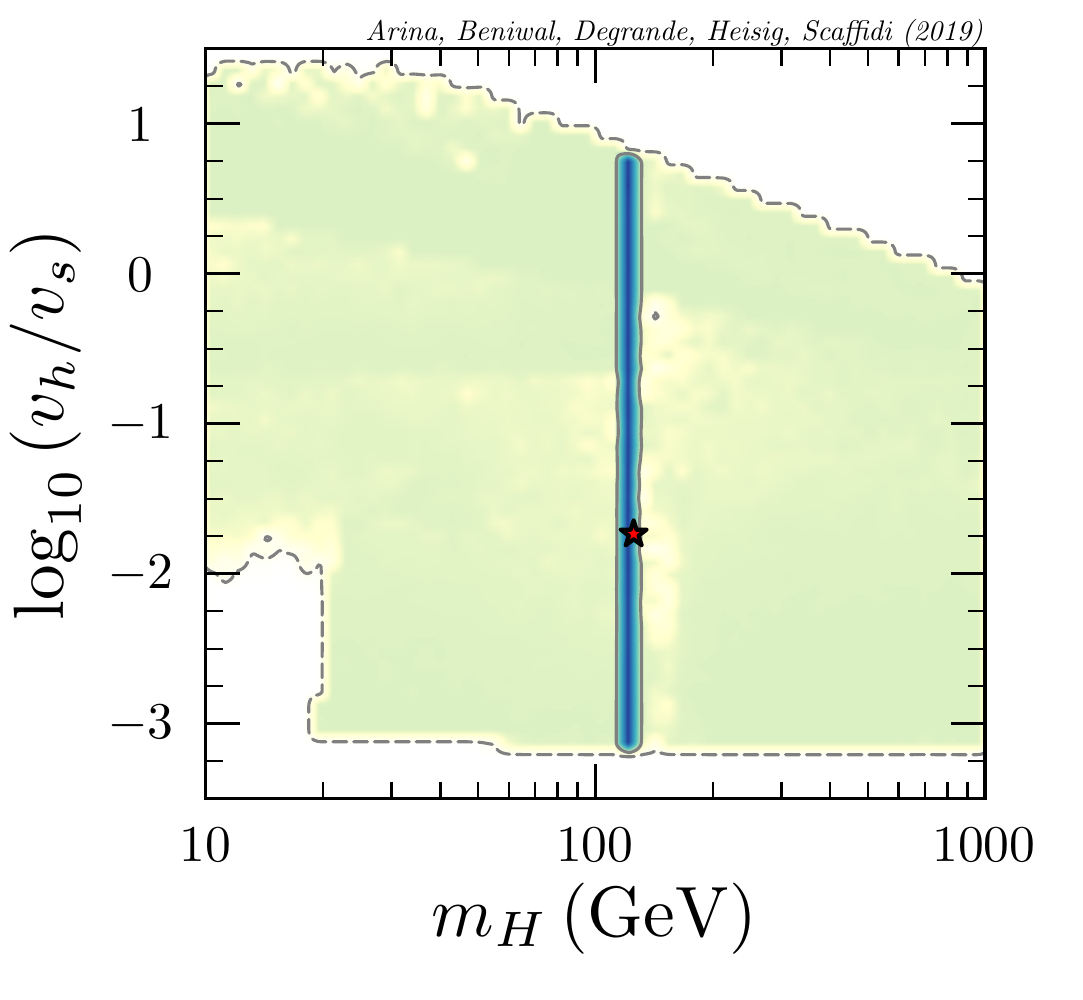}
	\hspace{-4.0mm}
	\includegraphics[width=0.301176\textwidth,trim={1.8cm  0 0 0.48cm},clip]{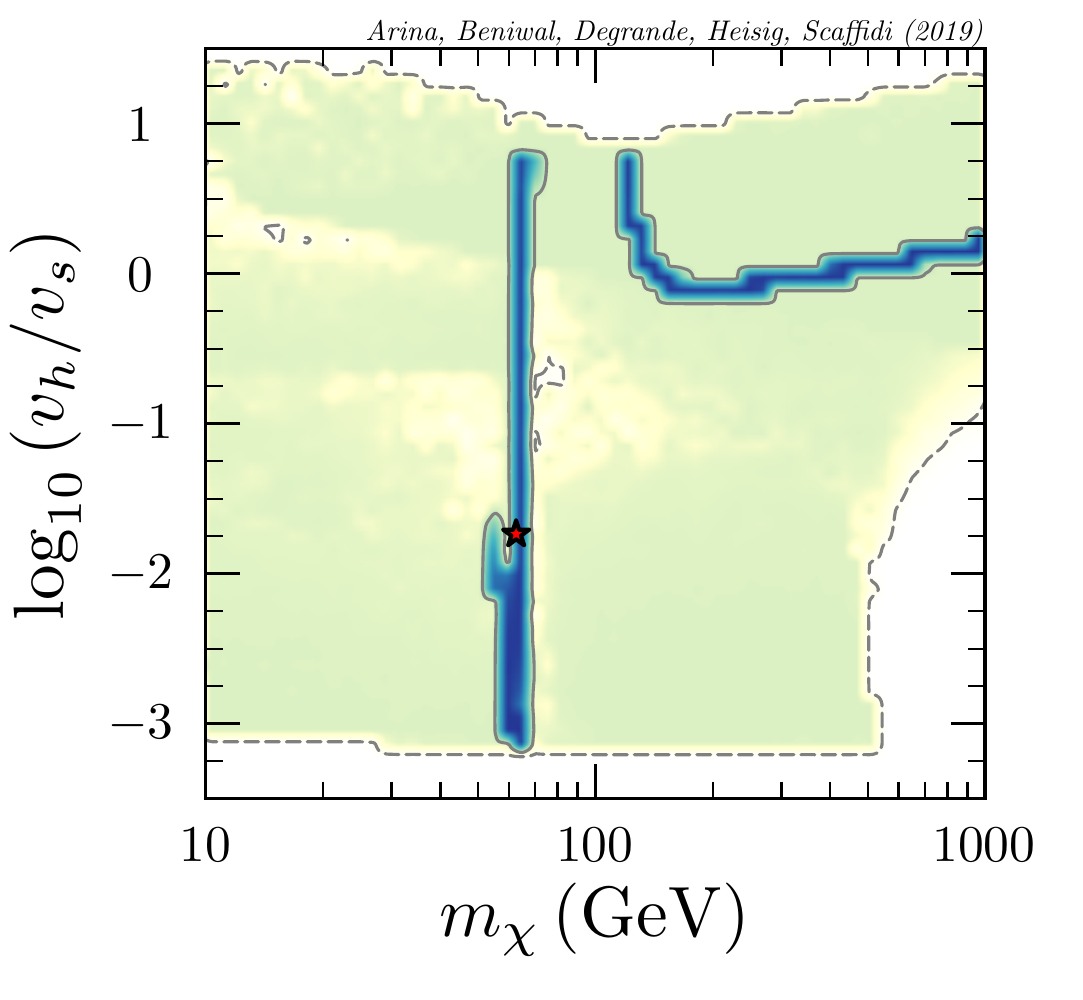}
	\hspace{-4.0mm}
	\includegraphics[width=0.34968\textwidth,  trim={1.8cm  0 0 0},clip]{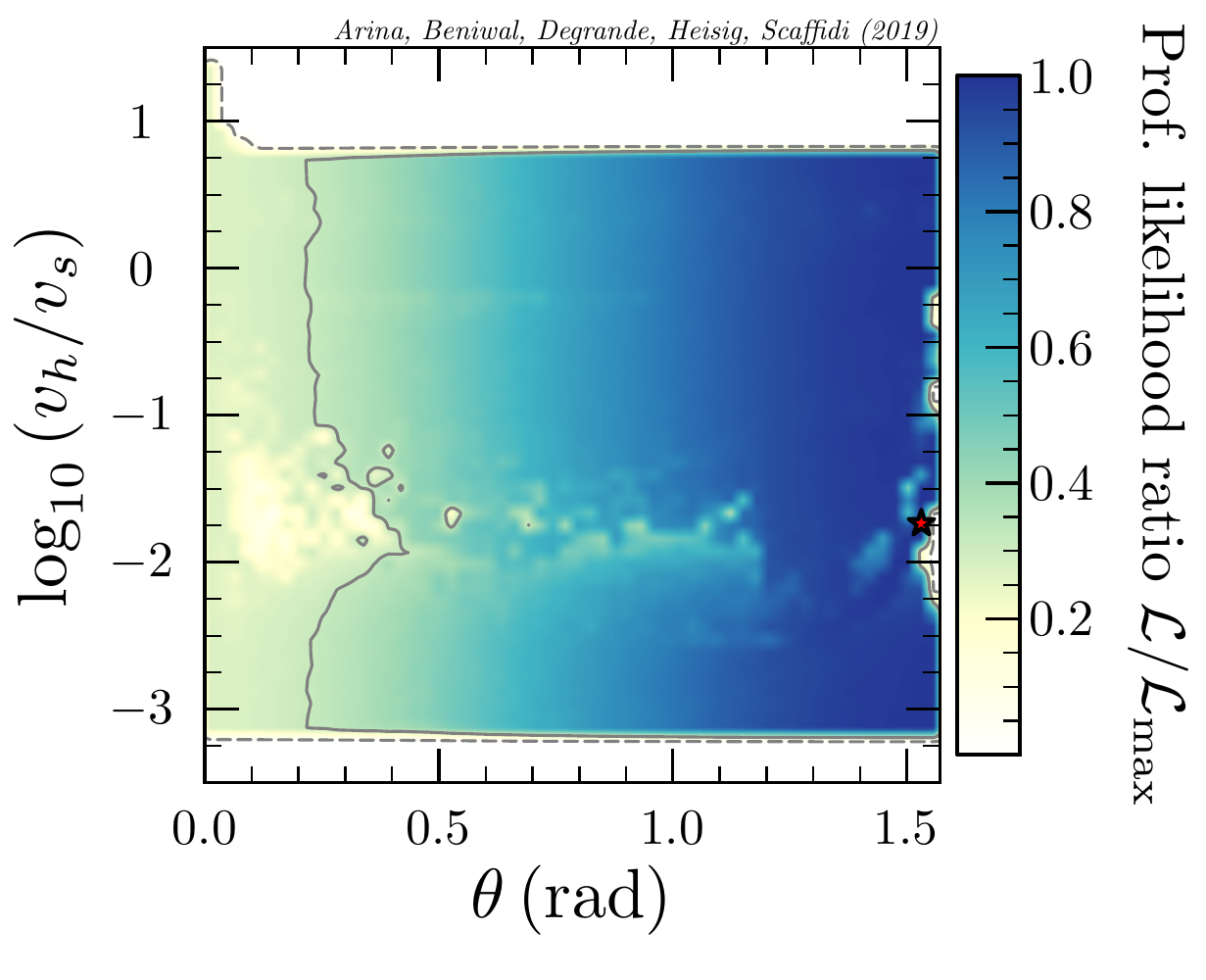}
    
    \caption{2D profile likelihood ratio (PLR) plots in the planes of pNG DM model parameters. The $1\sigma$ $(2\sigma)$ CL regions are marked by solid (dashed) lines. The best-fit point is marked by a red star; it is also summarised in column 1 of table~\ref{tab:bestfit}.}
    \label{fig:prof1}
\end{figure}

In figure~\ref{fig:prof1}, we show our PLR plots in six 2D planes spanned by all combinations of four model parameters.~These are generated using \pippi \cite{Scott:2012qh}.~In each plane, model parameters that are not shown are profiled over.~The $1\sigma$ $(2\sigma)$ CL contours are marked by solid (dashed) lines.~The best-fit point is shown as a red star; it is also summarised in column 1 of table~\ref{tab:bestfit}.\footnote{From our plots, it is evident that the exact position of the best-fit point is not significant, as the PLR $\like/\like_{\textnormal{max}}$ is mostly flat and close to 1 in a large portion of the $1\sigma$\,CL region.}

\begin{figure}
    \centering
    \includegraphics[width=0.48\textwidth]{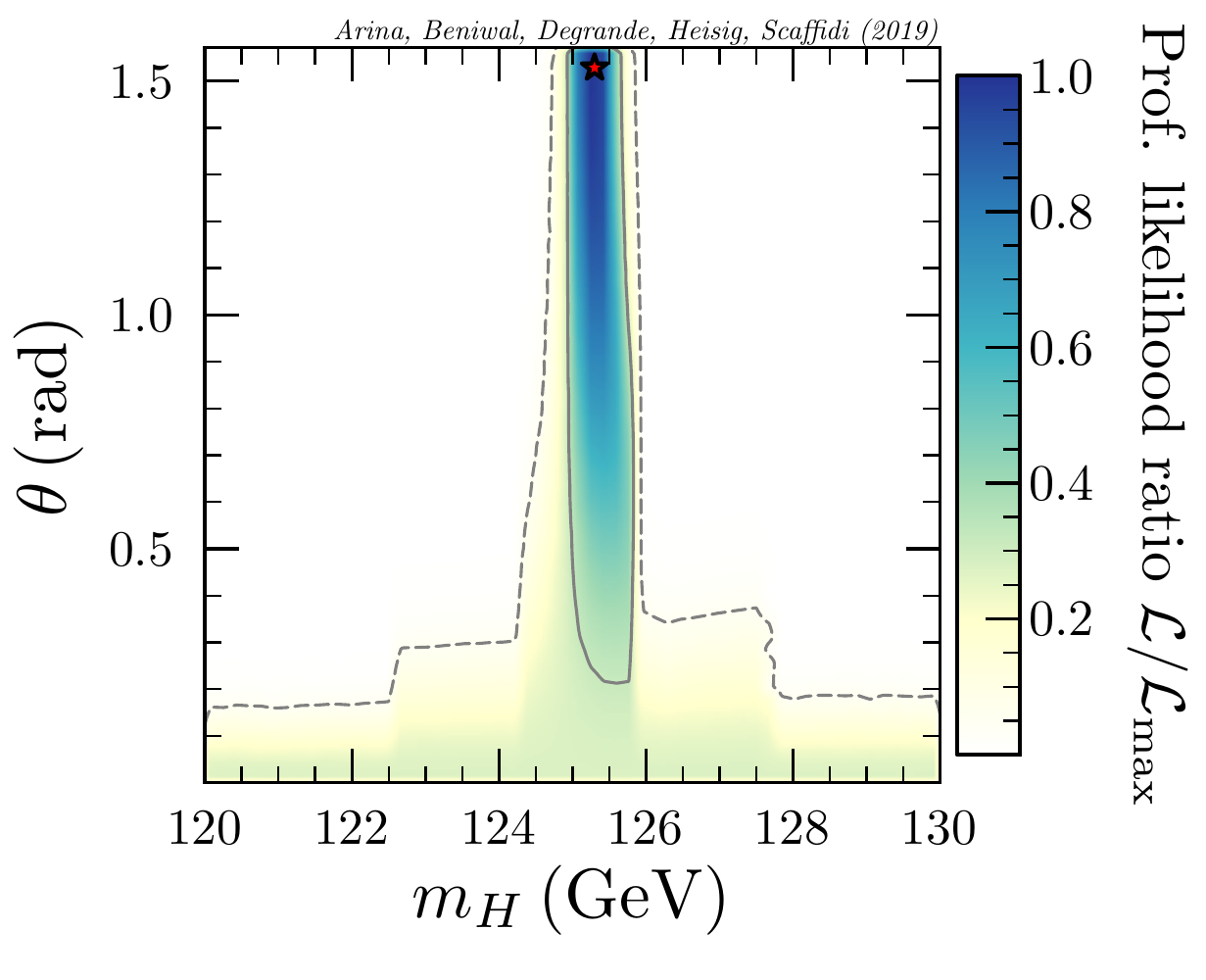} \quad
    \includegraphics[width=0.48\textwidth]{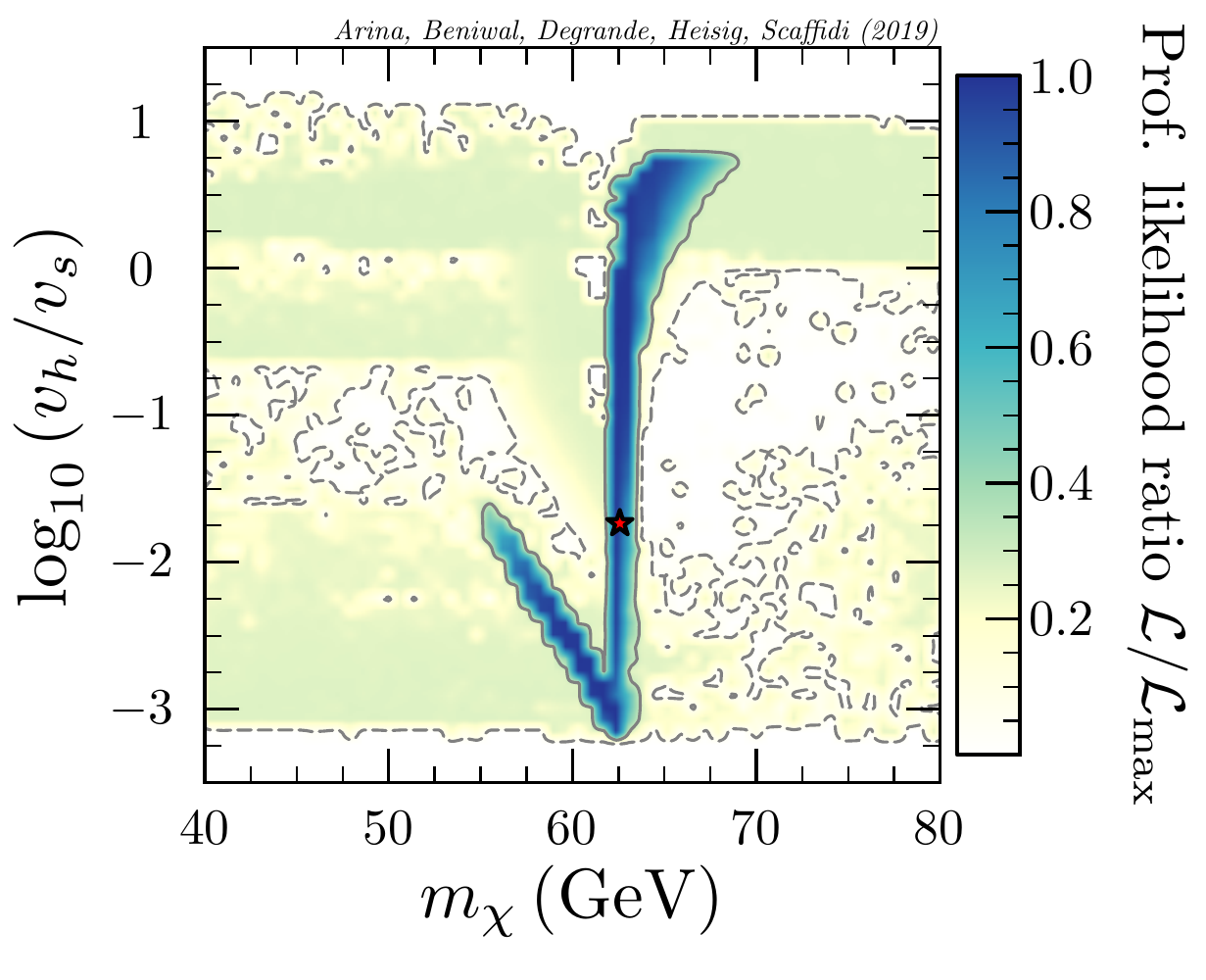}
    \vspace{-1cm}
    \caption{2D PLR plots from our global fit in the $(m_H,\,\theta)$- and $(m_\chi,\,v_h/v_s)$-planes after zooming into the region $m_H\sim m_h$ and the resonance region, $m_\chi \sim m_h/2$, respectively.} 
    \label{fig:prof2}
\end{figure}

A central constraint is imposed by the relic density selecting a thin slice in parameter space that provides a thermally averaged cross section $\langle \sigma v \rangle_\text{FO} \sim 3\times 10^{-26}$\,cm$^3$ s$^{-1}$. We find two phenomenologically distinct regions characterised by the type of annihilation channels relevant during freeze-out:

\begin{enumerate}
    \item Dominant annihilation via $s$-channel Higgs exchange ($h$ and/or $H$) into SM fermions and vector bosons.~Within this region, we encounter resonant and non-resonant annihilation.~In the former case (also known as the \emph{Higgs funnel}), the thermally averaged cross section has a sizeable contribution from the center-of-mass energy $\sqrt{s}=m_{h/H}$ providing a significant resonant enhancement.~According to the thermal momentum distribution, it is supported by DM mass in the range between somewhat below $m_{h/H}/2$ and $m_{h/H}/2$.~The point of maximal resonant enhancement is close to the upper boundary of this range.~The $H$- and $h$-resonance is visible as the (lower) diagonal stripe around $m_\chi\sim m_H/2$ and the 
    horizontal band around $m_\chi \sim m_h/2$, respectively, in the $(m_H,\,m_\chi)$-plane in figure~\ref{fig:prof1}.~Due to the small coupling involved (i.e.,~small $v_h/v_s$, see appendix~\ref{app:dm-nucleon}), the resonant regions are not subject to strong constraints from other observables.\footnote{As pointed out in ref.~\cite{Binder:2017rgn}, the assumption of local thermal equilibrium during freeze-out can break down near the resonances. This has the effect of changing the coupling value by a factor of order $\mathcal{O}(1)$. However, this part of the parameter space has small $v_h/v_s$ and is well beyond the sensitivity of current and future experiments. Thus, we employ the standard calculation of the DM relic density within \textsf{micrOMEGAs} assuming local thermal equilibrium during freeze-out.}
    
    Non-resonant annihilation via Higgs exchange only leads to allowed points in the range  $m_\chi>m_h/2$ and $m_\chi<m_H$. For points in the region $m_\chi\gtrsim m_H$, annihilation into Higgs pairs is dominant (see the next bullet point). Points below the resonance are (mostly) excluded by the perturbativity condition (see discussion further below) as the required coupling towards small DM masses quickly becomes too large. This can be understood from eq.~\eqref{eq:amp}, representing the amplitude of the respective annihilation process. For small center-of-mass energies compared to $m_h/2$ and $m_H/2$, a partial cancellation takes place and the amplitude is suppressed by $s$. This is in contrast to the singlet scalar Higgs portal model~\cite{Silveira:1985rk,McDonald:1993ex,Burgess:2000yq} where this suppression is not present and the region below the resonance can satisfy the relic density constraint for perturbative couplings.~The features at low $m_\chi$ are made more apparent by zooming into a small DM mass window around the Higgs mass resonance region, $m_\chi \sim m_h/2$, as shown in figure~\ref{fig:prof2}.

    \item Annihilation into Higgs pairs ($\chi\chi\to HH$, $hH$, $hh$) can be the dominant channel for $m_\chi> m_{h/H}$, and according to the thermal momentum distribution during freeze-out, for DM masses slightly below the Higgs threshold $m_\chi\lesssim m_{h/H}$. In our scan $\theta\sim 0$ is preferred except for $m_H\simeq m_h$ (see discussion further below). As the annihilation cross section into $HH$ ($hh$) is proportional to $\cos^4\theta$ ($\sin^4\theta$), we find that $\chi\chi\to HH$ dominates over $\chi\chi\to hh$ except for the region $m_H\simeq m_h$ where both are present. Consequently, annihilation into $HH$ and $hh$ leads to allowed points in the area above the diagonal band $m_\chi\sim m_H$ in the $(m_H,\,m_\chi)$-plane in figure~\ref{fig:prof1}.~Annihilation into $hH$ is only relevant for $m_H\simeq m_h$ and $\theta\simeq \pi/4$ as well as in a small region where $m_\chi>m_h$ and $m_\chi \lesssim m_H$.
\end{enumerate}

Mixing between $h$ and $H$ is highly constrained by several observations. First, to obtain a good agreement with the global electroweak fit results for the oblique parameters $S$, $T$ and $U$, according to eq.~\eqref{eqn:mod_oblique}, either a small mixing angle $\theta$ or $m_H \simeq m_h$ is required. Secondly, Higgs searches at LEP exclude most of the model parameter space for $m_H \lesssim 120$\,GeV and sizeable $\theta$. For larger $m_H$, the measured signal strengths at the LHC impose strong constraints on the parameter space.~In summary, a SM-like Higgs $h$ is compatible with $\theta \lesssim 0.1$~rad for all values of $m_H$, except for $m_H \simeq m_h$ where arbitrary values of $\theta$ are allowed, see the $(m_H,\, \theta)$-plane of figure~\ref{fig:prof1}.~A similar behaviour was found in ref.~\cite{Robens:2015gla}. On top of this, the observed signal strengths for a SM-like Higgs at the LHC exhibit a slight preference (around 1$\sigma$) for $m_H \simeq m_h$ and sizeable $\theta$ over the SM only prediction with $m_h=125\,$GeV. However, the presence of a second Higgs improves the fit only due to the freedom in $m_H$. In fact, keeping $m_h$ as a free parameter as well is expected to improve the fit and broaden the $1\sigma$ CL region beyond $m_H \simeq m_h$.\footnote{The observed signal strengths are sensitive to the exact value of the SM-like Higgs mass, $m_h$.~In a global fit, one could include $m_h$ as a nuisance parameter and associate a corresponding Gaussian likelihood function that can be profiled (marginalised) over in a frequentist (Bayesian) analysis.}

As mentioned above, constraints from perturbative unitarity are relevant in and exclude parts of the parameter space where
the measured relic density could only be matched with extremely large couplings. In the limit of small mixing, the perturbative unitarity limit in eq.~\eqref{eqn:pert_unit} translates to
\begin{equation}\label{eqn:pert-lim}
    \lambda_S v_h^2 \stackrel{\theta \, \rightarrow \, 0}{\approx} \frac{m_H^2}{v_s^2} v_h^2 < \frac{8\pi}{3} v_h^2 \implies \frac{v_h}{v_s} < \sqrt{\frac{8\pi}{3}} \frac{v_h}{m_H} \simeq \frac{713\,\text{GeV}}{m_H}.
\end{equation}
This limit is evident in the $(m_H,\,v_h/v_s)$-plane for $m_H \gtrsim 10$\,GeV, i.e., parameter points that lie outside the boundary of the $2\sigma$ CL implies $v_h/v_s > 713\,\text{GeV}/m_H$. Due to the strong constraints from perturbative unitarity towards small masses, limits from the invisible Higgs decay are less relevant than e.g.,~in the singlet scalar Higgs portal model. We found that dropping the likelihood from invisible Higgs decay does not significantly change the results shown in figure~\ref{fig:prof1}.

\begin{figure}[t]
    \centering
    
    \includegraphics[width=0.48\textwidth]{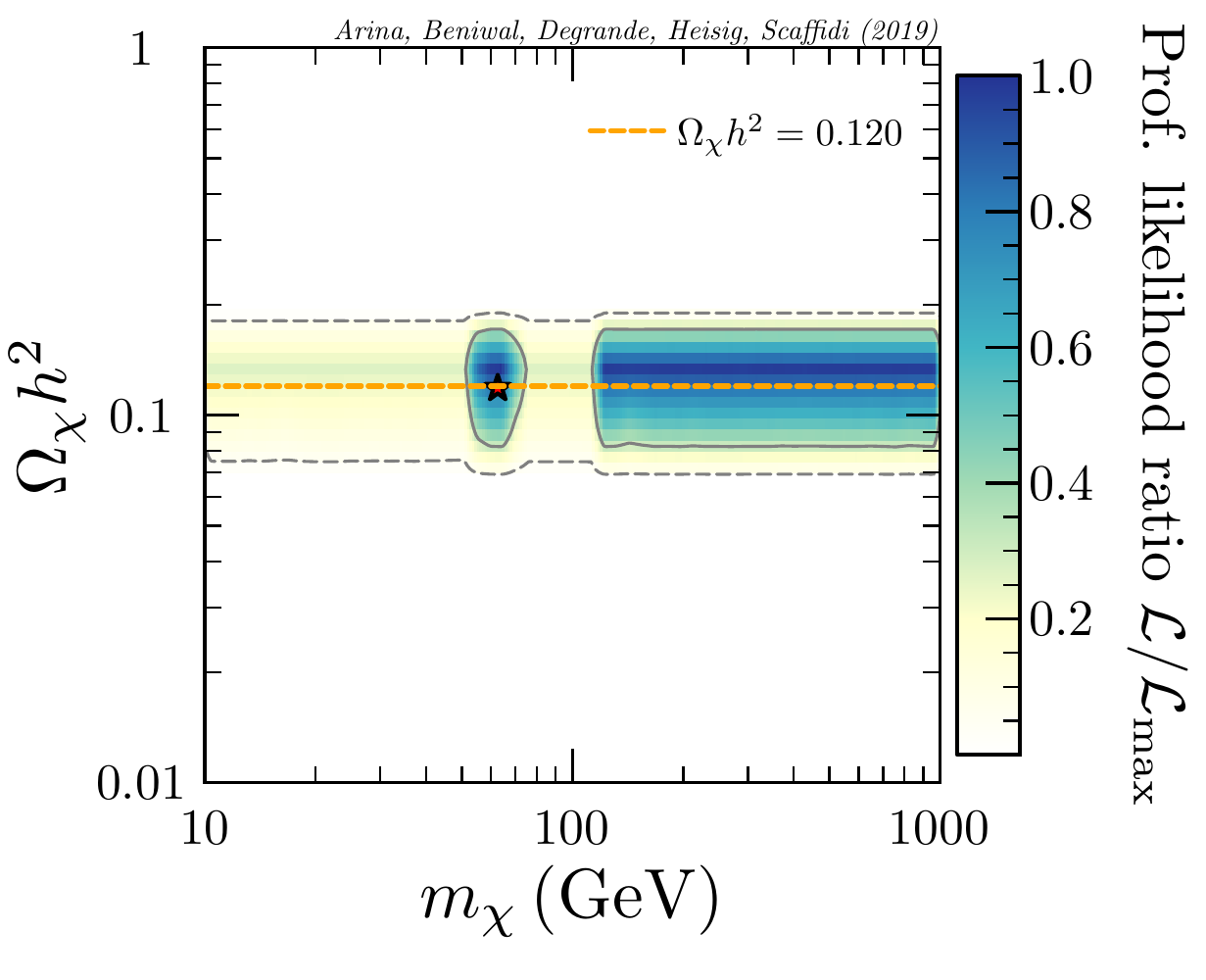} \quad
    \includegraphics[width=0.48\textwidth]{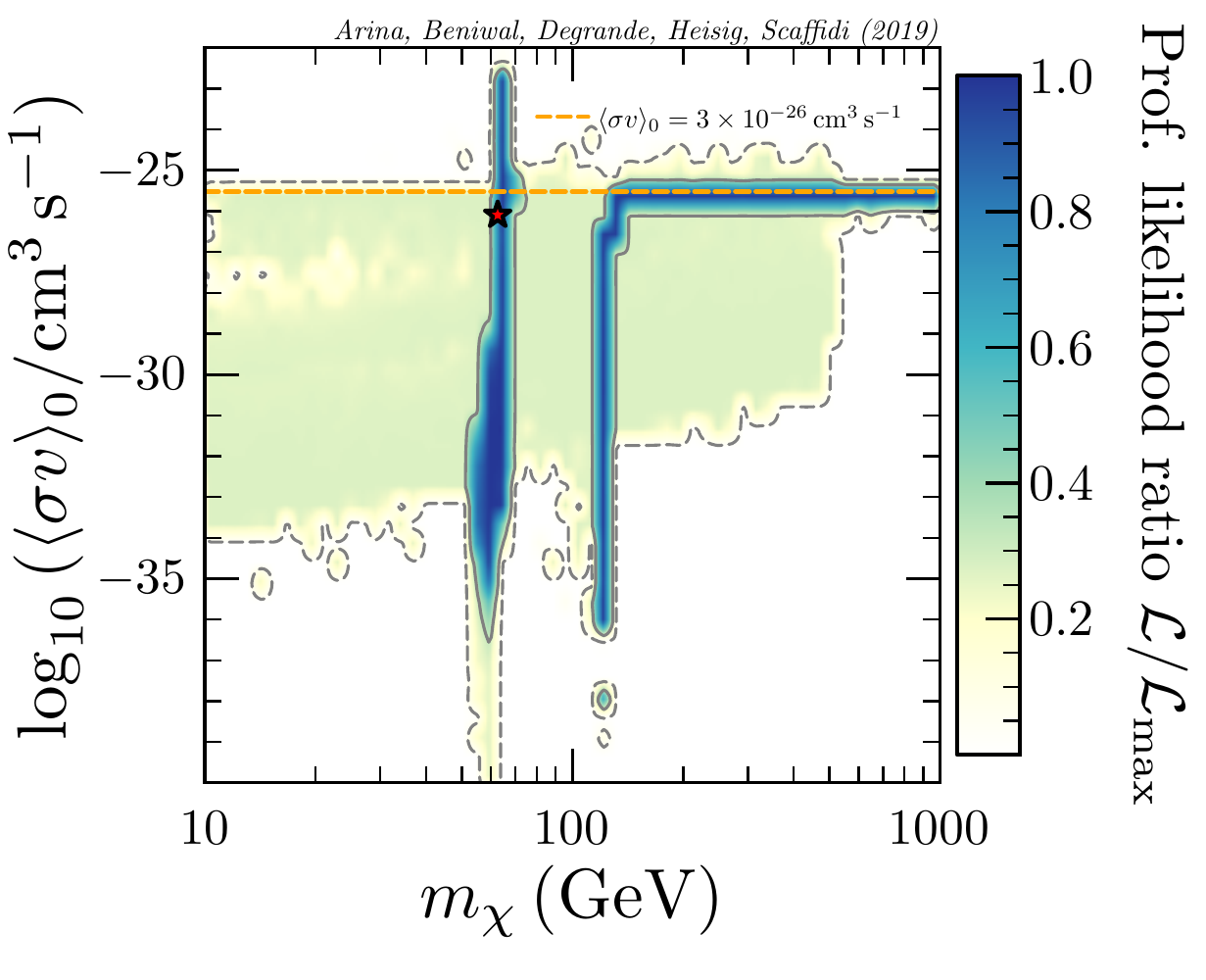} 
    \vspace{-1cm}
    \caption{2D PLR plots for key DM observables: the pNG DM relic abundance, $\Omega_\chi h^2$ (\emph{left panel}) and DM annihilation cross section today, $\langle \sigma v \rangle_0$ (\emph{right panel}).~The solid (dashed) contours and red star have the same meaning as in figure~\ref{fig:prof1}.~In the left panel, the orange dashed line shows the relic density measured by \planck; in the right panel, it shows the canonical freeze-out cross section, $\langle \sigma v \rangle_0 = 3 \times 10^{-26}$\,cm$^3$\,s$^{-1}$.}
    \label{fig:prof3}
\end{figure}

In figure~\ref{fig:prof3}, we show the PLR plots for key DM observables such as the pNG DM relic density, and its annihilation cross section into SM and non-SM particles today, $\langle \sigma v \rangle_0$. The $1\sigma$ CL region shows up as two disconnected islands.~The small island at $m_\chi \simeq m_h/2 = 62.5$\,GeV corresponds to the $h$ resonance, where $\chi\chi \rightarrow b\bar{b}$ channel is most dominant. The second island appears for $m_\chi \gtrsim 125$\,GeV. As $m_H$ is profiled over, and given that $m_H \simeq m_h$ is favoured, this island corresponds to the region where $\chi \chi \rightarrow hh,\,HH$ is dominant during freeze-out and sets the pNG DM relic abundance to the observed value today. Note that, although $\langle \sigma v \rangle_{\textnormal{FO}} \sim 3 \times 10^{-26}$\,cm$^{3}$\,s$^{-1}$ as required by the relic density constraint, the annihilation cross section today, $\langle \sigma v \rangle_0$ varies over many order of magnitude. This is due to the large velocity dependence of the annihilation cross section in the vicinity of a resonance and a threshold. Resonant annihilation can lead to both $\langle \sigma v \rangle_0$ smaller or larger than $\langle \sigma v \rangle_{\textnormal{FO}}$ depending on whether the DM mass is smaller or slightly larger than the point of maximal enhancement during freeze-out.~For the $1\sigma$ CL region, this behaviour can be seen in the right panel of figure~\ref{fig:prof3}.~For annihilation into Higgs pairs, in contrast, $\langle \sigma v \rangle_0$ can only be suppressed compared to $\langle \sigma v \rangle_{\textnormal{FO}}$ due to the smaller phase space around threshold today. Again, this behaviour can be seen for the $1\sigma$ CL region above 125\,GeV in the right panel of figure~\ref{fig:prof3}.

The best-fit point lies in the $h$-resonance region exhibiting large mixing and relatively small $v_h/v_s$.~The best-fit for the second Higgs mass is $m_H=125.3$\,GeV resulting from LHC signal strength measurements. The corresponding values are summarised in table~\ref{tab:bestfit}. Note, however, that the PLR is relatively flat within the $1\sigma$ CL region. Furthermore, as stated above, the $1\sigma$ preference for $m_H\simeq125.3$\,GeV to some extent is a result of our choice $m_h=125$\,GeV. This is in slight tension with the LHC Higgs signal strength and could be alleviated by treating the Higgs mass as a nuisance parameter in the fit. We therefore consider the entire $2\sigma$ CL region to be consistent with observation on a compatible level.

\subsubsection{Marginalised posteriors}
In Bayesian statistics, we rely on the Bayes' theorem:
\begin{equation}
    \mathcal{P}(\bm{\theta}|\bm{x}) = \frac{\,\like (\bm{x}\,|\bm{\theta} ) \, \pi (\bm{\theta})}{ \int d\bm{\theta}\,\like (\bm{x}\,|\bm{\theta} ) \, \pi (\bm{\theta})},
\end{equation}
where $\bm{\theta}$ are the free parameters of our model, $\bm{x}$ is the observed data, $\mathcal{P}(\bm{\theta}|\bm{x})$ is the posterior pdf, $\like (\bm{x}\,|\bm{\theta} )$ is the likelihood function and $\pi (\bm{\theta})$ is the prior pdf. The denominator involves an integral over the free model parameters and is known as the Bayesian evidence $\mathcal{Z}$. 

In our case of a multi-dimensional model, we are interested in 2D marginalised posterior (MP) distributions. These are constructed in the following way \cite{Trotta:2008qt}
\begin{equation}
    \mathcal{P}(\theta_i,\,\theta_j | \bm{x}) = \int_{l \, \neq \, i, \, j} d\theta_1,\ldots, d\theta_{l} \, \mathcal{P}(\bm{\theta} | \bm{x}),
\end{equation}
where we integrate over the irrelevant parameters $\{\theta_l | \, l \neq i,\,j\}$. The MP distribution above is used to define a Bayesian \textit{credible region} (CR) $\mathcal{\omega}$ in such a way that there is a probability $\alpha$ of containing the true values of model parameters:
\begin{equation}
    \int_\mathcal{\omega} d \theta_{i} \, d \theta_j \, \mathcal{P}(\theta_i,\, \theta_j | \bm{x}) = \alpha.
\end{equation}

\begin{figure}[t]
    \raggedright
    
    \hspace{-0.66mm}
    \includegraphics[width=0.35814\textwidth,trim={0 1.8cm 0 0},clip]{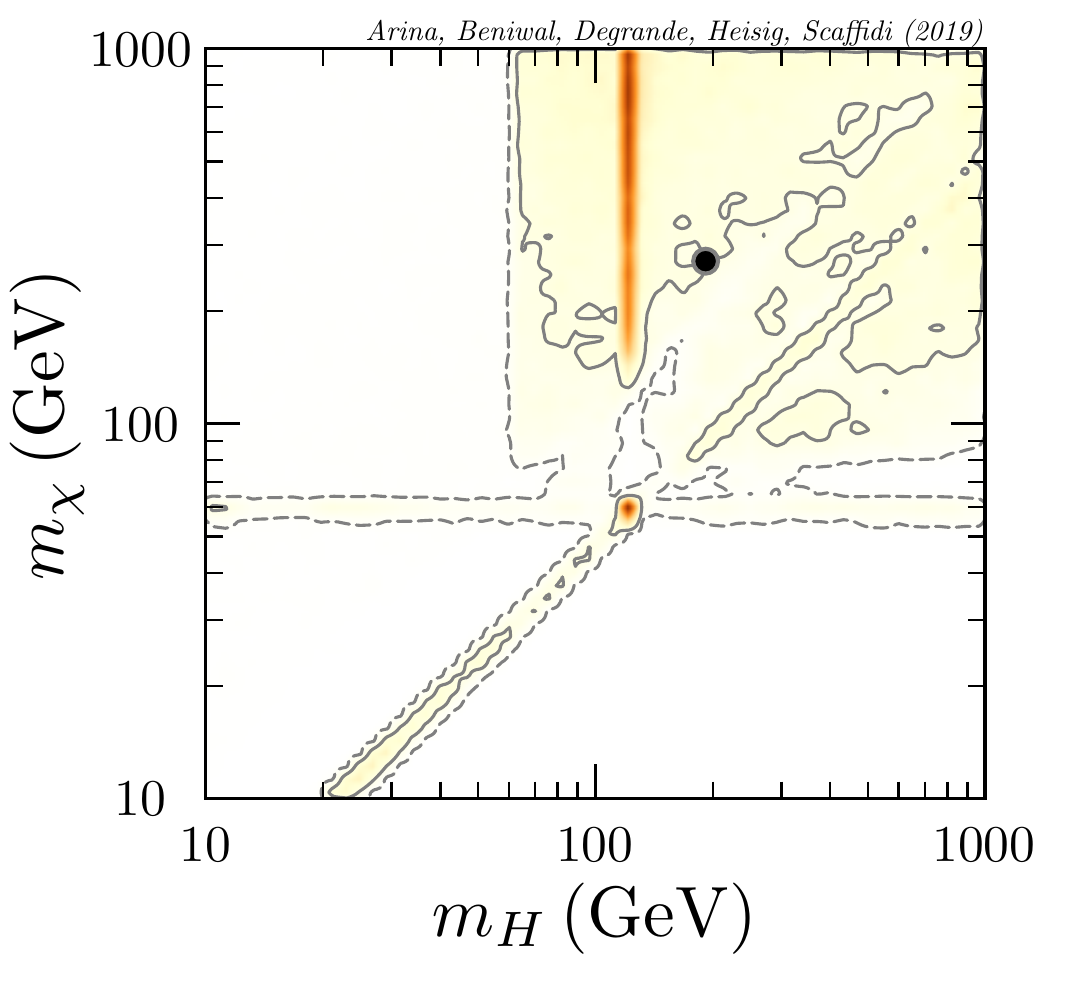} 
	
	\hspace{-0.66mm} 
	\includegraphics[width=0.35814\textwidth,trim={0 1.8cm 0 0.48cm},clip]{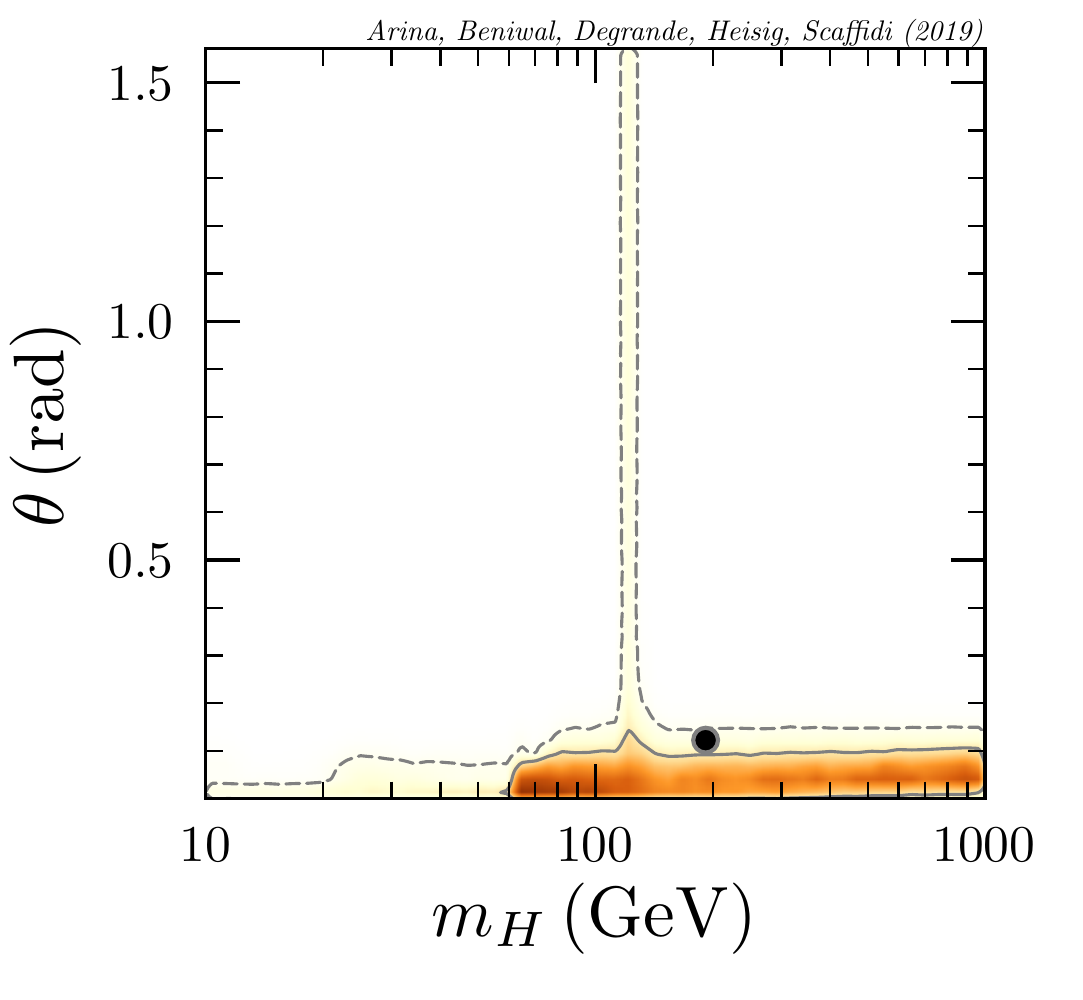} 
	\hspace{-4mm}
	\includegraphics[width=0.301176\textwidth,trim={1.8cm 1.8cm 0 0},clip]{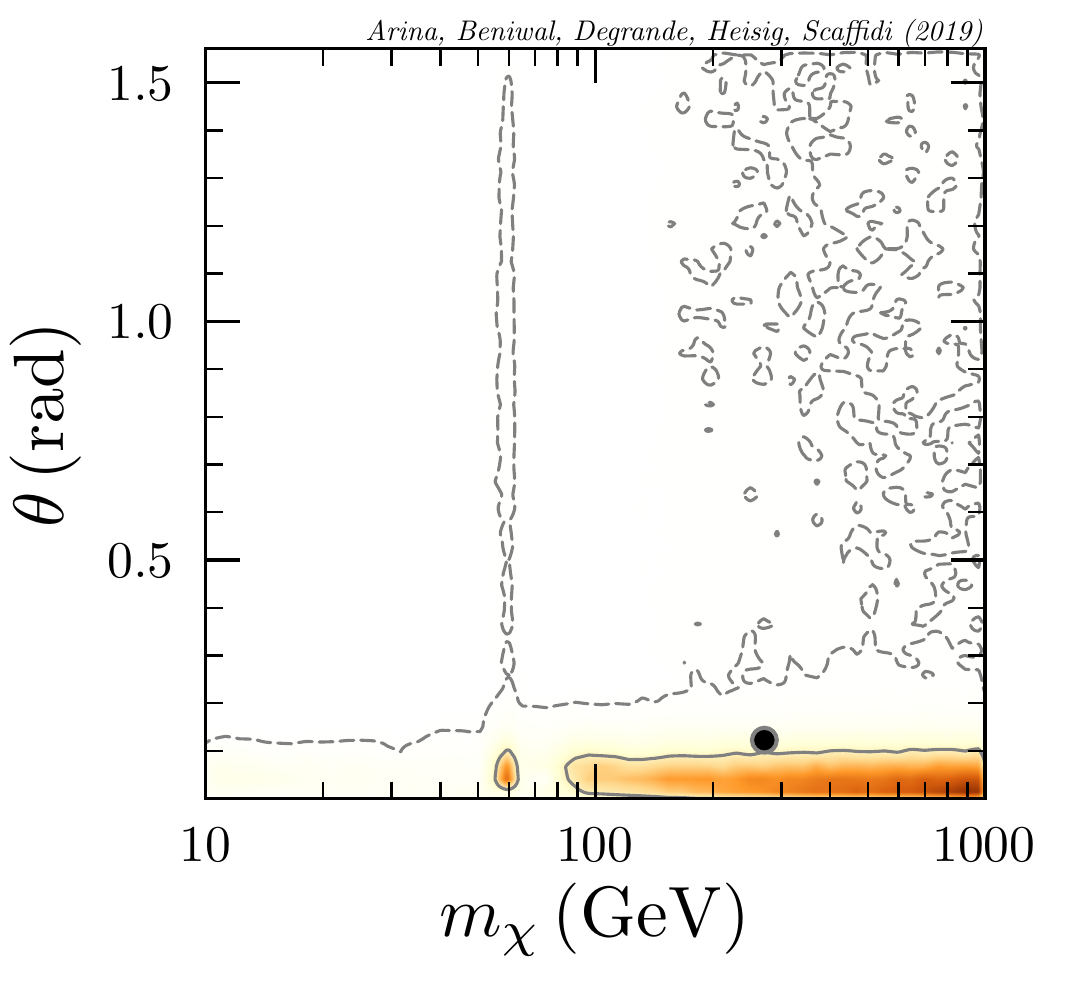}
	
	\hspace{-0.66mm} 
	\includegraphics[width=0.35814\textwidth,trim={0 0 0 0.48cm},clip]{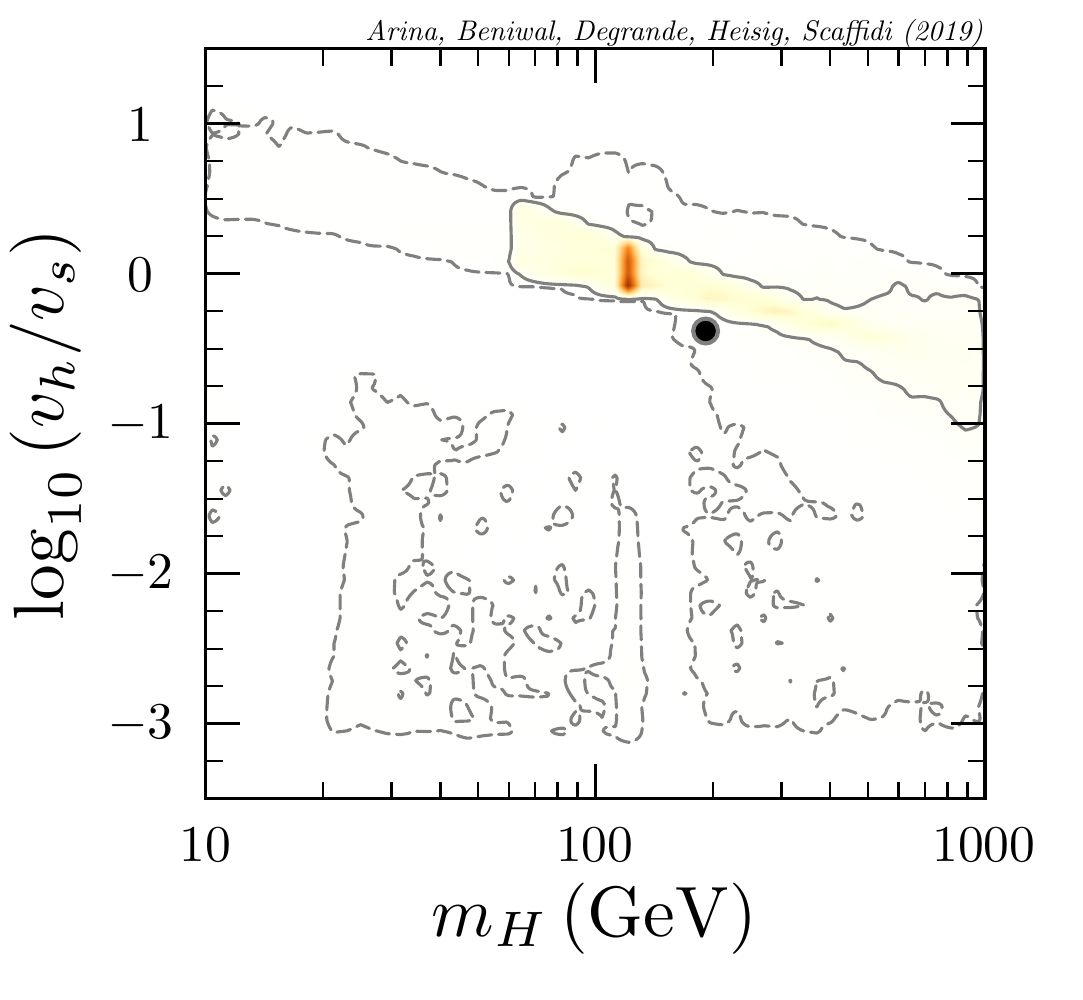}
	\hspace{-4.0mm}
	\includegraphics[width=0.301176\textwidth,trim={1.8cm  0 0 0.48cm},clip]{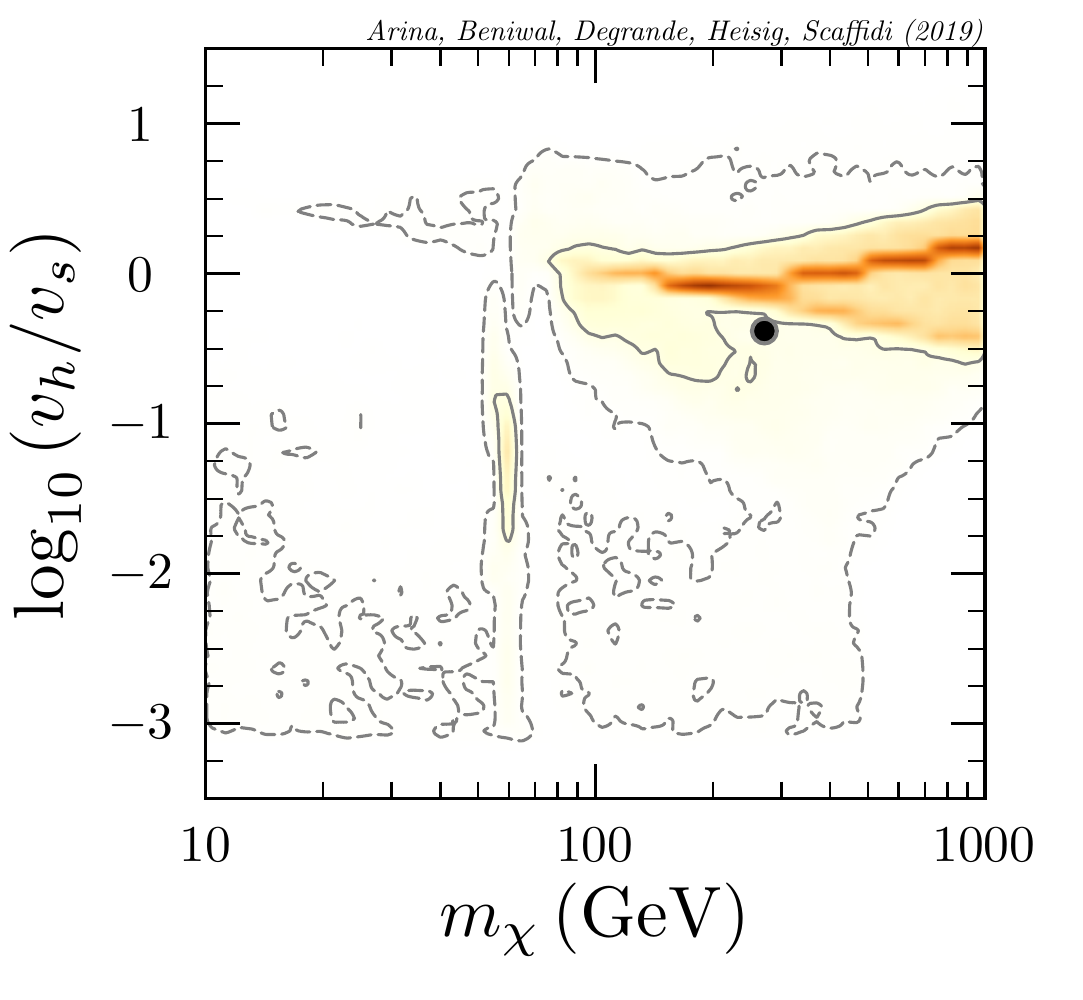}
	\hspace{-4.0mm}
	\includegraphics[width=0.34968\textwidth,  trim={1.8cm  0 0 0},clip]{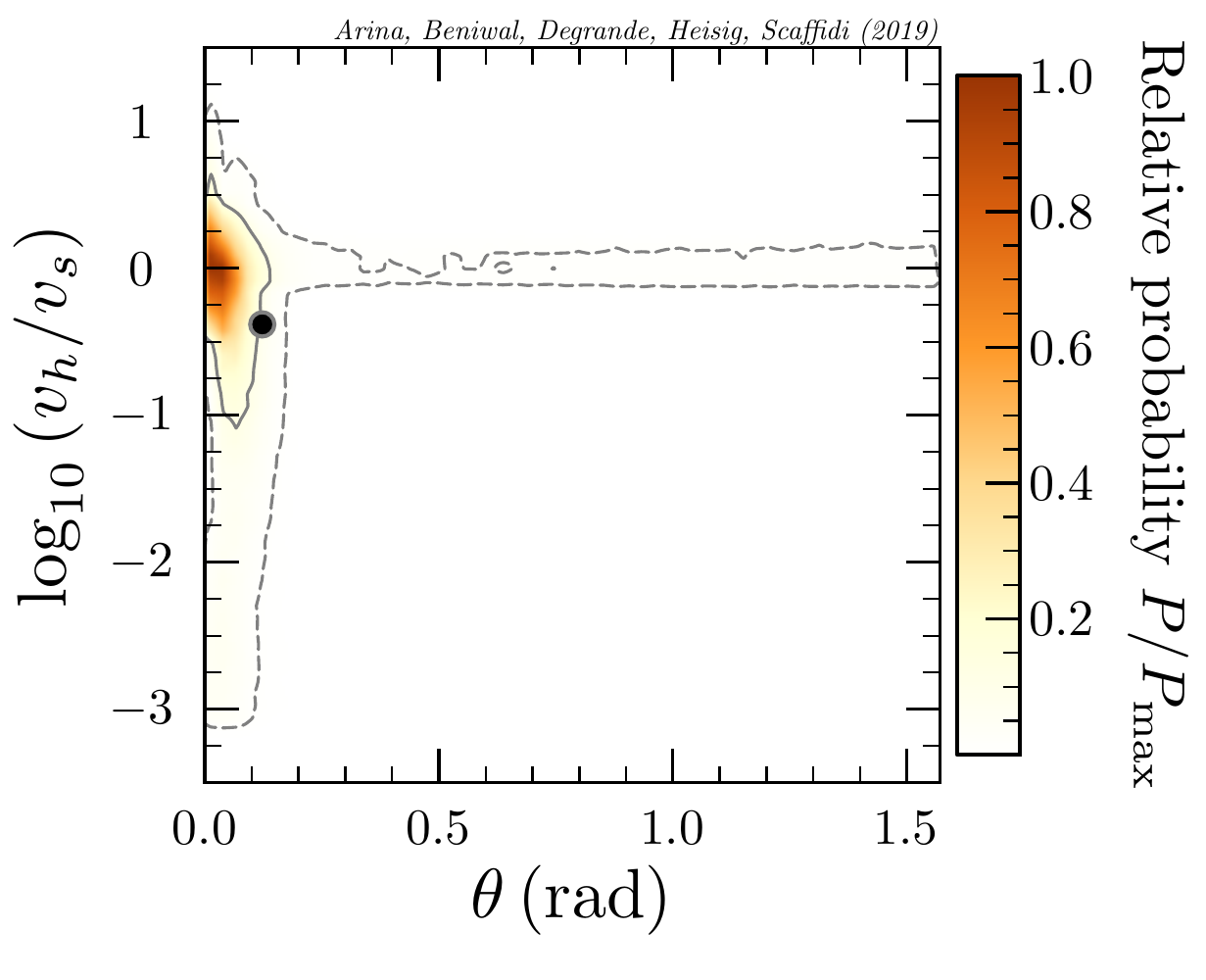}	
    \caption{2D marginalised posterior (MP) distributions in planes of the pNG DM model parameters.~The $1\sigma$ ($2\sigma$) credible intervals are marked by solid (dashed) lines. The posterior mean is shown as a black circle.}
    \label{fig:marg1}
\end{figure}

In figure~\ref{fig:marg1}, we show the MP distributions in various 2D planes of the model parameter space.~Similar to the PLR plots in figure~\ref{fig:prof1}, these are also generated using \pippi \cite{Scott:2012qh}. The $1\sigma$ $(2\sigma)$ credible intervals are marked by solid (dashed) lines.~The posterior mean is shown as a black circle.~In each panel, model parameters that are not shown are integrated/marginalised over.~Consequently, regions with a smaller ``volume of support'' \cite{Athron:2018hpc} are less favoured as they require an extra degree of tuning of model parameters to satisfy all of the included constraints.

In comparison to the PLR plots in figure~\ref{fig:prof1}, the allowed regions in the MP plots are more constrained, especially where a large degree of tuning is required from marginalising over the model parameters. Again, we see a vertical stripe in the $(m_H,\,m_\chi)$-plane. On the other hand, the second resonance region, $m_\chi \simeq m_{h,\,H}/2$, is less-favoured as it falls outside the $2\sigma$ credible interval due to an extra need for tuning over $v_h/v_s$. In addition, regions where $m_\chi > m_H$ also appears to be fine-tuned, especially after marginalising over $v_h/v_s$.

In the $(m_H,\,\theta)$- and $(m_\chi,\,\theta)$-planes, large values of $\theta$ fall inside the $2\sigma$ credible interval. On the other hand, regions with $\theta \lesssim 0.1$~rad have a larger volume of support, as is evident from a large posterior density. In the $(m_\chi,\,v_h/v_s)$-plane for $m_\chi \gtrsim 100$\,GeV, the $1 \sigma$ credible interval is larger than the $1\sigma$ CL region seen in the PLR plots.~On the other hand, $m_\chi \lesssim m_h/2$ region requires a large degree of fine-tuning in $v_s$ and $m_H$  to satisfy the relic density constraint, and thus is less favoured.~Lastly, in the $(\theta,\,v_h/v_s)$-plane, the posterior mass is large for $\theta \lesssim 0.1$~rad. However, large values of $\theta$ are still allowed as they fall within the $2\sigma$ credible interval.

\begin{figure}[t]
    \centering
    
    \includegraphics[width=0.48\textwidth]{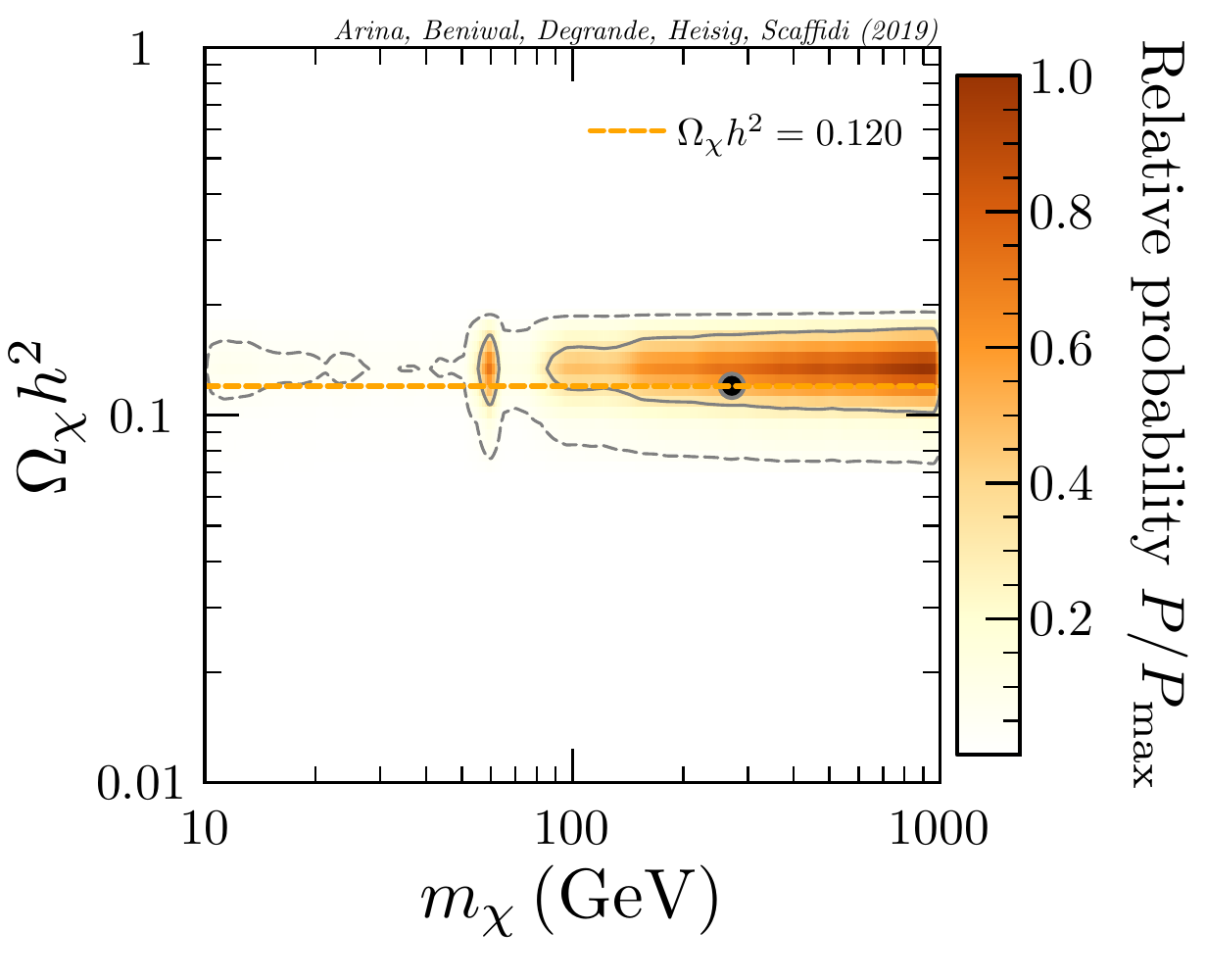} \quad
    \includegraphics[width=0.48\textwidth]{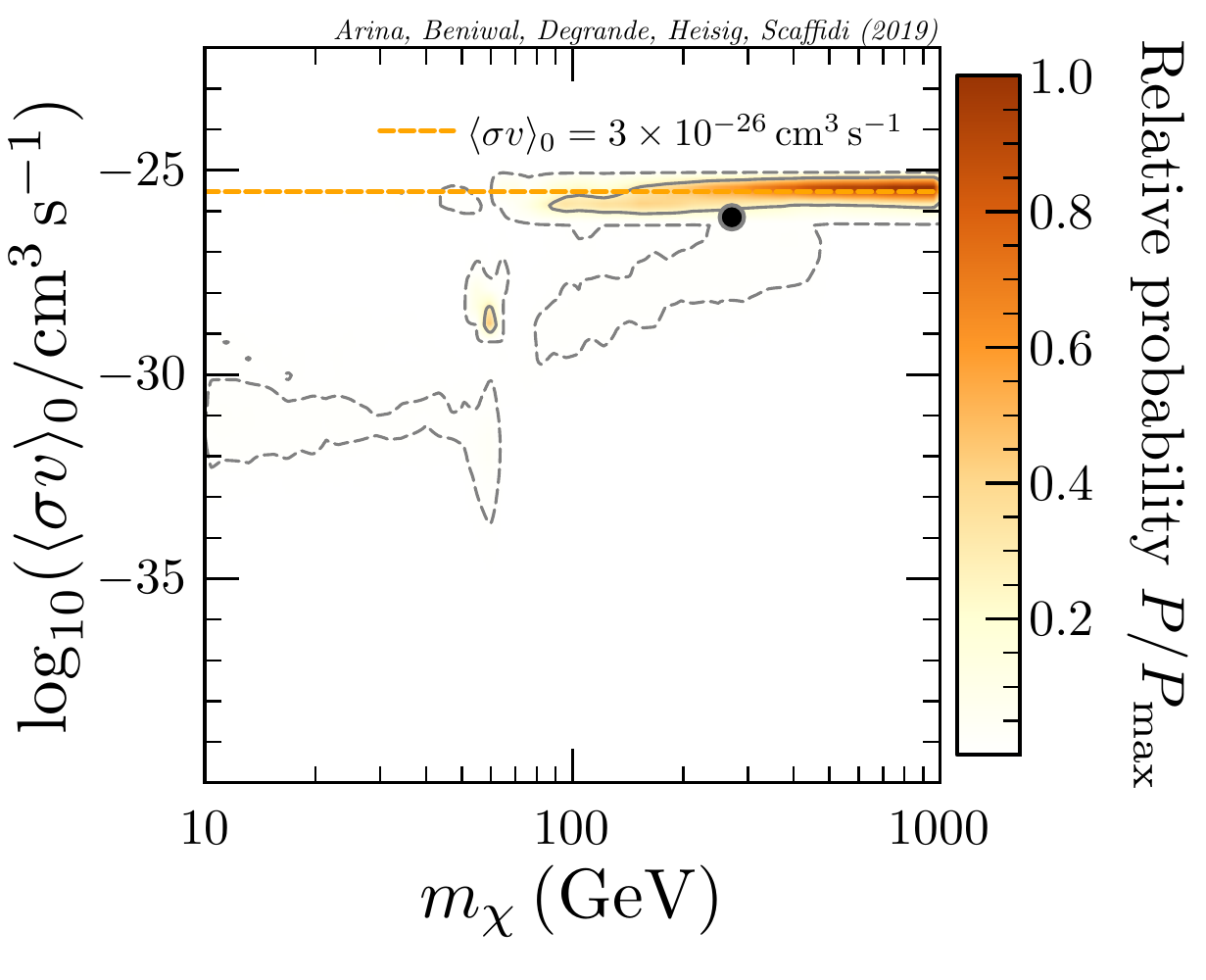} 
    \vspace{-1cm}
    \caption{2D MP distributions for key DM observables such as the pNG DM relic density (\emph{left panel}) and its annihilation rate today (\emph{right panel}).~The meaning of solid (dashed) lines and black circle is same as in figure~\ref{fig:marg1}.}
    \label{fig:marg2}
\end{figure}

In figure~\ref{fig:marg2}, we show the MP distributions for key DM observables. In contrast to the right panel of figure~\ref{fig:prof3}, we do not see at least 4 orders of magnitude smaller DM annihilation cross sections than the freeze-out value; the region with velocity suppressed annihilation cross section is somehow fine-tuned and less favoured after marginalising over $m_H$ \& $v_h/v_s$.

\subsection{Post-processing of samples}
In addition to the constraints included in our global fit, we consider indirect and direct detection constraints, see sections~\ref{sec:FermiCon} and~\ref{sec:direct}, respectively. For the computation of corresponding observables, we post-process our final samples.~This greatly reduces the computational time, in particular, for indirect detection constraints.~The \fermi likelihood is computationally intensive due to the generation of the annihilation spectra for $2 \to 2$ up to $2 \to 6$ processes (see section~\ref{sec:FermiCon}).~We nevertheless expect a sufficient coverage within the resulting $(1\!-\!2)\sigma$~CL contours after combining various scans.\footnote{As stated in footnote \ref{fn:sets1}, we combine results from several \multinest\ scans.~This is done with various specific priors to guarantee sufficient coverage in the resonant regions as well as in regions preferred by \fermi when considering 45 dSphs. The resulting chain contains more than 3 million points.\label{fn:sets2}}~Accordingly, for the post-processed samples, we provide a frequentist interpretation only.~While indirect detection constraints from \fermi\ observations of dSphs have a significant effect on the PLR, current direct detection experiments are not yet sensitive to our model, as we will show below.~We thus refrain from including a likelihood for the latter, and restrict ourselves to comparing the model prediction to the reach of current and future experiments for this case.

\subsubsection{Indirect detection}
As explained in section~\ref{sec:FermiCon}, we consider two cases regarding the set of dSphs included.~We take into account the likelihoods from all 45 dSphs considered in ref.~\cite{Fermi-LAT:2016uux} as well as excluding the four dSphs that show an excess, correspondingly including 41 dSphs. The latter choice only imposes an upper limit on the annihilation cross section and is described first.

\begin{figure}[t]
    \raggedright
	
	\includegraphics[width=0.35814\textwidth,trim={0 1.8cm 0 0},clip]{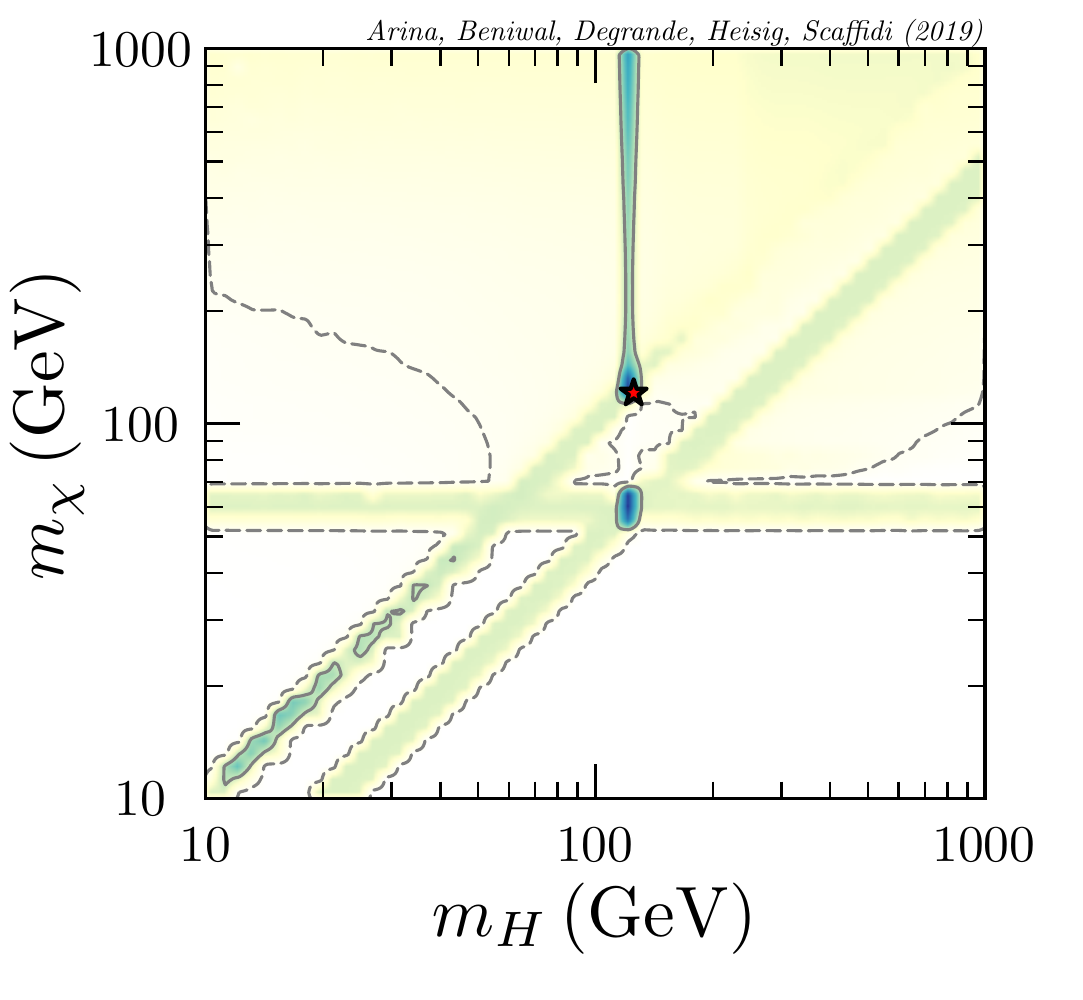}
	
	\hspace{-0.66mm} 
	\includegraphics[width=0.35814\textwidth,trim={0 1.8cm 0 0.48cm},clip]{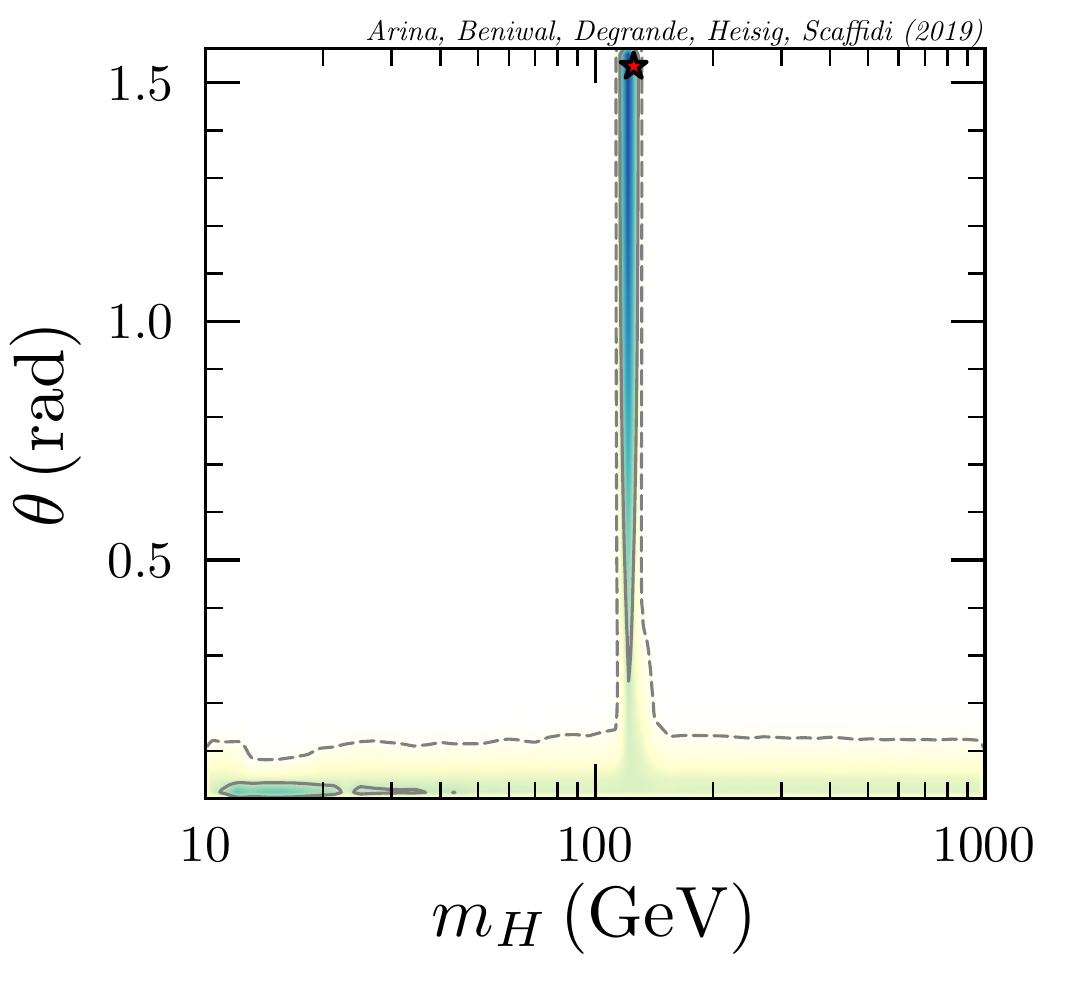} 
	\hspace{-4.0mm}
	\includegraphics[width=0.301176\textwidth,trim={1.8cm 1.8cm 0 0},clip]{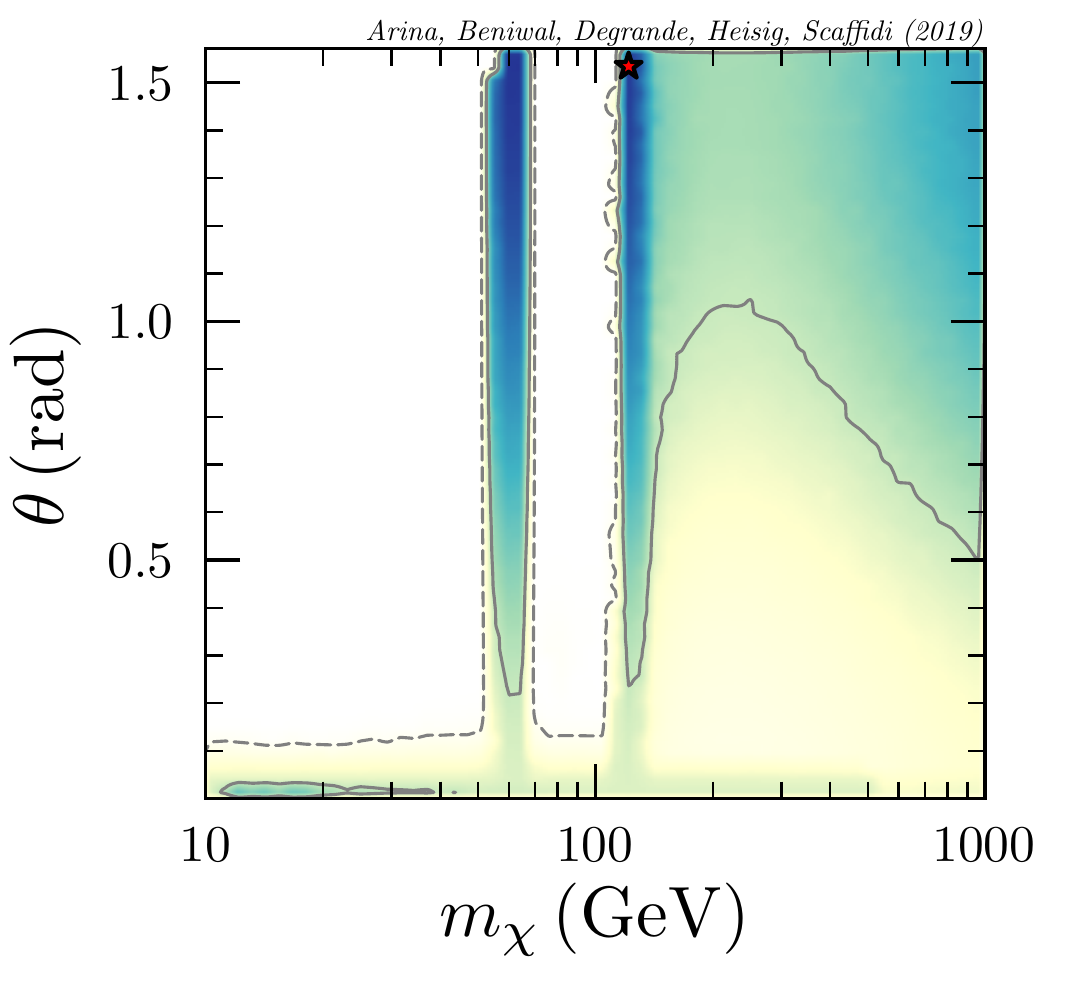}
	
	\hspace{-0.66mm} 
	\includegraphics[width=0.35814\textwidth,trim={0 0 0 0.48cm},clip]{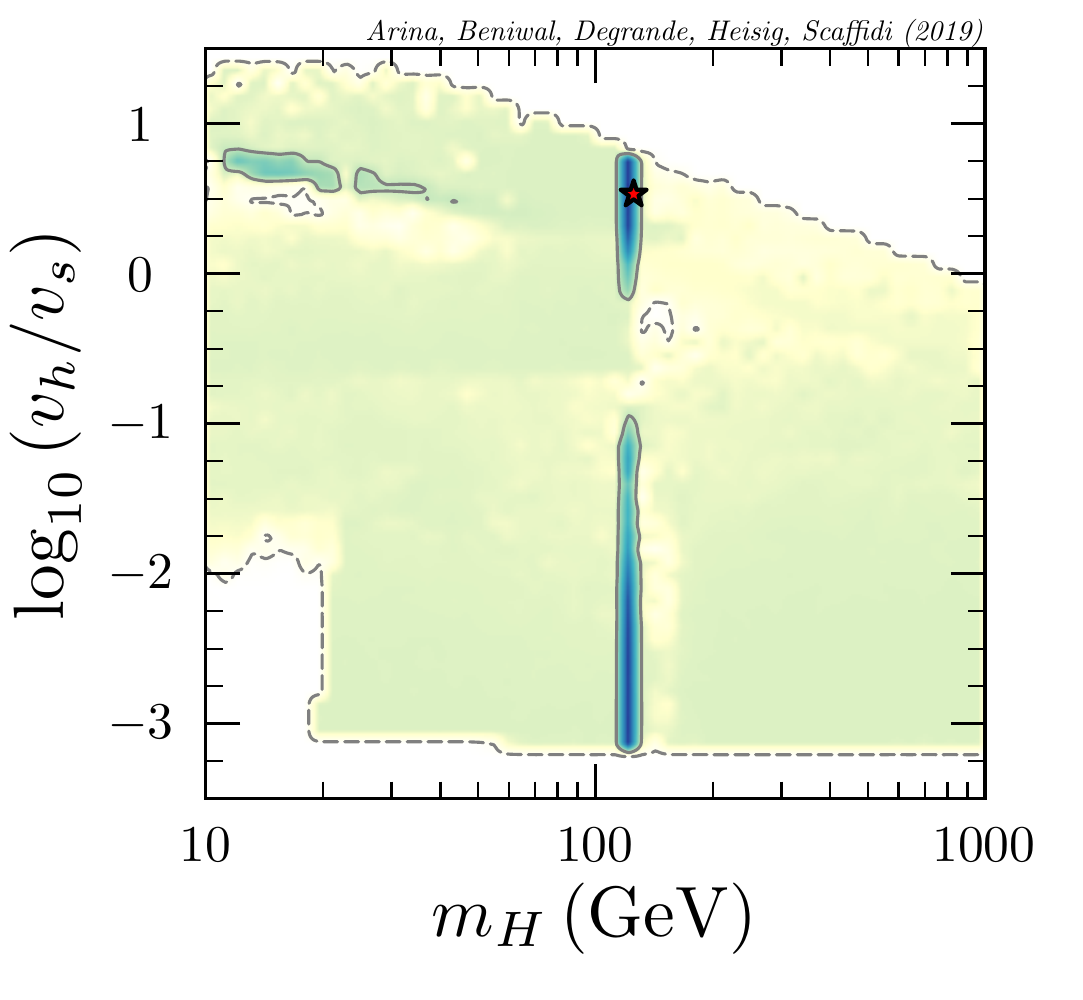}
	\hspace{-4.0mm}
	\includegraphics[width=0.301176\textwidth,trim={1.8cm  0 0 0.48cm},clip]{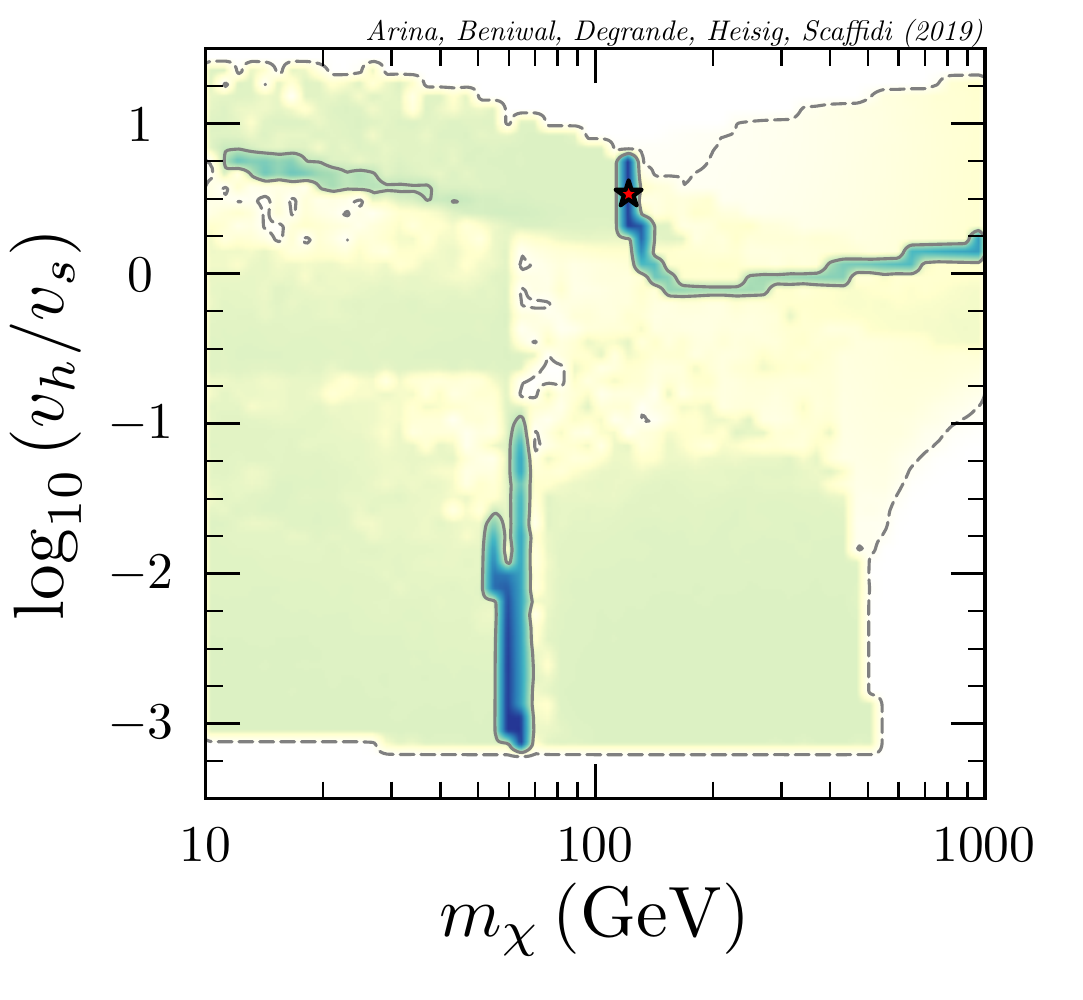}
	\hspace{-4.0mm}
	\includegraphics[width=0.34968\textwidth,trim={1.8cm  0 0 0},clip]{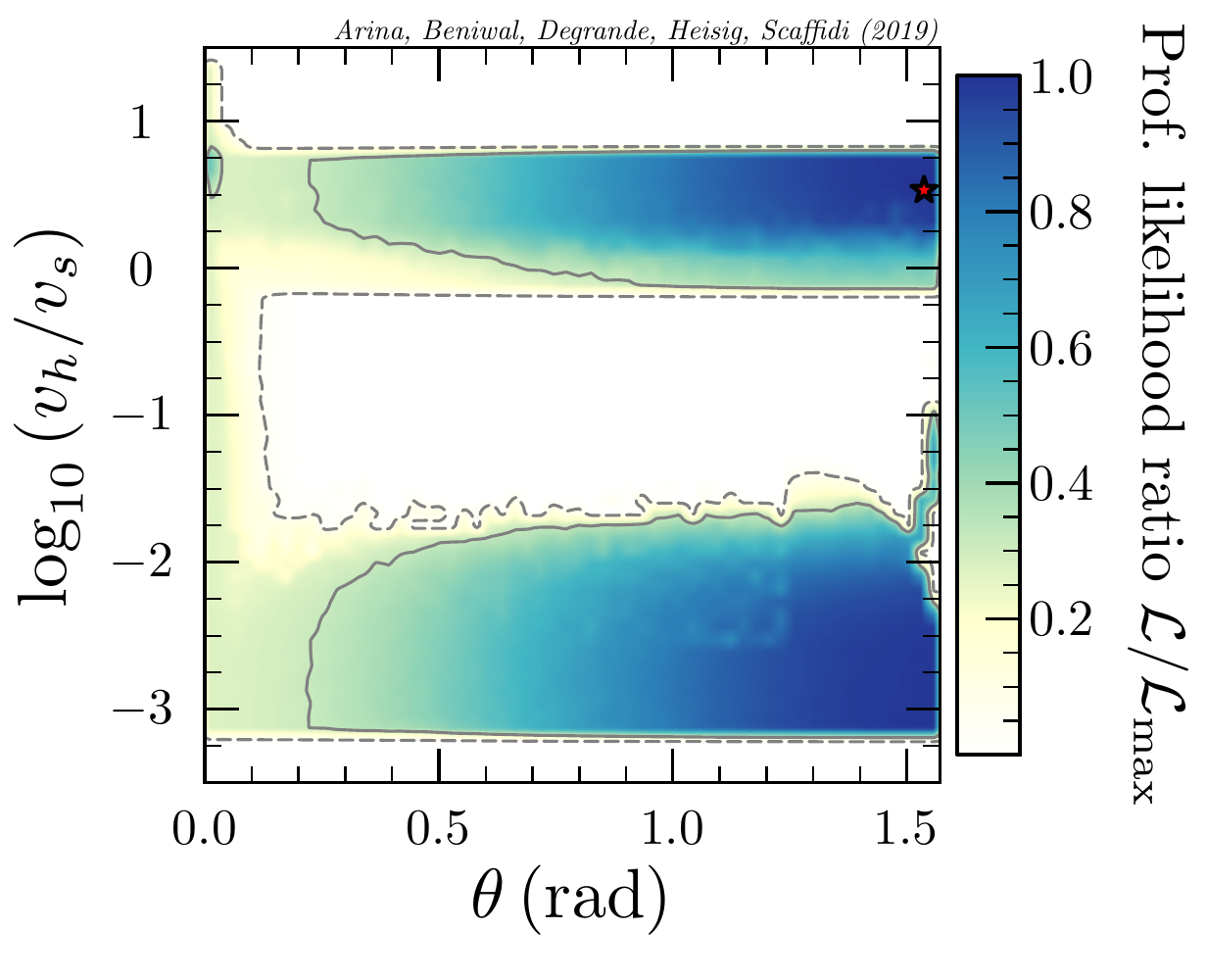}

    \caption{2D PLR plots for the pNG DM model parameters after post-processing our \multinest samples with \fermi likelihood from 41 dSphs. The best-fit point is shown as a red star and summarised in column 2 of table~\ref{tab:bestfit}.}
    \label{fig:post-process-41-dSph-1}
\end{figure}

\begin{figure}[t]    
    \centering 
    
    \includegraphics[width=0.48\textwidth]{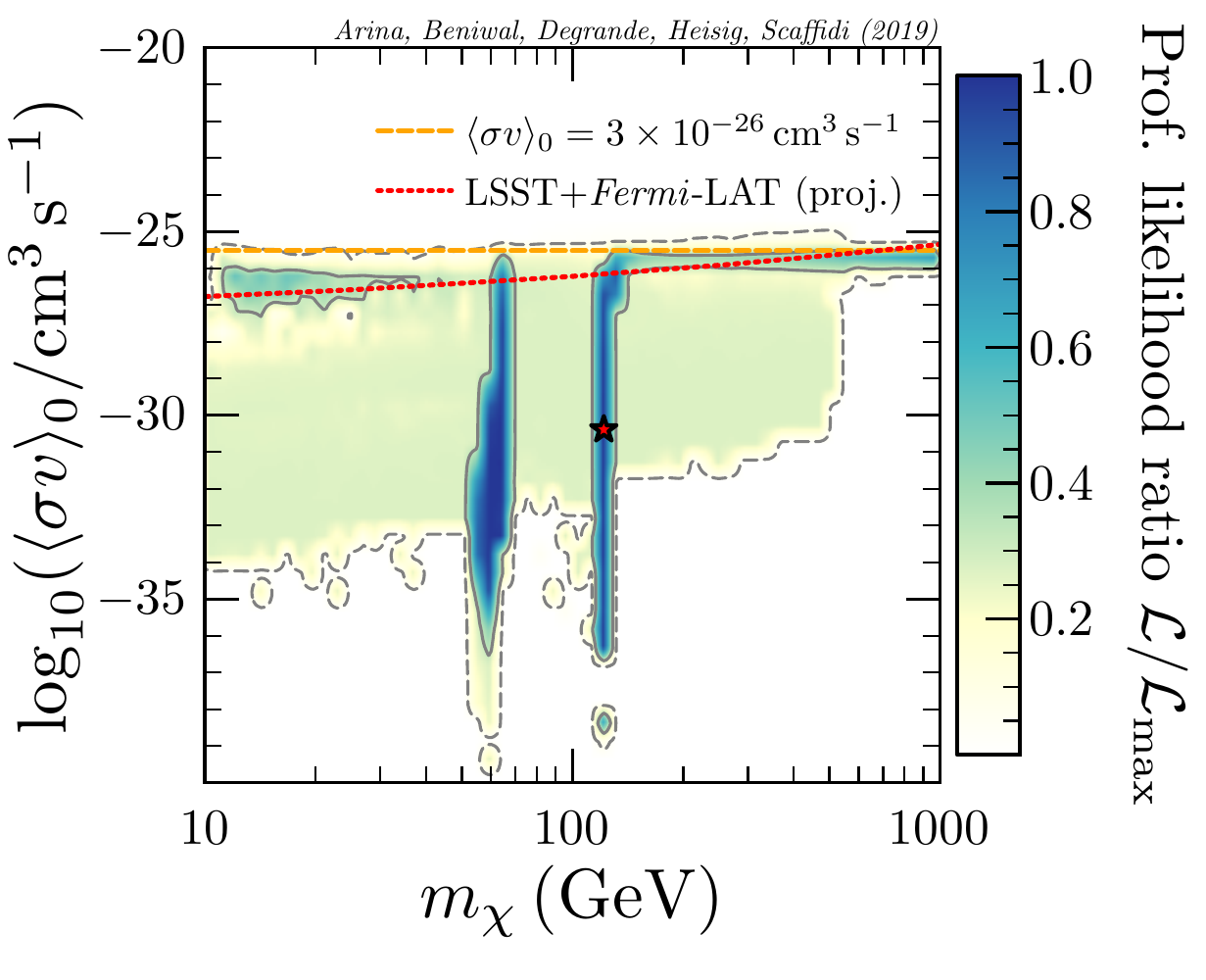} \quad
    \includegraphics[width=0.48\textwidth]{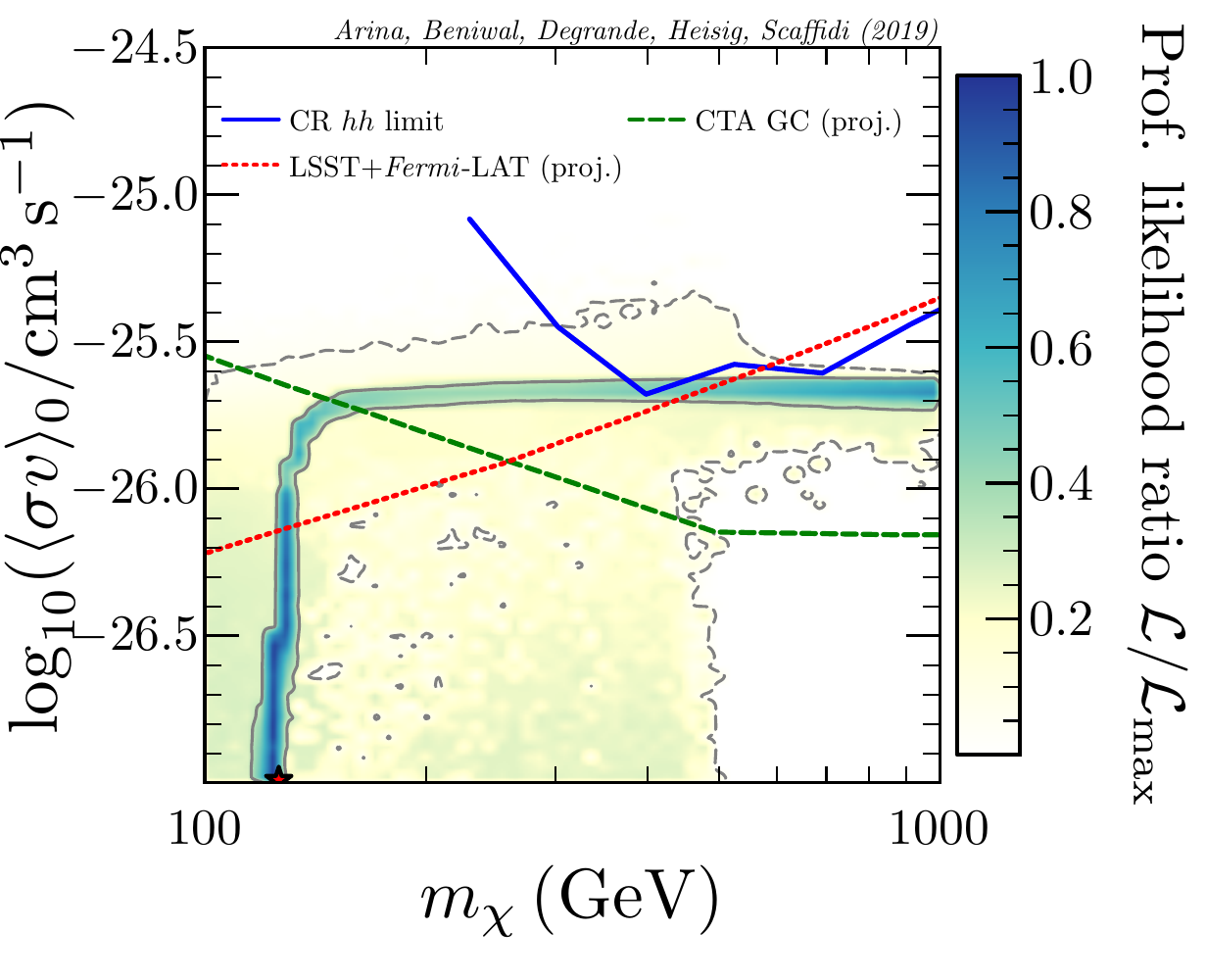}
    \vspace{-1cm}
    \caption{\emph{Left panel}: 2D PLR plots for the pNG DM annihilation cross section today after post-processing our samples with \fermi likelihood for 41 dSphs.~Projected limits from LSST+\fermi dSphs \cite{Drlica-Wagner:2019xan} are shown as dotted red curve.~\emph{Right panel}: Same as the left panel, except for $m_\chi \in [100, 1000]$\,GeV. The solid blue curve shows the current cosmic-ray (CR) antiproton limit for the $\chi \chi \rightarrow hh$ channel \cite{Cuoco:2017iax}. Projected limits from CTA Galactic Centre (GC) \cite{EcknerCTA2018} (dashed green), and LSST+\fermi dSphs~\cite{Drlica-Wagner:2019xan} (dotted red) for the $b\ovr{b}$ channel are also shown.}
    \label{fig:post-process-41-dSph-2}
\end{figure}

In figures~\ref{fig:post-process-41-dSph-1} and \ref{fig:post-process-41-dSph-2}, we show the PLR plots in the planes of pNG DM model parameters as well as the DM relic abundance and annihilation cross section today, respectively, after accounting for the likelihoods of 41 dSphs.~The implications on the parameter space compared to our global fit results (see figure~\ref{fig:prof1}) are moderate. 
However, for $m_\chi \lesssim 100$\,GeV, the \fermi\ limits exclude a large portion of the parameter space where the pNG DM annihilation today proceeds via $\chi\chi\to HH$ channel, i.e., where $m_\chi > m_H$.~The constraint becomes stronger for smaller DM masses as lighter DM requires a larger DM number density to match the same energy density. This enhances the annihilation rate.~The tendency is partly softened by the fact that for a given $m_H$, the spectrum becomes more peaked for larger $m_\chi$, which tends to strengthen the constraints. 

Taking into account the \fermi\ likelihood, the new best-fit point has moved to the region of dominant annihilation into Higgs pairs during freeze-out ($\chi\chi\to hh$ in this case), see column 2 in table~\ref{tab:bestfit}.~However, $m_\chi$ is slightly smaller than $m_h$ such that $\chi\chi\to hh$ is kinematically forbidden today. Thus, $\langle \sigma v \rangle_0$ is largely suppressed as now it proceeds via (a highly off-shell) Higgs propagator (dominantly into $WW^*$ final states).~Consequently, the best-fit point effectively evades any constraint from indirect detection.

Note that other indirect detection searches can impose further constraints on the parameter space.~Here we would like to comment on current constraints from cosmic-ray (CR) antiproton fluxes as measured by AMS-02~\cite{Aguilar:2016kjl}.~While the corresponding analyses are typically plagued by large CR propagation uncertainties, recent progress has been made by fitting propagation and DM parameters at the same time~\cite{Cuoco:2017iax}.~This analysis provides very strong bounds on the DM annihilation cross section for DM masses above 200\,GeV. While performing a respective, dedicated analysis for the considered model is beyond the scope of this work, we can, nevertheless, interpret the results of ref.~\cite{Cuoco:2017iax} in parts of our model parameter space. The analysis provides limits for annihilation into a pair of Higgses with $m_h=125$\,GeV. In our model, $m_H\sim m_h$ in the entire $1\sigma$ CL region.~Moreover, for $m_\chi>200\,$GeV (where the analysis becomes constraining), the dominant annihilation channels are $\chi\chi\to hh$, $hH$ and $HH$.
Hence, for the $1\sigma$ CL region, the result from ref.~\cite{Cuoco:2017iax} can be directly applied without approximation (except for neglecting the small difference between $m_H$ and $m_h$, which is however, not expected to have a noticeable effect on the gamma-ray energy spectrum).~We show the corresponding 95\% CL limit as solid blue curve in the right panel of figure~\ref{fig:post-process-41-dSph-2}. For the $2\sigma$ CL region, in general, $m_H\neq m_h$. Nevertheless, the solid blue curve is expected to provide an order of magnitude estimate of the sensitivity of CR antiproton searches.

With future experiments, the pNG DM parameter space can be tested with gamma-ray observation with improved sensitivity.~A large part of the region of dominant annihilation into Higgs pairs is expected to be probed in the near future by a combination of new dSphs discovered by the Large Synoptic Survey Telescope (LSST)~\cite{Zhan:2017uwu} with \fermi observations.~First, the inclusion of more satellite galaxies will augment the \fermi data~\cite{Drlica-Wagner:2019xan,Ando:2019rvr}.~Secondly, the LSST novel spectroscopic observations will provide precise measurements of $J$-factors, decreasing the associated astrophysical uncertainties. These improvements are expected to provide sensitivity to the thermal cross-section for DM masses up to around 600\,GeV. For illustration, we show the corresponding projected limit for annihilation into $b\bar b$ (assuming 18 years of observation) as dotted red curve in figure~\ref{fig:post-process-41-dSph-2}. Furthermore, in the right panel of figure~\ref{fig:post-process-41-dSph-2} only, we display the recent projection for Galactic centre gamma-ray observations with the Cherenkov Telescope Array (CTA)~\cite{EcknerCTA2018}, assuming 500 hours of exposure (no systematics), again using DM annihilation into $b\bar b$ as a benchmark channel, which is expected to provide a reasonable order of magnitude estimate for the sensitivity for our model.~It can probe cross sections down to $5 \times 10^{-27}$\,cm$^3$ s$^{-1}$ for masses above 500 GeV, i.e., a large portion of the allowed pNG DM parameter space characterised by dominant annihilation into Higgs pairs.

\begin{figure}[t]
    \raggedright
	
	\includegraphics[width=0.35814\textwidth,trim={0 1.8cm 0 0},clip]{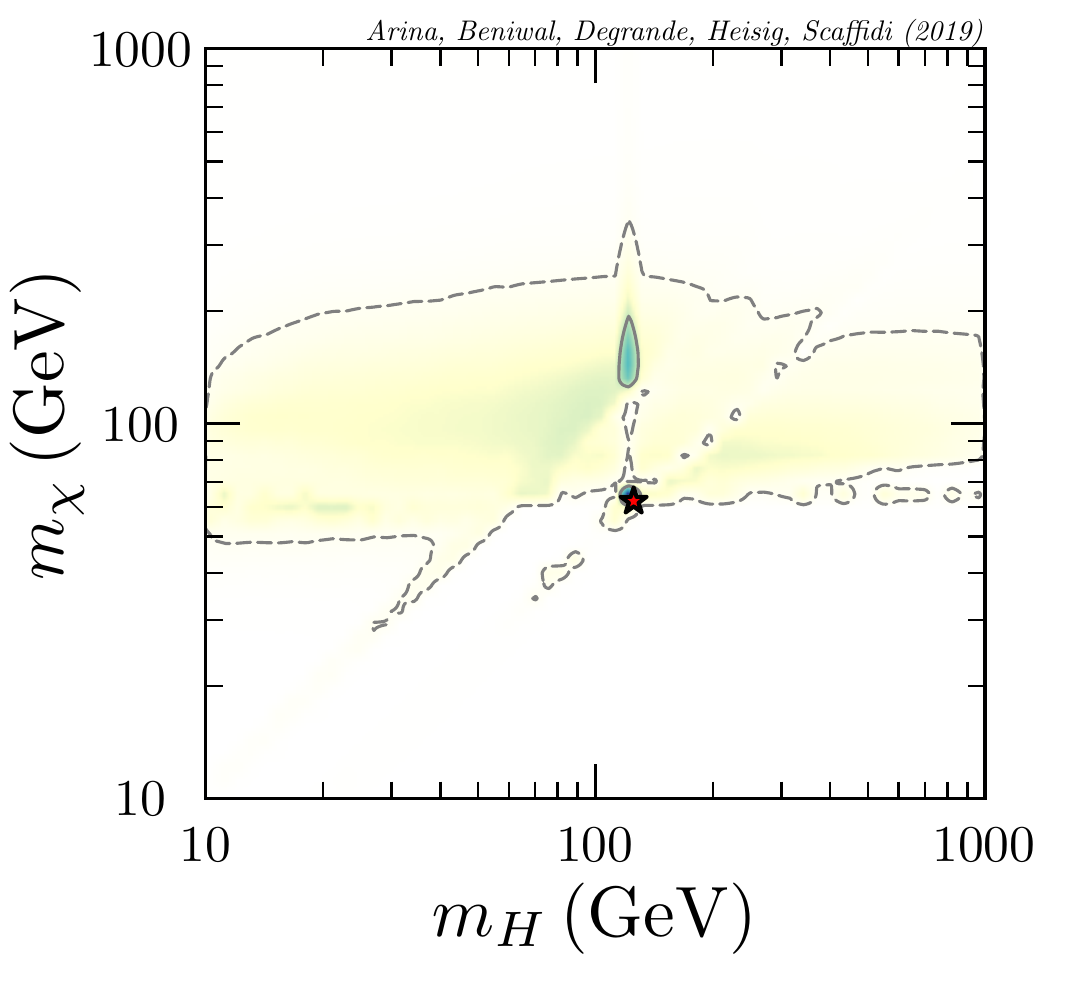} 
	
	\hspace{-0.66mm} 
	\includegraphics[width=0.35814\textwidth,trim={0 1.8cm 0 0.48cm},clip]{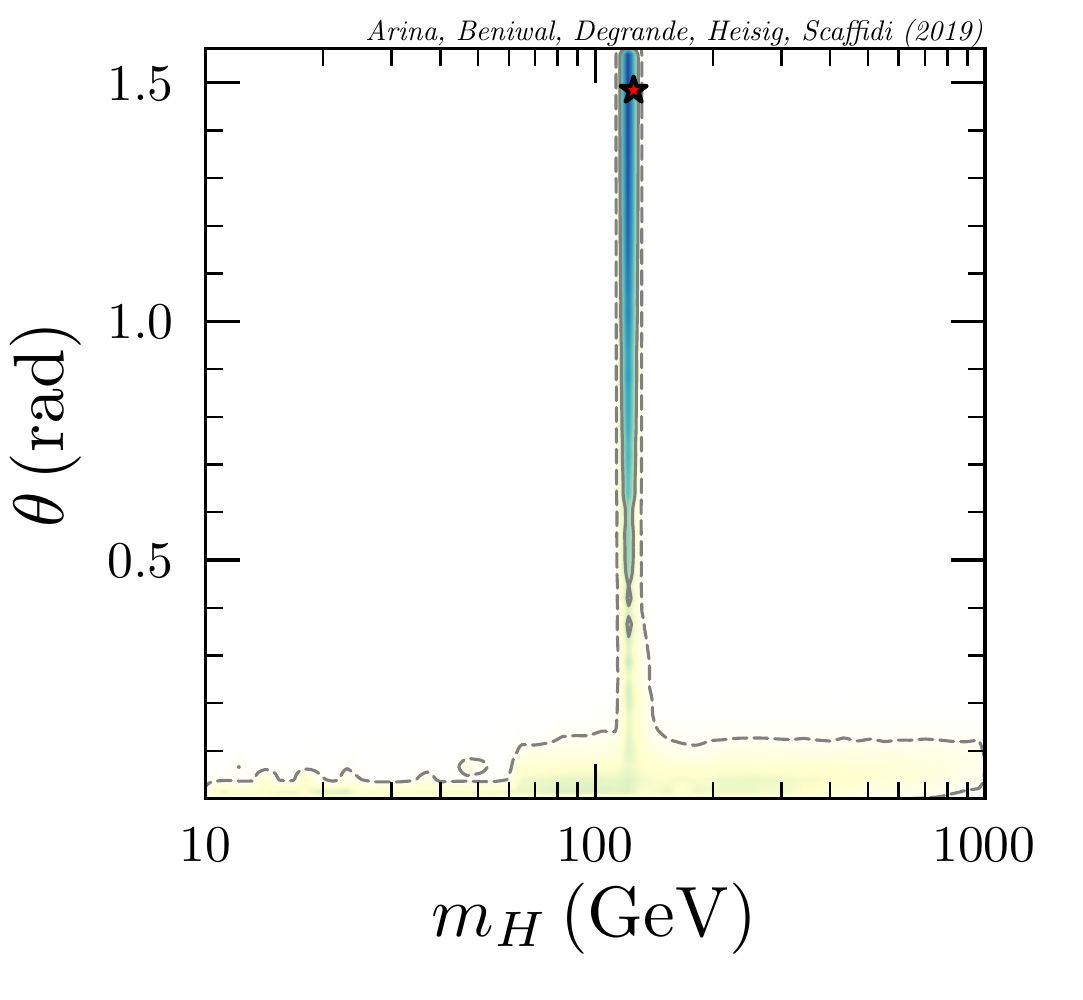} 
	\hspace{-4mm}
	\includegraphics[width=0.301176\textwidth,trim={1.8cm 1.8cm 0 0},clip]{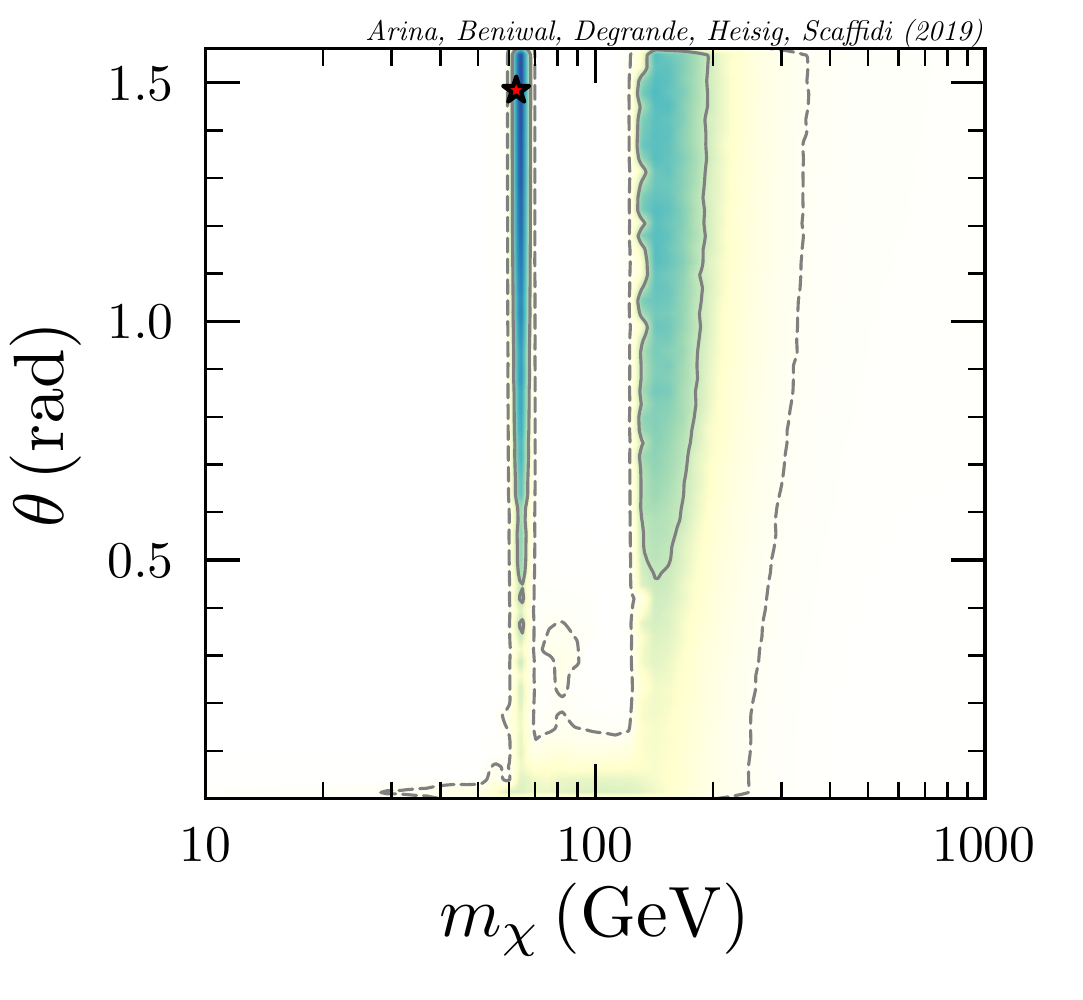}
	
	\hspace{-0.66mm} 
	\includegraphics[width=0.35814\textwidth,trim={0 0 0 0.48cm},clip]{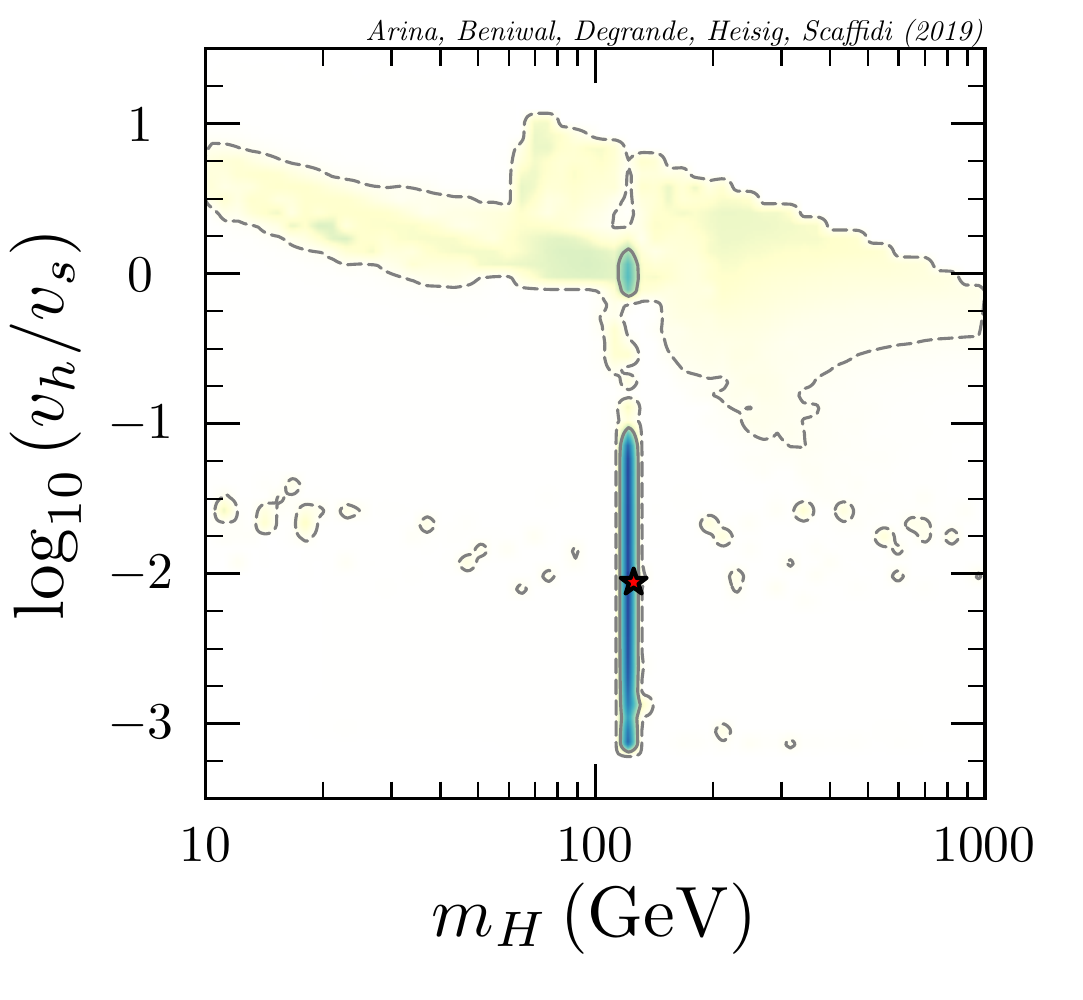}
	\hspace{-4.0mm}
	\includegraphics[width=0.301176\textwidth,trim={1.8cm  0 0 0.48cm},clip]{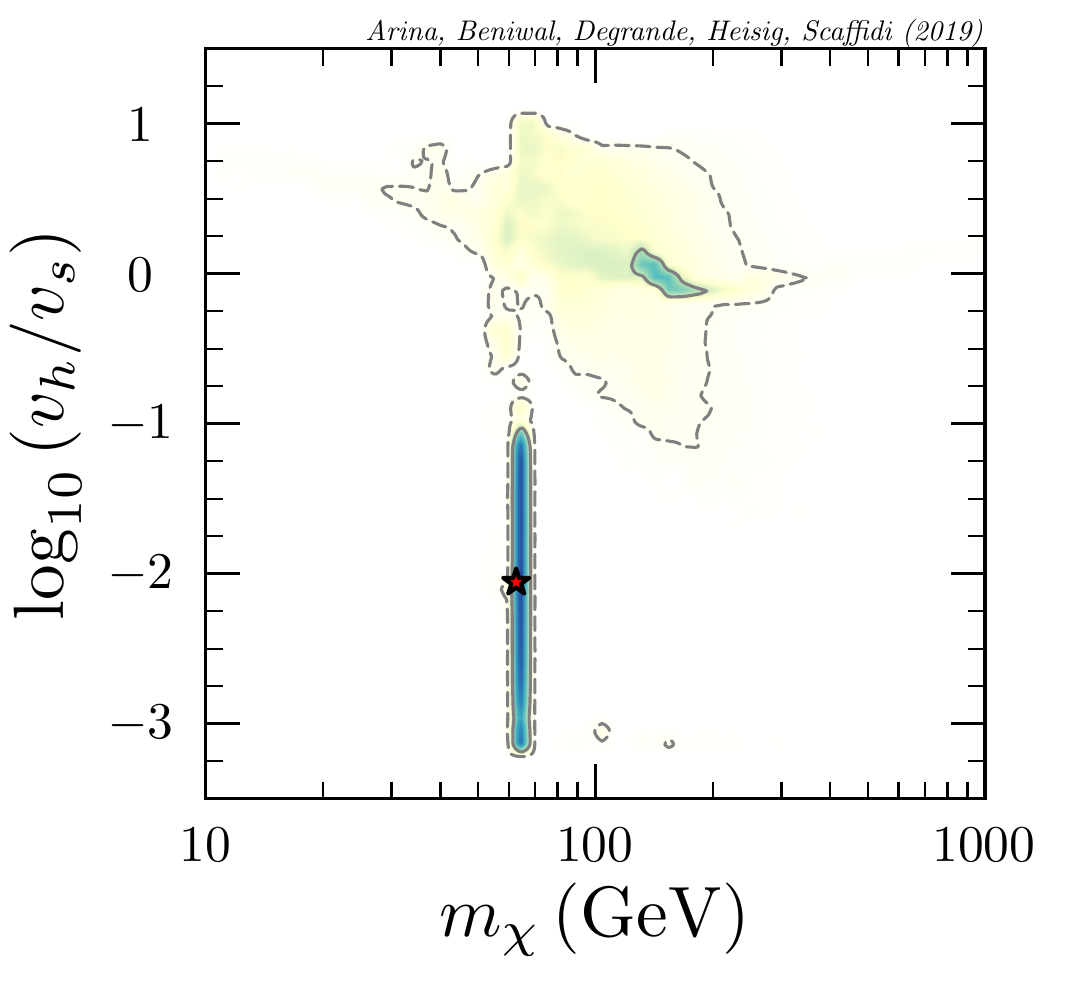}
	\hspace{-4.0mm}
	\includegraphics[width=0.34968\textwidth,trim={1.8cm  0 0 0},clip]{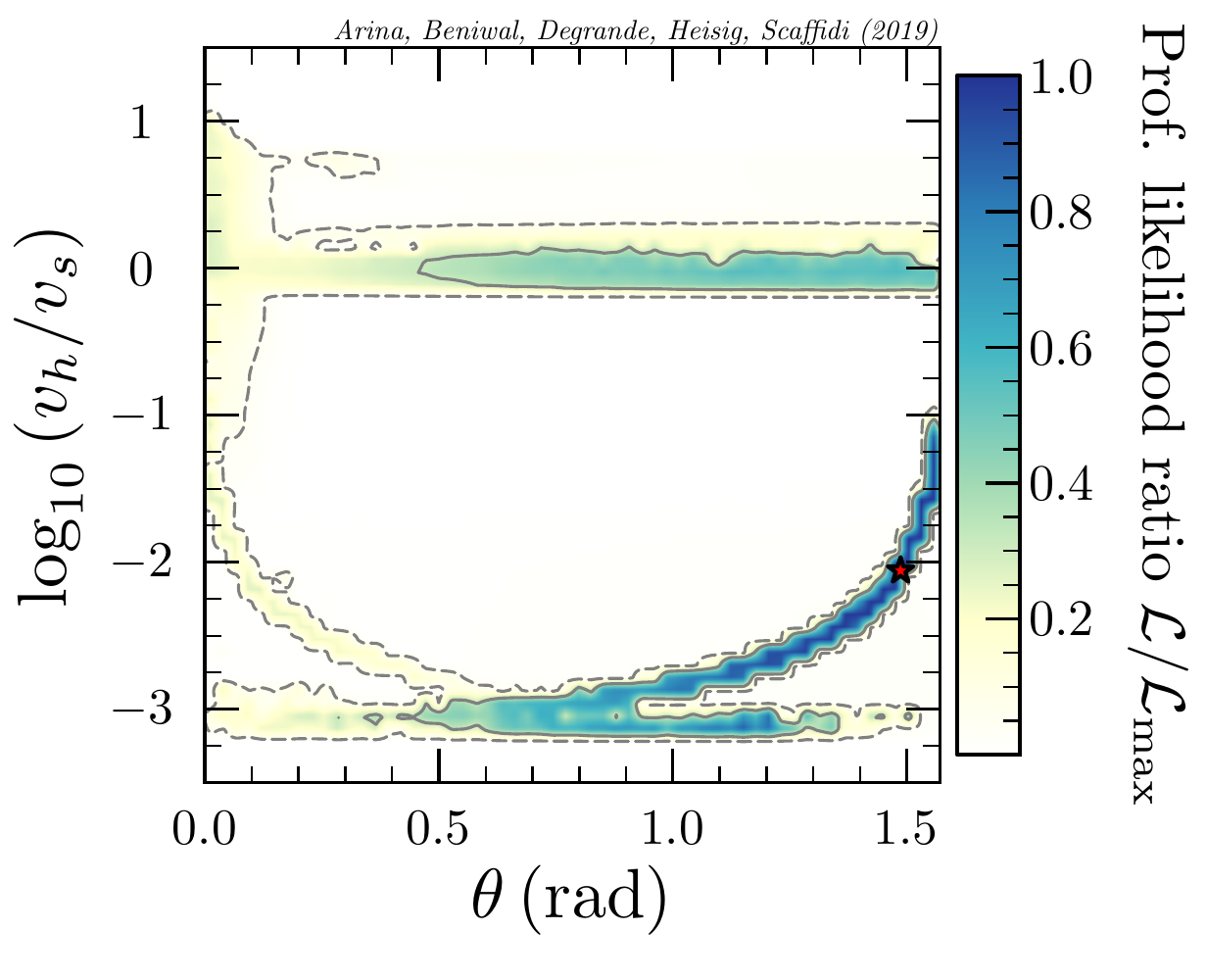} \vspace{0.6mm}
    
    \caption{2D PLR plots for the pNG DM model parameters after post-processing our samples with \fermi likelihood from 45 dSphs.}
    \label{fig:post-process-45-dSph-1}
\end{figure}

\begin{figure}[t]
    \centering
    
    \includegraphics[width=0.48\textwidth]{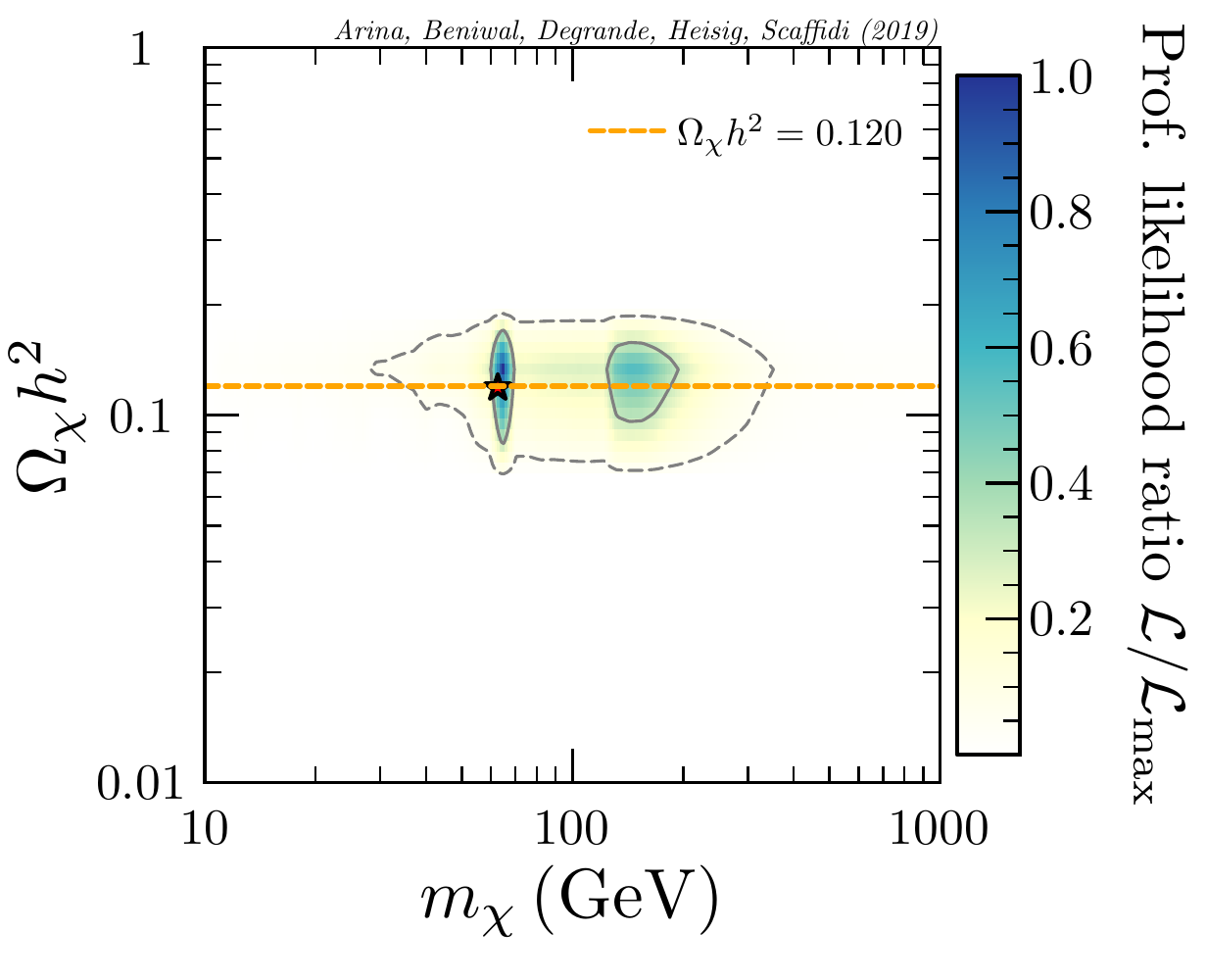} \quad
    \includegraphics[width=0.48\textwidth]{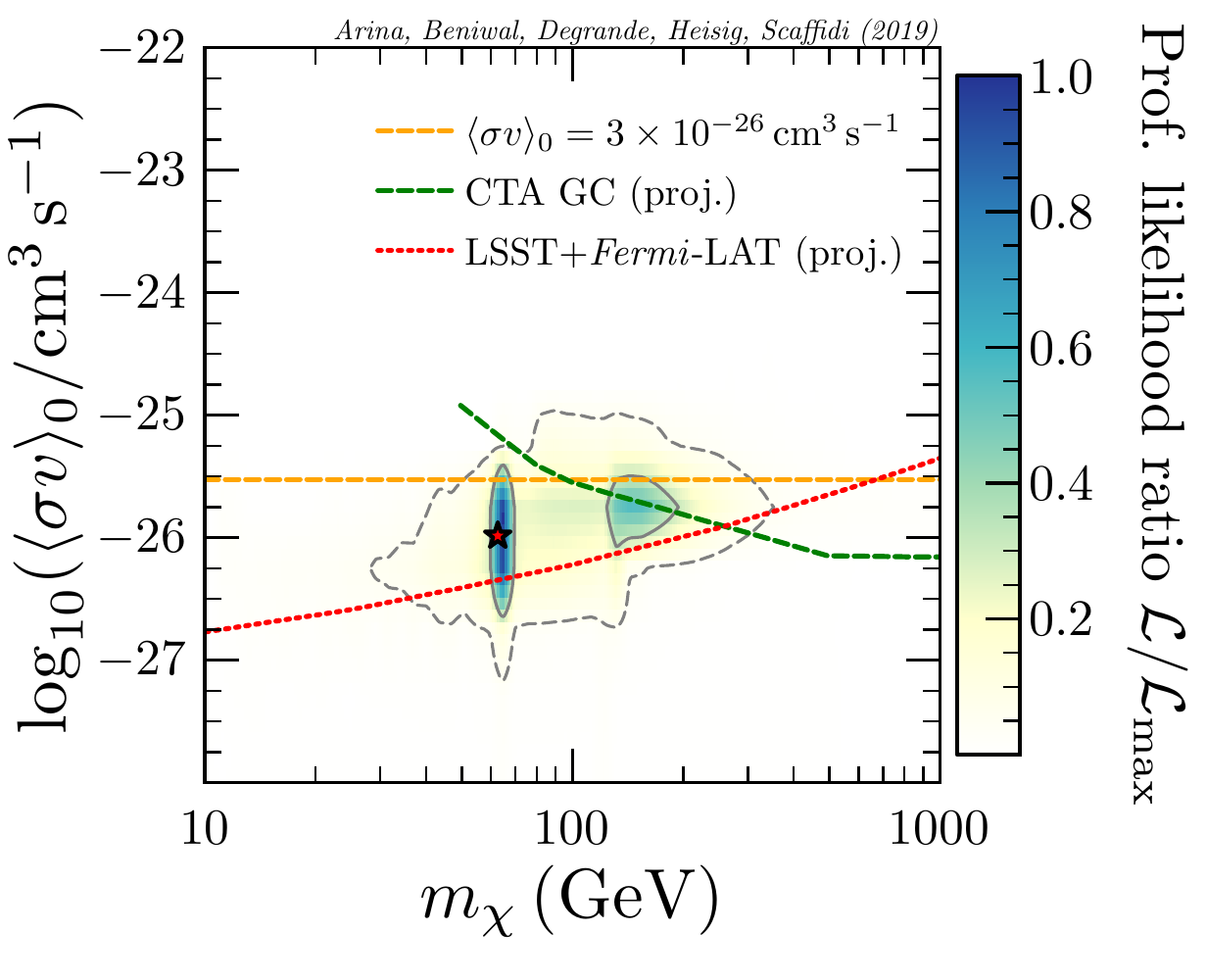}
    \vspace{-1cm}
    \caption{2D PLR plots for the pNG DM relic abundance and its annihilation cross section today after post-processing our samples with \fermi likelihood from 45 dSphs.~In the right panel, projected limits from CTA GC searches~\cite{EcknerCTA2018} and LSST+\fermi dSphs~\cite{Drlica-Wagner:2019xan} are also shown.}
    \label{fig:post-process-45-dSph-2}
\end{figure}

In figures~\ref{fig:post-process-45-dSph-1} and \ref{fig:post-process-45-dSph-2}, we show the respective results after accounting for the \fermi likelihood from all 45 dSphs, i.e.,~including Reticulum II, Tucana III, Tucana IV and Indus II, which exhibit slight excesses with a local significance of around $2\sigma$ each \cite{Geringer-Sameth:2015lua,Li:2015kag,Fermi-LAT:2016uux}.~Interestingly, the excess can be fitted by an annihilation cross section in the ballpark of the thermal one, $\langle\sigma v\rangle_0\sim 3\times 10^{-26}$\,cm$^3$ s$^{-1}$, i.e., in regions where the annihilation cross section today is similar to the one typically required during freeze-out.~This places additional constraints on the parameter space and excludes parts of the resonant region where the cross section is highly velocity dependent, and hence $\langle\sigma v\rangle_0$ deviates strongly from $\langle\sigma v\rangle_\text{FO}$.~More concretely, for DM masses below the point of maximal resonant enhancement, $\langle\sigma v\rangle_0<\langle\sigma v\rangle_\text{FO}$ and the flux today tends to be too low to fit the signal. In contrast, for DM masses above that point, the flux tends to be too high. Finally, in between, $\langle\sigma v\rangle_0\sim\langle\sigma v\rangle_\text{FO}$. This is approximately the point of maximal resonant enhancement of the thermally averaged cross section during freeze-out which allows for the smallest possible couplings in the scan. If the annihilation proceeds via an on-shell Higgs ($h$ or $H$), the respective cross section is proportional to
\begin{equation}
    \sigma \propto \frac{\cos^2\theta \sin^2\theta}{v_s^2\, \Gamma_{h/H}^\text{tot}}\,.
\label{eq:xsres}
\end{equation}
Here $\Gamma_{h/H}^\text{tot}$ is the total Higgs decay width, which is dominated by the partial width into SM particles if the DM mass is very close to $m_{h/H}/2$, such that the corresponding phase space is suppressed.~In this case, $\Gamma_{h/H}^\text{tot}$ is proportional to $\cos^2\theta$ ($\sin^2\theta$) for $h$ ($H$), thereby cancelling the respective factors in the numerator of eq.~\eqref{eq:xsres}. Consequently, the cross section for resonant annihilation via $h$ or $H$ is approximately proportional to $(\sin\theta/v_s)^2$ and $(\cos\theta/v_s)^2$, respectively.~The two regions 
can be recognised in the lower part of the $(\theta,\,v_h/v_s)$-plane as the two overlapping thin bands with decreasing ($\propto \sin^{-1} \theta$) and increasing ($\propto \cos^{-1}\theta$) slope. The best-fit point falls in the second band where the pNG DM annihilation proceeds dominantly via a resonant $H$ (see column 3 in table~\ref{tab:bestfit}).

The second region fitting the signal is characterized by annihilation into a pair of Higgses, where (unless very close to or below threshold) no strong velocity dependence of the annihilation cross section is present, and thus $\langle\sigma v\rangle_0\sim\langle\sigma v\rangle_\text{FO}$ naturally.~This region extends from the respective threshold up to around 200\,(300)\,GeV within $1\sigma$ ($2\sigma$) CL region away from the best-fit point.~Accordingly, the  allowed region spans roughly an order of magnitude in the pNG DM mass, i.e.,~around 30--300\,GeV. 

Note that in ref.~\cite{Cline:2019okt}, the pNG DM model has been considered
as an explanation of the gamma-ray galactic centre excess \cite{Goodenough:2009gk,Hooper:2011ti,Abazajian:2012pn,Hooper:2013rwa,Gordon:2013vta,Abazajian:2014fta,Daylan:2014rsa,Calore:2014xka,TheFermi-LAT:2015kwa} and the CR antiproton excess~\cite{Cuoco:2016eej,Cui:2016ppb,Cuoco:2017rxb,Cuoco:2019kuu,Cholis:2019ejx}.~While a discussion of the robustness of a DM explanation of these excesses as well as an explicit interpretation is beyond the scope of this work, we briefly comment on this possibility. In particular, we distinguish two cases.
\begin{enumerate}
    \item For pNG DM annihilation via a resonant $h$ or $H$ exchange in the $s$-channel, the composition of final states is the same as for the (singlet scalar) Higgs portal model, unless $m_\chi>m_{h,\,H}$.~For this case, explicit fits to the gamma-ray galactic centre excess~\cite{Cuoco:2016jqt} and the CR antiproton excess~\cite{Cuoco:2017rxb} have been performed. The latter analysis also provides a joint fit of both observations and the above-mentioned excess in the \fermi\ dSphs.~It reveals that all three observations, if arising from DM annihilation, are compatible and point to a DM mass of around (50--60)\,GeV and a velocity averaged annihilation cross section today of around $(1\!-\!2)\times10^{-26}$\,cm$^3$ s$^{-1}$. 
    
    \item The other case concerns dominant annihilation into $h$ or $H$. Due to their subsequent decays into lighter SM particles, their gamma-ray spectra are typically softer.~For instance, just above its threshold, the photon spectrum for $\chi\chi\to HH \to b\bar b b \bar b$ has the same shape as the one for $\chi\chi\to b\bar b$ but is shifted by a factor of 2 towards smaller energies. This is also reflected in the fact that the three observations can be fitted by DM annihilation into a SM-like Higgs for masses around the threshold~\cite{Cuoco:2017rxb}, i.e.,~roughly a factor of two larger than the DM mass providing the best fit for $\chi\chi\to b\bar b$.
\end{enumerate}
In conclusion, the regions that fit the gamma-ray galactic centre excess and the CR antiproton excess are very similar to those preferred by the 45 dSphs (see figures~\ref{fig:post-process-45-dSph-1} and \ref{fig:post-process-45-dSph-2}). 
We expect all three observations to be well fitted by a significant subset of the parameter points fitting the 45 dSphs.

Similar to the case of 41 dSphs (see figure~\ref{fig:post-process-41-dSph-2}), we also display the projected limits (for the $b \bar b$ channel) for future gamma-ray observations of dSphs from LSST+\fermi and for CTA in the right panel of figure~\ref{fig:post-process-45-dSph-2}. It is evident that LSST+\fermi will be able to test almost the entire $2\sigma$ CL region preferred by the 45 dSphs.

\subsubsection{Direct detection}
\begin{figure}[t]
    \centering
    
    \includegraphics[width=0.48\textwidth]{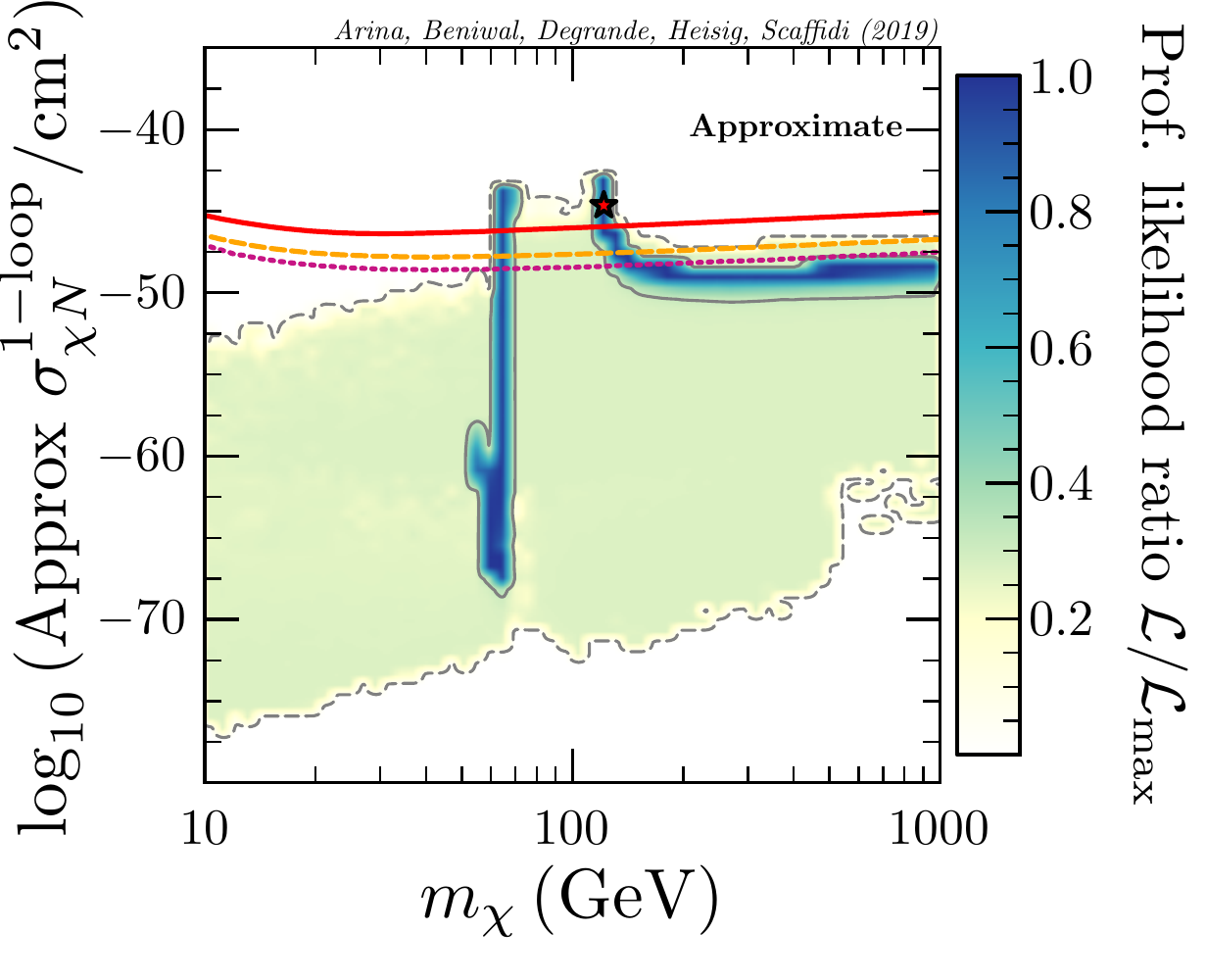} \quad
    \includegraphics[width=0.48\textwidth]{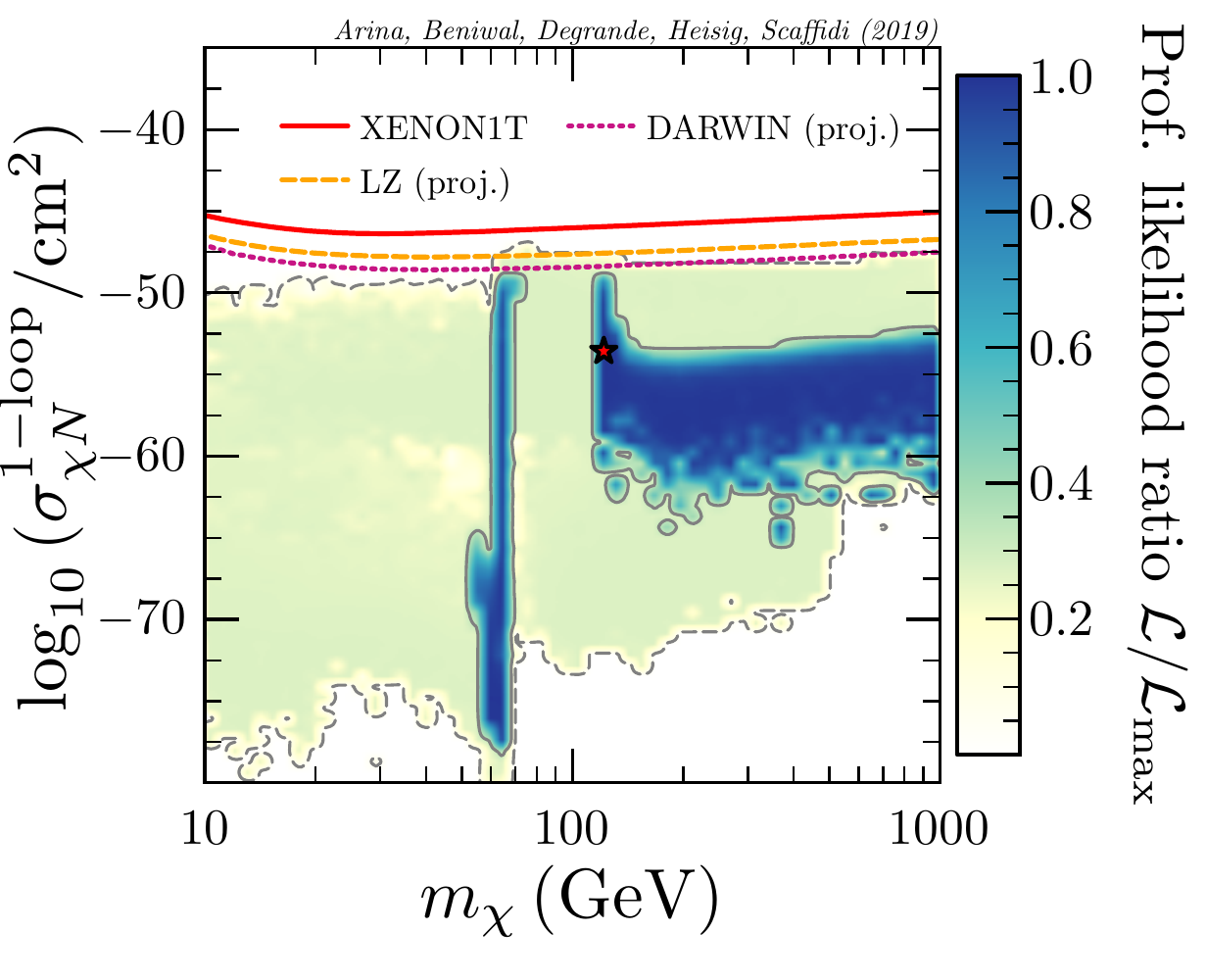} 
    \vspace{-1cm}
    \caption{2D PLR plots for the pNG DM-nucleon cross section at one-loop level using approximate expression (\emph{left panel}) and full computation (\emph{right panel}).~The solid red curve shows the current exclusion limit of XENON1T \cite{Aprile:2018dbl}, whereas the dashed orange and dotted magenta curves show the projected sensitivities of LUX-ZEPLIN (LZ) \cite{Akerib:2018lyp} and DARWIN \cite{Aalbers:2016jon}, respectively.}
    \label{fig:dd_plots}
\end{figure}

In figure~\ref{fig:dd_plots}, we show the PLR plots for the pNG DM-nucleon cross section at one-loop level after post-processing our \multinest samples.~These cross sections are based on the approximate expressions (\emph{left panel}) and full computations (\emph{right panel}), i.e.,~eqs.~\eqref{eqn:one-loop-approx} and \eqref{eqn:one-loop-exact}, respectively. The solid red curve shows the current sensitivity of XENON1T \cite{Aprile:2018dbl}, whereas the dashed orange and dotted magenta curves show the projected sensitivities of LZ \cite{Akerib:2018lyp} and DARWIN \cite{Aalbers:2016jon}, respectively.

The approximate cross section overestimates the full one-loop prediction up to several orders of magnitude. For instance, parts of the $1\sigma$ CL region in the left panel are already excluded by XENON1T, while they are currently allowed when considering the full one-loop computation.~In fact, we find that the entire $2\sigma$ CL region is not challenged by the current limits from XENON1T\@. Even the projected LZ and DARWIN experiments will probe only a small portion of the $2\sigma$ CL region.~The best-fit point lies completely out of reach of these experiments. In particular, the resonance region, $m_\chi \simeq m_h/2$, predicts a DM-nucleon cross section that is smaller than $\sim 10^{-50}$\,cm$^2$ and lies well below the proposed neutrino floor \cite{Boehm:2018sux}.~It is still interesting to see that upcoming generation of direct detection experiments are starting to probe models of DM with momentum-suppressed tree-level cross-section.

\section{Conclusions}\label{sec:conclusions}
We performed a global fit of the pNG DM model by combining constraints from the DM relic abundance, perturbative unitarity, Higgs invisible decay, electroweak precision observables and Higgs searches at colliders.~We presented our results in both frequentist and Bayesian statistical frameworks. In addition, we post-processed our samples by imposing indirect detection constraints from \fermi dwarf spheroidal galaxies within the former framework. Furthermore, we computed the one-loop pNG DM-nucleon cross sections, and compared the resulting values against the current limit from XENON1T (2018), and projected future limits from LUX-ZEPLIN (LZ) and DARWIN. 

In the frequentist analysis, we found two main regions with similar profile likelihood ratio that are compatible with all observations: the Higgs funnel region where DM annihilates resonantly via one of the two Higgs bosons, $m_\chi\sim m_{h,\,H}/2$, and the region of dominant annihilation into Higgs pairs, $m_\chi\gtrsim m_{h,\,H}$.~In contrast, the region of non-resonant annihilation into SM fermions and gauge bosons is highly constrained and mostly falls outside the $2\sigma$ CL region, in particular, for DM masses below the resonant region where the annihilation cross section is suppressed and requires non-perturbative couplings to match the measured relic density.

Electroweak precision observables, LEP searches and observed Higgs signal strengths at the LHC impose strong constraints on the mixing angle $\theta$ between the two Higgs bosons in our model. They require $\theta \lesssim 0.1$~rad except for the mass degenerate case $m_h\sim m_H$, where large mixing angles are allowed as well. In fact, the observed Higgs signal strength exhibit a slight preference (around $1\sigma$) for the latter choice. However, this preference arises from the fact that the LHC signal strengths are better fitted with a slightly heavier Higgs of around 125.3\,GeV while the SM Higgs mass is fixed at 125\,GeV in our scan. Hence, the fit prefers the second Higgs to have a mass of 125.3\,GeV and non-suppressed couplings to the SM particles.~We expect this preference to be alleviated if the SM Higgs mass is included as a nuisance parameter in the fit.

Our Bayesian results led to an even stronger constraint after marginalisation over the free model parameters. In particular, regions with a smaller volume of support fell outside the $2\sigma$ credible interval.~For instance, this concerns regions where the annihilation cross section today is much larger or smaller than the canonical freeze-out cross section $\langle \sigma v \rangle_\text{FO}\sim 3 \times 10^{-26}$\,cm$^3$ s$^{-1}$, arising very close to the resonant condition $m_\chi \sim m_{h,\,H}/2$ or the threshold $m_\chi \sim m_{h,\,H}$.~Similarly, our Bayesian results do not imply a preference for large mixing angles $\theta$ induced by Higgs signal strength observations, as it requires $m_H$ to be very close to $125.3$\,GeV, again, providing a small volume of support.

We computed the pNG DM-nucleon cross section at one-loop level for all of our samples after utilising the results of ref.~\cite{Azevedo:2018exj}.~We found that none of the points in our scan are challenged by current direct detection limits from XENON1T (2018).~Future based experiments (e.g., LUX-ZEPLIN, DARWIN) will only probe a small portion of the $2\sigma$ CL region.

We took into account the \fermi\ likelihood by considering two different sets of dSphs. On the one hand, we considered those imposing an upper limit on the annihilation cross section only (41 dSph). The effect on the parameter space is mild and the DM mass is not constrained towards large values within the consider range.~On the other hand, we considered all 45 dSphs analysed by \fermi, including the four dwarfs that show slight excesses at the level of $2\sigma$ each. These excesses can be well fitted within our model.~They favour a DM mass in range $30\,\text{GeV}\lesssim m_\chi\lesssim 300\,\text{GeV}$ at the $2\sigma$ CL.
We also expect a large part of this region to provide a good fit to the gamma-ray Galactic centre excess and the cosmic-ray antiproton excess seen in the AMS-02 data, if interpreted as a signal of DM annihilation.

Other indirect detection searches can further constrain our model. For instance, limits from AMS-02 antiprotons already exclude parts of the $1\sigma$ CL region in the 41-dSph fit with a DM mass around 400\,GeV.
Future gamma-ray observations by \fermi of newly discovered dSphs by LSST and CTA observations of the Galactic centre are expected to improve on the sensitivity and probe a significant portion of the allowed parameter space for DM masses above $m_h$ in the 41 dSphs fit. They are also expected to probe almost the entire $2\sigma$ CL region preferred by the current \fermi observations of all 45 dSphs.

\acknowledgments{We thank Peter Athron, Anders Kvellestad, Tim Stefaniak and Jonas Wittbrodt for helpful discussions regarding \textsf{HiggsBounds}/\textsf{HiggsSignals}.~We gratefully acknowledge Da Huang for providing access to his \textsf{C++} code for the calculation of one-loop pNG DM-nucleon cross section.~A.S.~thanks the CP3 (UCLouvain) staff for the hospitality offered during his visit where part of this work was completed.

Computational resources have been provided by the Consortium des \'{E}quipements de Calcul Intensif (C\'{E}CI), funded by the Fonds de la Recherche Scientifique de Belgique (F.R.S.-FNRS) under Grant No.~2.5020.11 and by the Walloon Region.~The work of A.B. and C.D. is supported by F.N.R.S. through the F.6001.19 convention.~A.S. and C.A.~are supported by ARC CoEPP (CE110001004) and Innoviris grant ATTRACT (2018) 104 BECAP 2, respectively. J.H.~acknowledges support from the F.R.S.-FNRS, of which he is a postdoctoral researcher.}

\appendix

\section{Dark matter-nucleon coupling}\label{app:dm-nucleon}
The dimensionful coupling between the mass eigenstates $(h,\,H)$ and pNG DM $\chi$ arises from eq.~\eqref{eqn:S_part}, namely
\begin{equation}\label{eqn:DM-h1-h2}
	\lagr_S \supset - \left(\lambda_{\Phi S}\, \Phi^\dagger \Phi |S|^2 + \frac{\lambda_S}{2} |S|^4 \right).
\end{equation}
After EWSB, this term expands to (keeping only terms proportional to $\phi \chi^2$ and $s \chi^2$)
\begin{align*}
	\lambda_{\Phi S} \, \Phi^\dagger \Phi |S|^2 &= \frac{\lambda_{\Phi S}}{4} (v_h + \phi)^2 \left[(v_s + s)^2 + \chi^2 \right] \supset \frac{\lambda_{\Phi S}}{4} (2 v_h \phi) \chi^2 = \frac{\lambda_{\Phi S} v_h}{2} \phi \chi^2, \\[1.5mm]
	\frac{\lambda_S}{2} |S|^4 &= \frac{\lambda_S}{8} \left[ (v_s + s)^2 + \chi^2 \right]^2 \supset \frac{\lambda_S}{8} \left[2 (v_s + s)^2 \chi^2 \right] \supset \frac{\lambda_S}{8} (4 v_s s) \chi^2 = \frac{\lambda_S v_s}{2} s \chi^2.
\end{align*}
Thus, eq.~\eqref{eqn:DM-h1-h2} can be expressed as
\begin{equation}\label{eqn:DM-h-s}
	\lagr_S \supset - \frac{1}{2} \chi^2 \left( \lambda_{\Phi S}  v_h \,\phi + \lambda_S v_s \,s \right).
\end{equation}
Using the following relation for the interaction eigenstates:
\begin{equation}
	\begin{pmatrix}
		\phi \\
		s 
	\end{pmatrix} = 
	\begin{pmatrix}
		\cos\theta & \sin\theta \\
		-\sin\theta & \cos\theta
	\end{pmatrix}
	\begin{pmatrix}
		h \\
		H
	\end{pmatrix},
\end{equation}
we can rewrite eq.~\eqref{eqn:DM-h-s} as
\begin{equation}\label{eqn:temp_DM_nucleon}
	\lagr_S \supset - \frac{1}{2} \chi^2 (\kappa_{\chi \chi h} \, h + \kappa_{\chi \chi H} \, H),
\end{equation} 
where the \emph{dimensionful} couplings $\{\kappa_{\chi \chi h},\,\kappa_{\chi \chi H}\}$ are \cite{Gross:2017dan}
{\small
    \begin{align}
	    \kappa_{\chi \chi h} = \left(\lambda_{\Phi S} v_h \cos\theta - \lambda_S v_s \sin\theta \right) &= \frac{1}{v_s} \left[ (m_H^2 - m_h^2) \sin \theta \cos^2 \theta - \sin\theta (m_h^2 \sin^2 \theta + m_H^2 \cos^2 \theta) \right] \nonumber \\
	    &= - \frac{m_h^2}{v_s} \sin\theta, \\[1.5mm]
	    \kappa_{\chi \chi H} = \left(\lambda_{\Phi S} v_h \sin\theta + \lambda_S v_s \cos\theta \right) &= \frac{1}{v_s} \left[(m_H^2 - m_h^2) \sin^2 \theta \cos \theta + \cos\theta (m_h^2 \sin^2 \theta + m_H^2 \cos^2 \theta) \right] \nonumber \\
	    &= + \frac{m_H^2}{v_s} \cos\theta.
    \end{align}
}
On the other hand, the Yukawa interaction term in the SM Lagrangian reads
\begin{equation}
	\lagr_{\textrm{Yukawa}} \supset - \phi \sum_f \frac{m_f}{v_h} \ovr{f} f = - \sum_f \kappa_{h f\ovr{f}} \, h \ovr{f} f + \kappa_{H f\ovr{f}} \, H \ovr{f} f,
\end{equation}
where 
\begin{equation}
	\kappa_{h f\ovr{f}} = \frac{m_f}{v_h} \cos\theta, \quad \kappa_{H f\ovr{f}} = \frac{m_f}{v_h} \sin \theta,
\end{equation}
are the \emph{dimensionless} couplings between $h/H$ and SM quarks/leptons.~Finally, the pNG DM-nucleon interaction Lagrangian can be written as
\begin{equation}
	\lagr_{\chi f} \supset - \frac{1}{2} \chi^2 (\kappa_{\chi \chi h} \, h + \kappa_{\chi \chi H} \, H) - \sum_f \kappa_{h f\ovr{f}} \, h \ovr{f} f + \kappa_{H f\ovr{f}} \, H \ovr{f} f.
\end{equation}

For a $\chi f \rightarrow \chi f$ scattering process via an $h/H$ exchange in $t$-channel, the tree-level direct detection (DD) scattering amplitude is proportional to
\begin{align*}
	\mathcal{A}_{\textrm{DD}}(q^2) \propto \frac{\kappa_{\chi \chi h} \kappa_{h f\ovr{f}}}{q^2 - m_h^2} + \frac{\kappa_{\chi \chi H} \kappa_{H f\ovr{f}}}{q^2 - m_H^2} \propto \sin\theta \cos\theta \left(\frac{m_H^2}{q^2 - m_H^2} - \frac{m_h^2}{q^2 - m_h^2}\right),
\end{align*}
where $q \equiv \sqrt{2 M E}$ is the momentum transfer and $M$ is the nucleus mass.~In the limit of $q^2 \ll m_{h,H}^2$, the above expression becomes
\begin{equation}
	\mathcal{A}_{\textrm{DD}}(q^2) \propto \sin\theta \cos\theta \left(\frac{1}{1 - q^2/m_h^2} - \frac{1}{1 - q^2/m_H^2}\right).
\end{equation}
For $x \ll 1$, the Taylor expansion for $(1 - x)^{-1} = 1 + x + \mathcal{O}(x^2)$. Thus, the above expression expands to
\begin{equation}
	\mathcal{A}_{\textrm{DD}}(q^2) \propto q^2 \sin\theta \cos\theta \left(\frac{1}{m_h^2} - \frac{1}{m_H^2}\right).
\end{equation}
As $q \sim \mathcal{O}(10)$\,MeV, the pNG DM-nucleon cross section is momentum-suppressed at tree-level.

\section{The $S$, $T$ and $U$ parameters}\label{app:oblique_pars}
In our model, the oblique parameters are shifted from their SM values by \cite{Grimus:2008nb,Beniwal:2018hyi}
\begin{align}
    \Delta T &= \frac{3}{16\pi s_W^2} \left[\cos^2\theta \left \{f_T    \left(\frac{m_h^2}{m_W^2}\right) - \frac{1}{c_W^2}f_T\left(\frac{m_h^2}{m_Z^2}\right) \right\} + \sin^2\theta \left\{f_T\left(\frac{m_H^2}{m_W^2} \right) \right. \right. \nonumber \\
    &\hspace{4mm} \left. \left. - \frac{1}{c_W^2} f_T\left(\frac{m_H^2}{m_Z^2} \right) \right\} - \left\{f_T\left(\frac{m_h^2}{m_W^2}\right) - \frac{1}{c_W^2}f_T\left(\frac{m_h^2}{m_Z^2}\right) \right\} \right], \label{eqn:del_T} \\[1.5mm]
    \Delta S &= \frac{1}{2\pi} \left[\cos^2\theta f_S \left(\frac{m_h^2}{m_Z^2}\right) + \sin^2\theta f_S \left(\frac{m_H^2}{m_Z^2}\right) - f_S \left(\frac{m_h^2}{m_Z^2}\right) \right], \label{eqn:del_S} \\[1.5mm]    
    \Delta U &= \frac{1}{2\pi} \left[\cos^2\theta f_S \left(\frac{m_h^2}{m_W^2}\right) + \sin^2\theta f_S \left(\frac{m_H^2}{m_W^2}\right) - f_S \left(\frac{m_h^2}{m_W^2}\right) \right] - \Delta S, \label{eqn:del_U}
\end{align}
where $\Delta \mathcal{O} \equiv \mathcal{O} - \mathcal{O}_{\textnormal{SM}}$ for $\mathcal{O} \in (S, T, U)$, $m_W\,(m_Z)$ is the $W\,(Z)$ boson mass, $c_W^2 = m_W^2/m_Z^2$ and $s_W^2 = 1 - c_W^2$. The loop functions $f_T(x)$ and $f_S(x)$ are given by
\begin{align}
    f_T(x) &= \frac{x\log x}{x-1}, \\[1.5mm]
    f_S(x) &= 
    \begin{dcases}
        \frac{1}{12} \left[ -2 x^2 + 9 x + \left((x-3) \left(x^2-4 x+12\right)+\frac{1-x}{x}\right) f_T(x) \right. \\ 
        \left. + 2 \sqrt{(4-x) x} \left(x^2-4 x+12\right)    
        \tan ^{-1}\left(\sqrt{\frac{4 - x}{x}}\right) \right], \quad 0 < x < 4, \\
        \frac{1}{12} \left[-2 x^2 + 9 x + \left((x-3) \left(x^2-4 x+12\right)+\frac{1-x}{x}\right)f_T(x) \right. \\
        \left. + \sqrt{(x-4) x} \left(x^2 - 4x + 12\right)  \log \left( \frac{x - \sqrt{(x-4)x}}{x + \sqrt{(x-4)x}} \right)\right], \quad x \geq 4. 
    \end{dcases} 
\end{align} 

\bibliographystyle{JHEP}
\bibliography{pNG_DM}

\end{document}